\documentclass[aps,prx,showpacs,twocolumn,amsmath,amssymb]{revtex4-1}
\usepackage{graphicx}
\usepackage{dcolumn}
\usepackage{bm}
\usepackage{amsmath}
\usepackage{amsfonts}
\usepackage{amssymb}
\usepackage{multirow}

\usepackage{amsthm}
\usepackage{pinlabel}
\usepackage{natbib}

\usepackage{epstopdf}
\usepackage[abs]{overpic}
\usepackage[dvipsnames]{xcolor}

\usepackage[dvipsnames]{xcolor}

\usepackage{braket}

\definecolor{specialgray}{HTML}{505050}
\definecolor{col10K}{HTML}{FFA000}
\definecolor{col300K}{HTML}{924FA4}
\definecolor{colMu}{HTML}{5278BD}
\definecolor{colMuI}{HTML}{924FA4}

\definecolor{specialgray}{HTML}{505050}
\definecolor{col10K}{HTML}{FFA000}
\definecolor{col300K}{HTML}{924FA4}
\definecolor{colMu}{HTML}{5278BD}
\definecolor{colMuI}{HTML}{924FA4}
\definecolor{newred}{HTML}{D53E4F}
\definecolor{newblue}{HTML}{5278BD}
\definecolor{newcyan}{HTML}{1EA0A0}
\definecolor{newgreen}{HTML}{5CB14E}
\definecolor{newpurple}{HTML}{924FA4}
\definecolor{newyellow}{HTML}{D1C72E}
\definecolor{neworange}{HTML}{D6923C}

\begin{document}

\title{Eliashberg theory for spin-fluctuations mediated superconductivity
-- Application to bulk and monolayer FeSe}
\author{Fabian Schrodi}\email{fabian.schrodi@physics.uu.se}
\author{Alex Aperis}\email{alex.aperis@physics.uu.se}
\author{Peter M. Oppeneer}\email{peter.oppeneer@physics.uu.se}
\affiliation{Department of Physics and Astronomy, Uppsala University, P.\ O.\ Box 516, SE-75120 Uppsala, Sweden}

\vskip 0.4cm
\date{\today}

\begin{abstract}
\noindent 
We present a novel method for embedding spin and charge fluctuations in an anisotropic, multi-band and full-bandwidth Eliashberg treatment of superconductivity.
Our analytical framework, based on the random phase approximation,
allows for a selfconsistent calculation of material specific characteristics in the interacting, and more specifically, the superconducting state. 
We apply this approach to bulk FeSe as representative for the iron-based superconductors  
and successfully solve for the superconducting transition temperature $T_c$, the gap symmetry and the gap magnitude. We obtain $T_c \approx 6$ K, consistent with experiment ($T_c \approx 8$ K), as well as other quantities in good agreement with experimental observations, thus supporting spin fluctuations mediated pairing in bulk FeSe.
On the contrary, applying our approach to monolayer FeSe on SrTiO$_3$ we find that spin fluctuations within the full Eliashberg framework give a $d$-wave gap with $T_c\le 11$ K and therefore cannot provide an explanation for a critical temperature as high as observed experimentally ($T_c \approx 70$ K). Our results hence point towards interfacial electron-phonon coupling as the dominant Cooper pairing mediator in this system.
\end{abstract}

\maketitle


\section{Introduction}\label{scIntroduction}

Ever since the discovery of superconductivity in iron-based compounds \cite{Kamihara2008,Hsu2008,Rotter2008,Medvedev2009} an enormous effort has been made to understand the prevailing mechanism responsible for Cooper pairing in these materials, both experimentally and theoretically (see Refs.\ \cite{Mazin2009,Johnston2010,Stewart2011,Wang2011,Dai2015,Si2016} and references therein).
For most members of this family the superconducting transition temperature is rather small compared to the high-$T_c$ cuprates, with a few exceptions such as monolayer FeSe on SrTiO$_3$ (FeSe/STO). There, the onset of superconductivity has been reported at temperatures as large as $60-100\,\mathrm{K}$ \cite{Qing-Yan2012,Liu2012,He2013,Tan2013,Peng2014,Lee2014,Ge2015}, which is an order of magnitude higher than in bulk FeSe ($\sim8\,\mathrm{K}$) \cite{Hsu2008,Jiao2017}. For many of the iron-based compounds there is consensus about an underlying unconventional mechanism responsible for superconductivity, although many theoretical investigations are so far based on the linearized Bardeen-Cooper-Schrieffer (BCS) equations at the Fermi level \cite{Scalapino1986,Carbotte2003,Mazin2009,Kubo2007,Graser2009,Kemper2010}. 
In this way the superconducting gap symmetry can presumably be obtained correctly, but other important experimental aspects, such as $T_c$ or the gap magnitude, remain partially elusive depending on the level of approximation, hampering hence also the unambiguous identification of the pairing mechanism. Thus there is an extensive need for a microscopic theory directly applicable to unconventional pairing, which can naturally provide the experimentally accessible characteristics of the system.

Superconductivity in bulk FeSe has been intensively studied since its discovery \cite{Hsu2008}.
As one of the intriguing properties of this material, a small nematic distortion of the tetragonal unit cell at low temperatures has been observed \cite{Margadonna2008} and its role to the complex superconductivity has been much discussed 
\cite{Fernandes2012,Boehmer2013,Fernandes2014,Kuo2016,Chubukov2016,Boehmer2018,Baek2020}. 
A second important feature of FeSe is that there is no long range magnetic order at ambient pressure \cite{Johnston2010,Stewart2011} yet there are enhanced spin fluctuations observed that signal a proximity to a magnetic phase transition \cite{Imai2009,Rahn2015,Wang2016NatMat,Wang2016NatCom}. A similar behavior is observed 
in several related Fe-based superconductors \cite{delaCruz2008,Christianson2009,Lumsden2009,Dai2015} but the spin fluctuations without local ordered moment are particularly strong in FeSe \cite{Wang2016NatCom}.
The spin fluctuations in Fe-based bulk superconductors have consequently been investigated theoretically with several approaches \cite{Yin2011NatPhys,Yin2011NatMat,Yin2014, Scherer2017}.
A further hint for an unconventional Cooper pairing mechanism  comes from \textit{ab initio} calculations  that predict a small effect of electron-phonon coupling in bulk iron pnictides \cite{Boeri2008,Nomura2014}.
Moreover, the appearance of superconductivity with an unusual $s_{\pm}$ gap symmetry in the Fe-based superconductors has drawn much theoretical attention (see, e.g.\ \cite{Mazin2008,Kuroki2008,Platt2011,Aperis2011}) and has in particular strengthened the picture of a close connection between spin fluctuations and superconductivity \cite{Mazin2009,Ikeda2010,Scalapino2012,Essenberger2012,Lischner2015,Essenberger2016}.
The most plausible scenario for bulk FeSe is dominant spin-fluctuations mediated pairing while nematicity in itself only modifies the magnetic fluctuations and thereby modifies the superconductivity indirectly
\cite{Boehmer2018,Baek2020}.

The case of monolayer FeSe on STO is markedly different from bulk FeSe. Due to substrate doping the Fermi surface of FeSe/STO does not exhibit the same nesting properties as FeSe \cite{He2013,Tan2013,Lee2014,Rebec2017} and the gap symmetry is plain $s$-wave instead of $s_{\pm}$ 
\cite{Fan2015}. As a consequence, there has been an intensive discussion
for FeSe/STO recently about the role of spin fluctuations and substrate phonons for the superconducting state \cite{Huang2017_2,Sadovskii2016}. On the one hand, it has been argued that a high-energy interfacial phonon mode can give rise to an enhanced coupling to FeSe electrons \cite{Lee2014,Rademaker2016,Song2019}. By imposing this assumption in a multiband full-bandwidth Eliashberg formalism many experimentally measured quantities can indeed be explained \cite{Aperis2018,Schrodi2018}. On the other hand, arguments have been presented 
in favor of an unconventional, 
spin-fluctuations mechanism \cite{Gao2017,Shishidou2018,Jandke2017}, e.g.\ via an incipient band scenario \cite{Linscheid2016} or orbital selective modifications in quasiparticle weights \cite{Kreisel2017}.
However, these previous theories have so far been formulated on the basis of approximations that are tailored to address one specific aspect of the problem.  For example, predictions within the incipient band scenario were based on solving isotropic two-band Eliashberg equations where only interband coupling was assumed and the $s_\pm$ symmetry of the gap was 
imposed as the only selfconsistent solution, which resulted in a very high $T_c$. 
In the case of the orbital selective scenario, the focus was on explaining the momentum anisotropy of the superconducting gap on the Fermi surface. This was achieved through a combination of static random phase approximation (RPA) and linearized BCS theory calculations \cite{Kreisel2017}. Evidently, to tackle the multiorbital spin-fluctuation problem at its full capacity, there is a need for a more generally applicable theoretical framework.

We develop here a full Eliashberg theory generalization of the multiorbital Hubbard-type model and show that it provides the amplitude, symmetry and momentum dependence of the superconducting gap over multiple bands, the renormalization of the electron mass and energy for all Brillouin zone (BZ) momenta, electronic energies and temperatures and therefore also the superconducting $T_c$; all calculated on the same footing. Our theory hence opens the door for treating both phononic and electronic pairing mechanisms on the same footing to settle the question about the dominant pairing mediator.


In the following 
we introduce a generic way of selfconsistently solving the anisotropic, full-bandwidth and multi-band Eliashberg equations for spin and charge fluctuations on an RPA level. Subsequently,
we apply our microscopic theory to bulk FeSe as a representative example for the iron-based superconductors. 
The only ingredient that is needed to selfconsistently solve for the 
critical temperature, the gap magnitude and its associated symmetry, is a tight-binding model reliably reproducing density functional theory (DFT) calculations for the electronic dispersions \cite{Eschrig2009}. In addition we examine the case of monolayer FeSe using modified electronic energies \cite{Hao2014}, while neglecting any possible influence of the substrate phonon. Our results for this material reveal a strong mismatch to experimental findings, in particular, a computed $T_c$ similar only to that of bulk FeSe. This leads us to the conclusion that spin fluctuations play a minor
role in FeSe/STO only for temperatures characteristic for superconductivity in the parent compound. We hence attribute the high $T_c$ to the characteristic interfacial electron-phonon coupling as extensively studied in previous works \cite{Lee2014,Rademaker2016,Zhang2017,Aperis2018,Schrodi2018}.


In the following we introduce in Sec.\ \ref{scMethod} the methodology used to compute spin- and charge-fluctuations mediated superconductivity in a full Eliashberg framework.  We then directly apply the method to bulk FeSe and compute several properties representative of its  superconducting state. In Sec.\ \ref{scMonolayer} we apply the same methodology to monolayer FeSe on STO. Finding that spin-fluctuations mediated pairing can explain superconductivity in bulk FeSe but not in FeSe/STO, we analyze deeper the origin of this result and compare to simplified approaches within BCS theory. 
Our conclusions on the plausible mechanisms for superconductivity in bulk FeSe and FeSe/STO are given in Sec.\  \ref{scDiscussion}.

\section{Methodology}\label{scMethod}

In this section we present a recipe of how to embed spin and charge fluctuations in a full-bandwidth, multiband and anisotropic Eliashberg theory. As a representative example for the iron-based superconductors we investigate bulk FeSe and solve for the main characteristics of this system in the superconducting state.

\subsection{Electronic energies of bulk FeSe}\label{scElectronicEnergies}

\begin{figure}[t!]
	\centering
	\includegraphics[width=1\columnwidth]{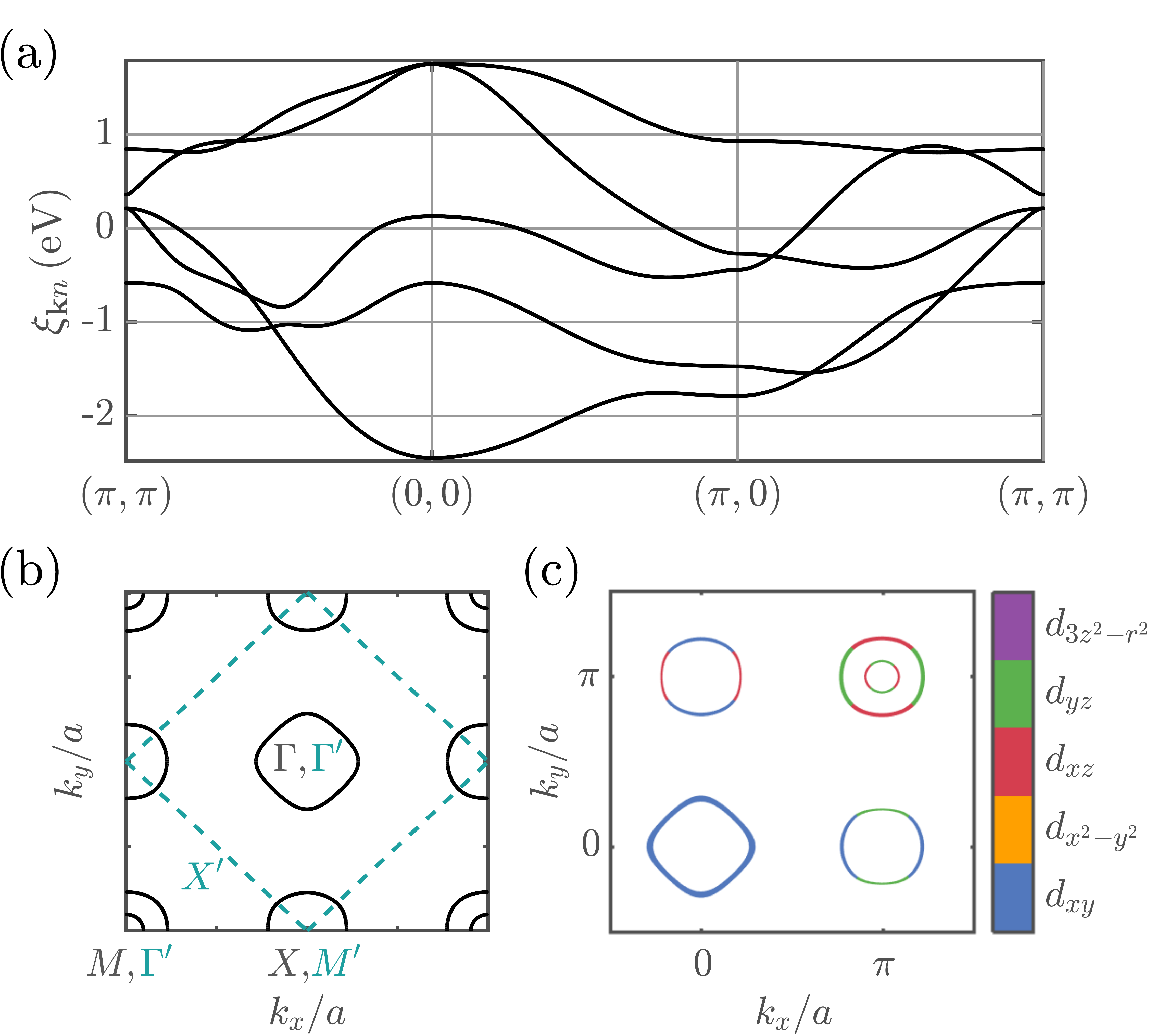}
	\caption{(a) Electronic bands $\xi_{\mathbf{k}n}$ of bulk FeSe along high-symmetry lines of the unfolded Brillouin zone. (b),(c) Fermi surface of $\xi_{\mathbf{k}n}$ in the unfolded BZ. (b) Connection to the folded BZ, drawn in cyan, and associated high-symmetry points. (c) Dominant orbital character depicted by different colors on the FS pockets.}
	\label{bulk_dispersion}
\end{figure}

The full system is modeled as $\hat{H}=\hat{H}_0+\hat{H}_{\mathrm{int}}$, where the interacting part $\hat{H}_{\mathrm{int}}$ is explained below in Section \ref{scSuscept}.
For the noninteracting part we consider a tight-binding model as introduced in Ref.\,\cite{Eschrig2009} that describes the electron band energies for the five iron $d$-orbitals
\begin{align}
 \hat{H}_0=\sum_{{\bf k}p\sigma}\xi_{{\bf k}p} \hat{c}^{\dagger}_{{\bf k}p\sigma}\hat{c}_{{\bf k}p\sigma}^{~} ,
\end{align}
where we use $\mathbf{k}$, $p$ and $\sigma$ as labels for momentum, orbital character and spin, respectively. $\hat{c}_{\mathbf{k}p\sigma}^{\dagger}$ ($\hat{c}_{\mathbf{k}p\sigma}$) are electronic creation (annihilation) operators and $\xi_{\mathbf{k}p}$ describes the dispersion in orbital space. From the diagonalization of $\hat{H}_0$ we find the band dependent energies $\xi_{\mathbf{k}n}$ as shown in Fig.\,\ref{bulk_dispersion}(a), and retrieve the matrix elements $a_{\mathbf{k}n}^p$, which serve as connection between band and orbital space. The tight-binding model used here has tetragonal symmetry, hence we assume that the influence of nematicity on the superconducting state is to first order negligible (cf.\ Ref.\ \cite{Baek2020}).
This assumption is to some extend justified because superconductivity does not compete with orthorhombicity in FeSe, as has been shown via hydrostatic pressure measurements \cite{Boehmer2013}. Here we work in the unfolded Brillouin zone corresponding to the one-FeSe unit cell in real space. The mapping to the folded BZ is indicated in Fig.\,\ref{bulk_dispersion}(b) by cyan dashed lines and the corresponding high-symmetry points. As explicitly shown in Fig.\,\ref{bulk_dispersion}(c) the hole band at $\Gamma$ has pure $d_{xy}$ orbital character, while $d_{xz}$ and $d_{yz}$ play important roles on the remaining FS sheets. 

With the just discussed dispersions $\xi_{\mathbf{k}n}$ and orbital weights $a_{\mathbf{k}n}^p$ we have the necessary tools at hand to calculate, within linear-response theory, the bare susceptibility of the system which is the same for both spin and charge channels. This in turn will be used for the calculation of the RPA interacting susceptibilities as we show in the next section.  Note that the basis vectors fulfill the orthonormality condition $\sum_{p}a_{\mathbf{k}n}^{p \, *}a_{\mathbf{k}n'}^p=\delta_{n,n'}$. 

\subsection{Linear response in the RPA approximation}\label{scSuscept}

\begin{figure}[t!]
	\centering
	\includegraphics[width=1\columnwidth]{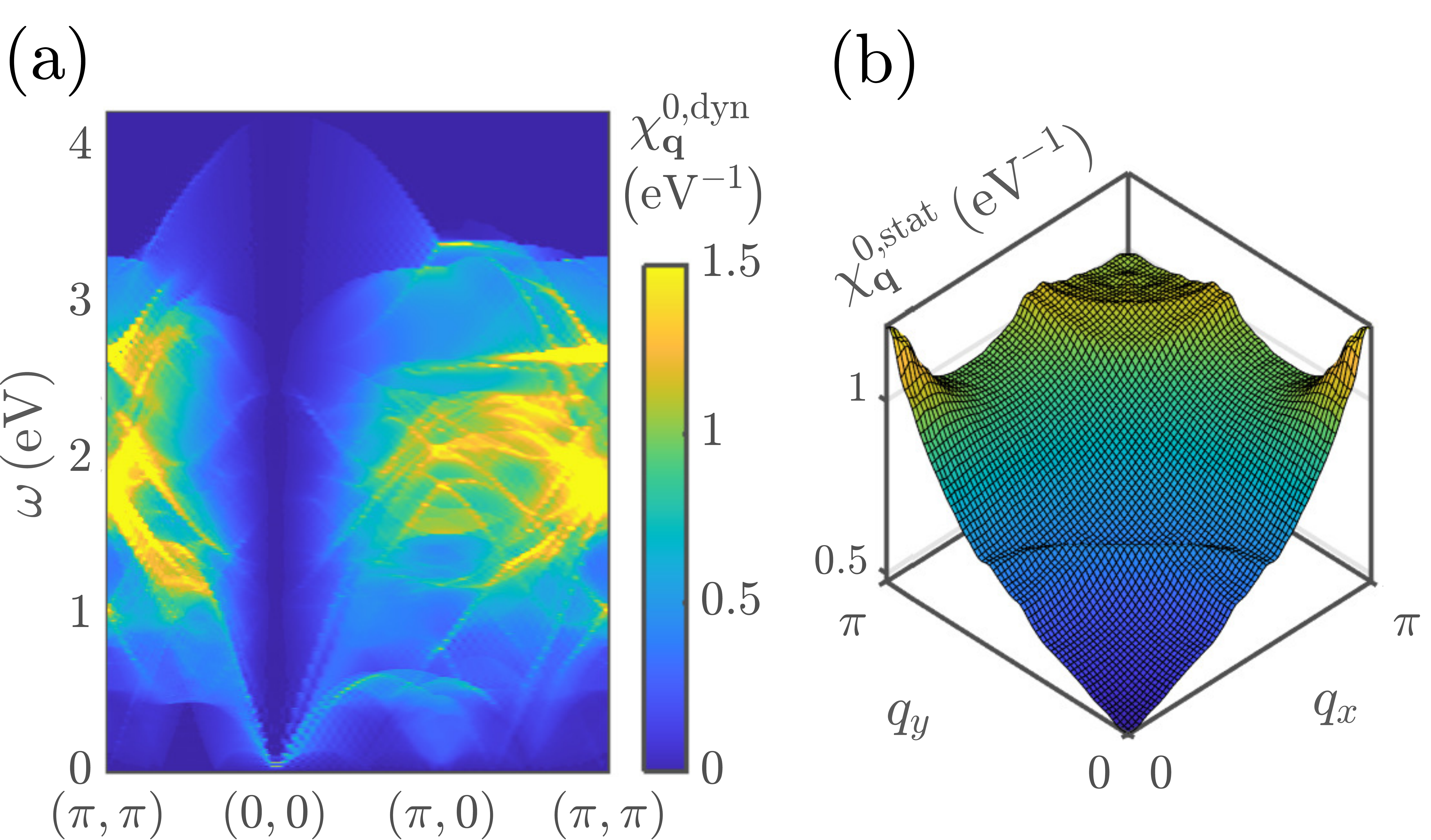}
	\caption{Bare susceptibility of FeSe as calculated from Eqs.\,(\ref{bubbleimag})-(\ref{statbubble}). (a) Dynamic bare susceptibility along high-symmetry lines and frequencies. (b) Static bare susceptibility plotted in the first quadrant of the BZ.}
	\label{bulk_dynbubble}
\end{figure}

The interaction part of our Hamiltonian is given by the intrasite Hubbard-type terms \cite{Oles1983, Takimoto2004, Kubo2007} 
\begin{align}
\hat{H}_{\mathrm{int}} = U\sum_{i,s}\hat{n}_{is\uparrow}\hat{n}_{is\downarrow} + \frac{V'}{2}\sum_{i,s,t\neq s}\hat{n}_{is}\hat{n}_{it} ~~~~~~~~~~~~~ \nonumber\\
-\frac{J}{2}\sum_{i,s,t\neq s}\hat{\vec{S}}_{is}\cdot\hat{\vec{S}}_{it} + \frac{J'}{2}\sum_{i,s,t\neq s,\sigma} \hat{c}^{\dagger}_{is\sigma}\hat{c}^{\dagger}_{is\bar{\sigma}} \hat{c}_{it\bar{\sigma}}^{~}\hat{c}_{it\sigma}^{~} ~, \label{hamiltonian}
\end{align}
where $\hat{\vec{S}}_{is}$ is the spin operator for orbital index $s$ at site $i$, compare Ref.\,\cite{Graser2009,Kuroki2009,Scalapino2012}. The occupation at site $i$ with electrons of orbital  index $s$,  for spin $\sigma$, is $\hat{n}_{is\sigma}=\hat{c}^{\dagger}_{is\sigma}\hat{c}_{is\sigma}^{~}$, which can be used to get $\hat{n}_{is}=\sum_{\sigma}\hat{n}_{is\sigma}$. Above we use $U$ ($V'$) as intraorbital (interorbital) onsite interaction, $J$ is the Hund's rule coupling and $J'$  the pair hopping energy. The interactions are related via $J'=J/2$ and $V'=U-3J/4-J'$, a choice consistent with related works \cite{Graser2009,Kemper2010,Kubo2007}. 
We start by calculating the imaginary part of the bare susceptibilities for real frequency $\omega$
\begin{align}\nonumber
{\rm Im}\left(\big[\chi^0_{{\bf q}}(\omega)\big]_{st}^{pq}\right) = -\pi \sum_{n,n',\mathbf{k}} a_{{\bf k}n}^s a_{{\bf k}n}^{p \,*} a_{{\bf k}+{\bf q}n'}^q a_{{\bf k}+{\bf q}n'}^{t \,*} ~~~~~  \\\label{bubbleimag}
\times\left[n_{\rm F}(\xi_{{\bf k}n}) - n_{\rm F}(\xi_{{\bf k}+{\bf q}n'})\right] \delta\big( \xi_{{\bf k}+{\bf q}n'} - \xi_{{\bf k}n} + \omega \big) , 
\end{align}
where we set $T=5\,$K to evaluate the Fermi-Dirac functions $n_F(\cdot)$, and keep this temperature from here on unless noted otherwise. As briefly discussed in Appendix \ref{appMethodDetails} it is a reasonable assumption to keep $T$ fixed with respect to bare susceptibilities. The delta function in Eq.\,(\ref{bubbleimag}) is treated by using an adaptive smearing method \cite{Yates2007}, see Appendix \ref{appMethodDetails} for details. The intraorbital sum 
\begin{align}
\chi^{0,\mathrm{dyn}}_{\mathbf{q}}(\omega) = \frac{1}{2}\sum_{p,s} {\rm Im}\left(\big[\chi^0_{\mathbf{q}}(\omega)\big]^{pp}_{ss}\right)  \label{dynbubble}
\end{align}
is shown in Fig.\,\ref{bulk_dynbubble}(a) as function of momenta and frequencies. It is apparent that the elementary excitations of our system range to rather large frequencies, that introduce a scale inappropriate to our low-energy theory. As we show below and in Section \ref{scSuperconducting}, in the interacting case the high-energy region gives rise to the Stoner continuum, which generally suppresses Cooper pair formation.

Next, we want to use the above results to obtain the real part of the bare susceptibility, which can be done by using the Kramers-Kronig relation for spectral functions,
\begin{align}
{\rm Re}\left(\big[\chi^0_{{\bf q}}(\omega)\big]_{st}^{pq}\right) = \frac{1}{\pi} \mathcal{P} \int_{-\infty}^{\infty} \frac{\mathrm{d}\omega'}{\omega'-\omega} {\rm Im}\left(\big[\chi^0_{{\bf q}}(\omega')\big]_{st}^{pq}\right) , \label{bubblereal}
\end{align}
where $\mathcal{P}$ denotes the principal value. From here we define the static bare susceptibility as
\begin{align}
\chi^{0,\mathrm{stat}}_{\mathbf{q}} = \frac{1}{2}\sum_{p,s} {\rm Re}\left(\big[\chi^0_{{\bf q}}(0)\big]_{ss}^{pp}\right) , \label{statbubble}
\end{align}
which,  as in Eq.\,(\ref{dynbubble}), is calculated from the intra-orbital components only. 

In Fig.\,\ref{bulk_dynbubble}(b) we observe dominant peaks at $\mathbf{q}=X$  (=$(0, \pi)$) when plotting $\chi^{0,\mathrm{stat}}_{\mathbf{q}}$ in the first quadrant of the BZ. This is easily explained by the relatively enhanced nesting between electron and hole FS pockets, compare Fig.\,\ref{bulk_dispersion}. Similarly, hole-hole and electron-electron nesting features at momenta slightly smaller than $(\pi,\pi)$  give rise to the two rings around $M$ (=$(\pi, \pi)$) in panel \ref{bulk_dynbubble}(b). We note that such a bare  susceptibility as shown in Fig.\,\ref{bulk_dynbubble} is rather generic for the family of Fe-based superconductors, see for example Refs.\,\cite{Graser2009,Kreisel2017}, and that qualitatively comparable results have been obtained by DFT calculations, too\,\cite{Heil2014}.

Within the RPA  the spin and charge susceptibilities are defined via Dyson equations,
\begin{align}
\big[\chi^S_{\mathbf{q}}(\omega)\big]^{pq}_{st} = \big[\chi^0_{\mathbf{q}}(\omega)\big]^{pq}_{st} ~~~~~~~~~~~~~~~~~~~~~~~~~~~~~ \nonumber \\
+ \sum_{uvwz} \big[\chi^S_{\mathbf{q}}(\omega)\big]^{pq}_{uv}\big[U^S\big]^{uv}_{wz}\big[\chi^0_{\mathbf{q}}(\omega)\big]^{wz}_{st} , \label{rpaspin} \\
\big[\chi^C_{\mathbf{q}}(\omega)\big]^{pq}_{st} = \big[\chi^0_{\mathbf{q}}(\omega)\big]^{pq}_{st} ~~~~~~~~~~~~~~~~~~~~~~~~~~~~~ \nonumber \\
- \sum_{uvwz} \big[\chi^C_{\mathbf{q}}(\omega)\big]^{pq}_{uv}\big[U^C\big]^{uv}_{wz}\big[\chi^0_{\mathbf{q}}(\omega)\big]^{wz}_{st}, \label{rpacharge}
\end{align}
with Stoner tensors $U^S$ for spin and $U^C$ for charge. The nonzero elements are given by
\begin{align}
&[U^{S}]_{aa}^{aa}=U ~,~ [U^S]_{bb}^{aa}=\frac{J}{2} ~, [U^S]_{ab}^{ab}=\frac{J}{4}+V' , \nonumber \\
&[U^{S}]_{ab}^{ba}=J' ~,~ [U^{C}]_{aa}^{aa}=U ~,~ [U^C]_{bb}^{aa}=2V' ~, \nonumber \\
&[U^C]_{ab}^{ab}=\frac{3J}{4}-V' ~,~ [U^{C}]_{ab}^{ba}=J' ~. \label{stonerparam}
\end{align}
We solve Eqs.\,(\ref{rpaspin}) and (\ref{rpacharge}) by mapping all four-rank tensors involved to usual (two-rank) matrices, for example $\big[U^S\big]^{pq}_{st}\rightarrow\big[U^S\big]_{ij}\equiv\hat{U}^S$. This leads to matrix equations
\begin{align}
\hat{\chi}^S_{\mathbf{q}}(\omega) = \hat{\chi}^0_{\mathbf{q}}(\omega) \big[\hat{1}  - \hat{U}^S\hat{\chi}^0_{\mathbf{q}}(\omega) \big]^{-1} , \label{rpaMatS} \\
~ \hat{\chi}^C_{\mathbf{q}}(\omega)  = \hat{\chi}^0_{\mathbf{q}}(\omega) \big[\hat{1} + \hat{U}^C\hat{\chi}^0_{\mathbf{q}}(\omega)\big]^{-1} , \label{rpaMatC}
\end{align}
which allow us to directly identify the Stoner instabilities. More explicitly, we define static susceptibilities,  in analogy to Eq.\,(\ref{statbubble}), as
\begin{align}
\chi^{S,\mathrm{stat}}_{\mathbf{q}} = \frac{1}{2}\sum_{p,s} {\rm Re}\left(\big[\chi^S_{{\bf q}}(0)\big]_{ss}^{pp}\right) , \\
\chi^{C,\mathrm{stat}}_{\mathbf{q}} = \frac{1}{2}\sum_{p,s} {\rm Re}\left(\big[\chi^C_{{\bf q}}(0)\big]_{ss}^{pp}\right) ,
\end{align}
where the aforementioned mapping is inverted to retrieve back the four-rank tensors corresponding to outcomes of Eqs.\,(\ref{rpaMatS}) and (\ref{rpaMatC}). 

\begin{figure}[b!]
	\centering
	\includegraphics[width=1\columnwidth]{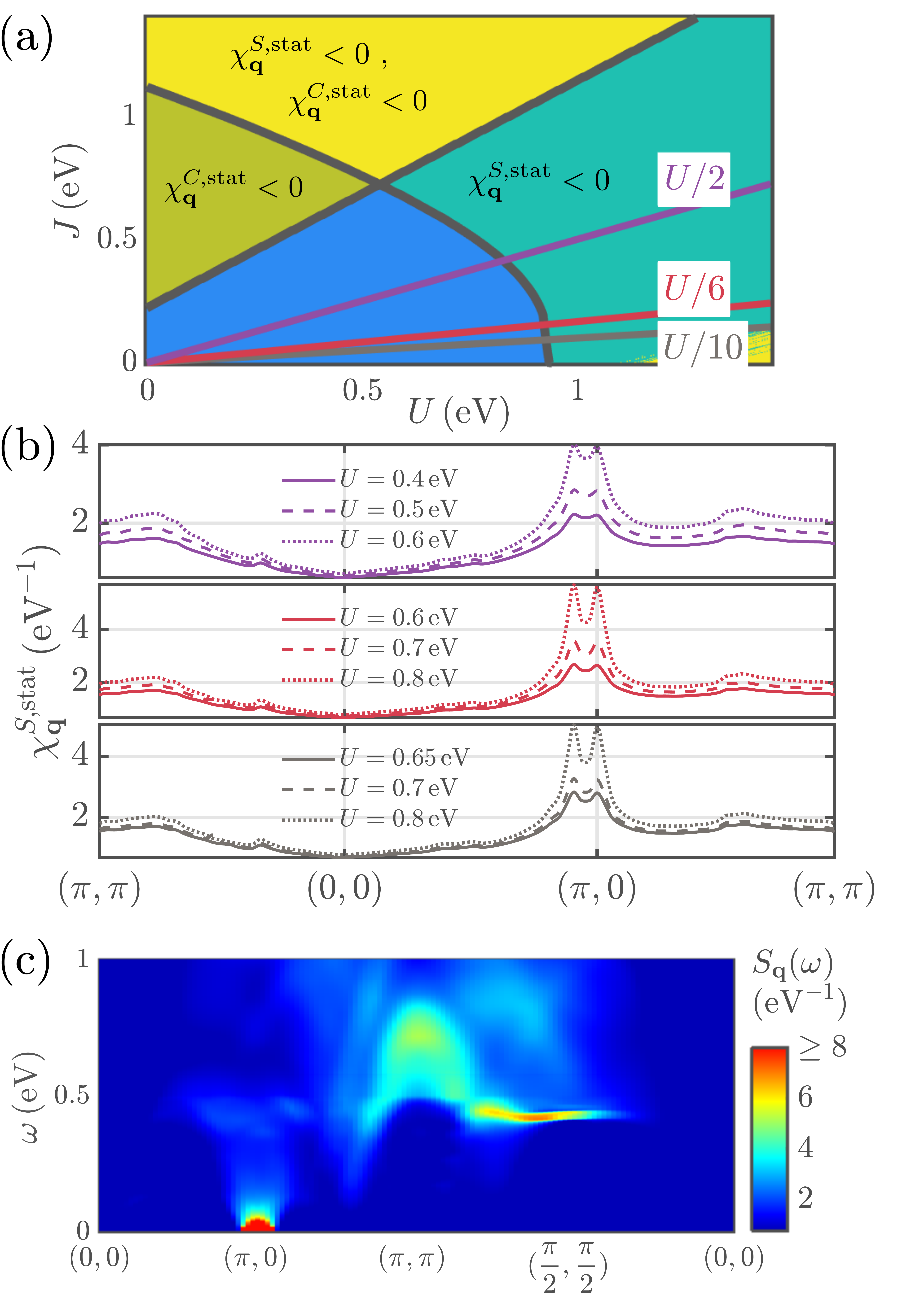}
	\caption{(a) Calculated phase diagram of allowed values for $(U,J)$ considering the Stoner criterion. The blue area denotes the allowed phase space, all remaining parts are forbidden (reason explicitly written). The purple, red and gray lines refer to representative ratios $J=U/2$, $J=U/6$, and $J=U/10$, which we consider here (see text). (b) Static spin susceptibilities for increasing values of $U$ along high-symmetry lines of the BZ. The upper, middle and bottom panels show our results for decreasing Hund's rule coupling, with similar $U/J$ ratios and color code as in (a). (c) Spin structure factor for bulk FeSe as function of frequencies and momenta, calculated from Eqs.\,(\ref{dynamicSpin}) and (\ref{structureFactor}) at $T=5\,$K, $J=U/2$ and $U=0.827\,\mathrm{eV}$.}
	\label{bulk_statRPA}
\end{figure}

The requirement $\chi^{S,\mathrm{stat}}_{\mathbf{q}}>0$ and $\chi^{C,\mathrm{stat}}_{\mathbf{q}}>0$ $\forall\mathbf{q}$, i.e.\ staying below the first Stoner instability, defines a finite region of allowed values in $(U,J)$-space.
In Fig.\,\ref{bulk_statRPA}(a) we plot the resulting phase diagram, where the allowed region is drawn in blue. For all remaining parts the Stoner criterion is violated due to spin (cyan), charge (green) or spin {and} charge (yellow). We test three different ratios for bulk FeSe, each indicated by a solid line in Fig.\,\ref{bulk_statRPA}(a): $J=U/10$ (gray line) describes strongly localized electrons, $J=U/2$ (purple) resembles a Hund-metal situation, and $J=U/6$ (red) is a reasonable intermediate choice.

The RPA susceptibilities do not change appreciably with $(U,J)$. This can be explicitly seen in Fig.\,\ref{bulk_statRPA}(b) where we show spin results for all three ratios and varying distance from the border of the allowed region in panel (a), using similar color code for $U/J$. In all three cases we observe two peaks near the $X$ point. These have their origin in the bare susceptibility (compare Fig.\,\ref{bulk_dynbubble}(b)) and are enhanced when approaching the Stoner instability. From this behavior one can directly conclude that momenta around $\mathbf{q}=X$ give rise to the leading instability and will approximately become delta-peaks in the vicinity of the spin-border in $(U,J)$-space. Further we find increased susceptibilities with growing $U$, while a change in $J$ leads barely to noticeable modifications. Since we are not interested in cases where $J>U$ there is no need of explicitly discussing the RPA charge susceptibility. The changes are minor in this quantity because we stay always well separated from $\chi_{\mathbf{q}}^{C,\mathrm{stat}}<0$.

In the following we want to look at the frequency dependence of Eqs.\,(\ref{rpaMatS}) and (\ref{rpaMatC}). Let us therefore define the dynamic spin and charge susceptibilities as 
\begin{align}
\chi^{S,\mathrm{dyn}}_{\mathbf{q}}(\omega) &= \frac{1}{2}\sum_{p,s} {\rm Im}\left(\big[\chi^S_{\mathbf{q}}(\omega)\big]^{pp}_{ss}\right) , \label{dynamicSpin}\\ \chi^{C,\mathrm{dyn}}_{\mathbf{q}}(\omega) &= \frac{1}{2}\sum_{p,s} {\rm Im}\left(\big[\chi^C_{\mathbf{q}}(\omega)\big]^{pp}_{ss}\right) . \label{dynamicCharge}
\end{align}
Note, that these dynamical susceptibilities  always pertain to the imaginary parts in our notation.
For $J=U/2$ and a rather critical value of $U=0.827\,\mathrm{eV}$ we plot in Fig.\,\ref{bulk_statRPA}(c) the spin structure factor
\begin{align}
S_{\mathbf{q}}(\omega) = \frac{\chi^{S,\mathrm{dyn}}_{\mathbf{q}}(\omega)}{1 - e^{-\hbar\omega/k_BT}} ~, \label{structureFactor}
\end{align}
calculated from Eq.\,(\ref{dynamicSpin}).

We can compare the result of our calculations with
the outcome of DFT-dynamical mean field theory (DFT-DMFT) calculations carried out in Ref.\,\cite{Yin2014}. Although the two approaches are rather different, 
the main characteristics of $S_{\mathbf{q}}(\omega)$ are actually similar. Starting with the maximum value of $S_{\mathbf{q}}(\omega)$, which is $22\,\mathrm{eV}^{-1}$ in Fig.\,\ref{bulk_statRPA}(c) and $16\,\mathrm{eV}^{-1}$ as obtained in Ref.\,\cite{Yin2014}. Note that our results are scalable with respect to criticality, hence we could fine-tune $U$ to achieve the same maximum structure factor. Both our calculation and that of Ref.\,\cite{Yin2014} reveal an enhanced
contribution at $X$ for small frequencies, as well as substantial values for $S_{\mathbf{q}}(\omega)$ at $(\pi,\pi)$ and $(\pi/2,\pi/2)$ for larger $\omega$. There are differences along the frequency axis that can be attributed partially to deviating choices for $T$ and $U/J$, but mainly stems from the different ways how $S_{\mathbf{q}}(\omega)$ is calculated\,\cite{Yin2014,Haule2010}. From the above comparison we can conclude that our results for the spin and charge susceptibilities show the correct main features.

\subsection{Coupling via the spin and charge sectors}\label{scCoupling}

\begin{figure}[b!]
	\centering
	\includegraphics[width=1\columnwidth]{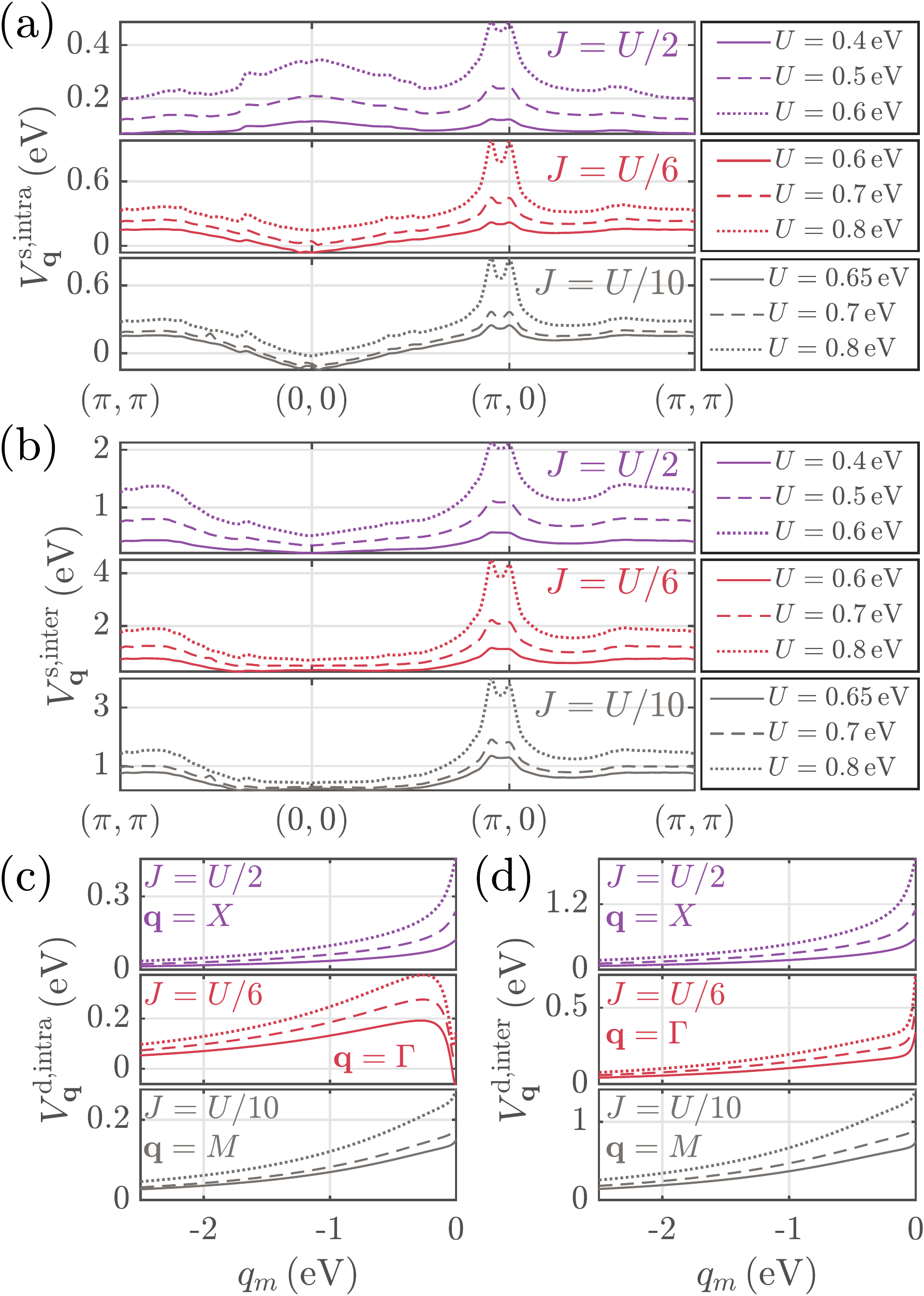}
	\caption{(a),(b) Computed static kernels in the superconducting channel for growing values of $U$. In both panels, the upper, middle and bottom graphs show the outcomes for $J=U/2$, $J=U/6$ and $J=U/10$, respectively. (a) Intra-band kernels from Eq.\,(\ref{matkernSIntra}). (b) Inter-band kernels from Eq.\,(\ref{matkernSInter}). (c),(d) Dynamic kernels as function of bosonic Matsubara frequencies for different $\bf q$ points. First row: $\mathbf{q}=X$ and $J=U/2$; second row: $\mathbf{q}=\Gamma$ and $J=U/6$; third row: $\mathbf{q}=M$ and $J=U/10$. (c) Intra-band kernels from Eq.\,(\ref{matkernDIntra}). (d) Inter-band kernels from Eq.\,(\ref{matkernDInter}).}
	\label{bulk_dynKernel}
\end{figure}

Before turning to the full Eliashberg problem we need to calculate band dependent interaction kernels in Matsubara space. To achieve this we first define
\begin{align}
\big[V^{(+)}_{{\bf q}}(\omega)\big]^{pq}_{st} =& \Big[\frac{3}{2}U^S\chi_{{\bf q}}^{S}(\omega)U^S + \frac{1}{2}U^C\chi^C_{{\bf q}}(\omega)U^C \Big]^{tq}_{ps} , \label{orbitalkernP} \\
\big[V^{(-)}_{{\bf q}}(\omega)\big]^{pq}_{st} =& \Big[\frac{3}{2}U^S\chi_{{\bf q}}^{S}(\omega)U^S + \frac{1}{2}U^S \nonumber\\
&- \frac{1}{2}U^C\chi^C_{{\bf q}}(\omega)U^C + \frac{1}{2}U^C \Big]^{tq}_{ps}   \label{orbitalkernM} ,
\end{align}
where we distinguish between a kernel for electron mass and energy dispersion renormalization $(+)$ and the superconducting pairing $(-)$, corresponding to diagonal and off-diagonal elements of the Green's function, respectively. Setting $\mathbf{q}=\mathbf{k}-\mathbf{k}'$ and averaging over $\mathbf{k}$ we transform these kernels from orbital into band space:
\begin{align}
\big[V^{(\pm)}_{{\bf q}}(\omega)\big]_{nn'} = \! \sum_{\mathbf{k}}\sum_{stpq} a_{{\bf k}n}^{t \,*} a_{{\bf k}n}^{s \,*} \big[V^{(\pm)}_{{\bf q}}(\omega)\big]^{pq}_{st} a^p_{{\bf k}-\mathbf{q}n'}a^q_{{\bf k}-\mathbf{q}n'} .\label{bandkern}
\end{align}
Since we are interested in imaginary frequencies on the Matsubara axis, we transform the result of Eq.\,(\ref{bandkern}) via the Kramers-Kronig relation
\begin{align}
\big[V^{(\pm)}_{{\bf q}}(iq_m)\big]_{nn'} = \frac{1}{\pi}\mathcal{P}\!  \int_{-\infty}^{\infty} \frac{\mathrm{d}\omega}{\omega-iq_m} {\rm Im} \! \left( \big[V^{(\pm)}_{{\bf q}}(\omega)\big]_{nn'} \right) , \label{matsubarakern}
\end{align}
where $q_m=2\pi Tm$ is a bosonic frequency. In analogy to Fig.\,\ref{bulk_statRPA}(b) we want to understand the influence of $U$ and $J$ on the interaction kernels. To this end we define static intra- and inter-band contributions as
\begin{align}
V^{\mathrm{s,intra}}_{\mathbf{q}} &= \frac{1}{2} \sum_n \big[V^{(-)}_{{\bf q}}(0)\big]_{nn} ,\label{matkernSIntra} \\
V^{\mathrm{s,inter}}_{\mathbf{q}} &= \frac{1}{2} \sum_{{n\neq n'}} \! \big[V^{(-)}_{{\bf q}}(0)\big]_{nn'} , \label{matkernSInter}
\end{align}
where the focus lies on the kernel of the superconducting channel. For similar choices of $(U,J)$ as in Fig.\,\ref{bulk_statRPA}(b) (same color code) we show the outcome of Eqs.\,(\ref{matkernSIntra}) and (\ref{matkernSInter}) in Fig.\,\ref{bulk_dynKernel}, panels (a) and (b), respectively. 

As an overall trend one observes increasing kernel values with growing $U$. In contrast to the spin susceptibilities plotted in Section \ref{scSuscept}, this increase applies to values throughout the whole BZ. While peaks at $X$ are still dominant in all panels of both figures, $V^{\mathrm{s,intra}}_{\mathbf{q}}$ grows large also at $\Gamma$ with increasing $J$ and $U$. Furthermore, a nonnegligible background develops as we slowly approach the Stoner instability (the values shown correspond to noncritical regions). This is caused by taking  the full Stoner continuum into account when transforming from real to Matsubara frequencies and causes potential problems for Cooper pair formation due to e.g.\ frustration effects, see Section \ref{scSuperconducting} for details.

\begin{figure*}[ht!]
	\centering
	\includegraphics[width=1\textwidth]{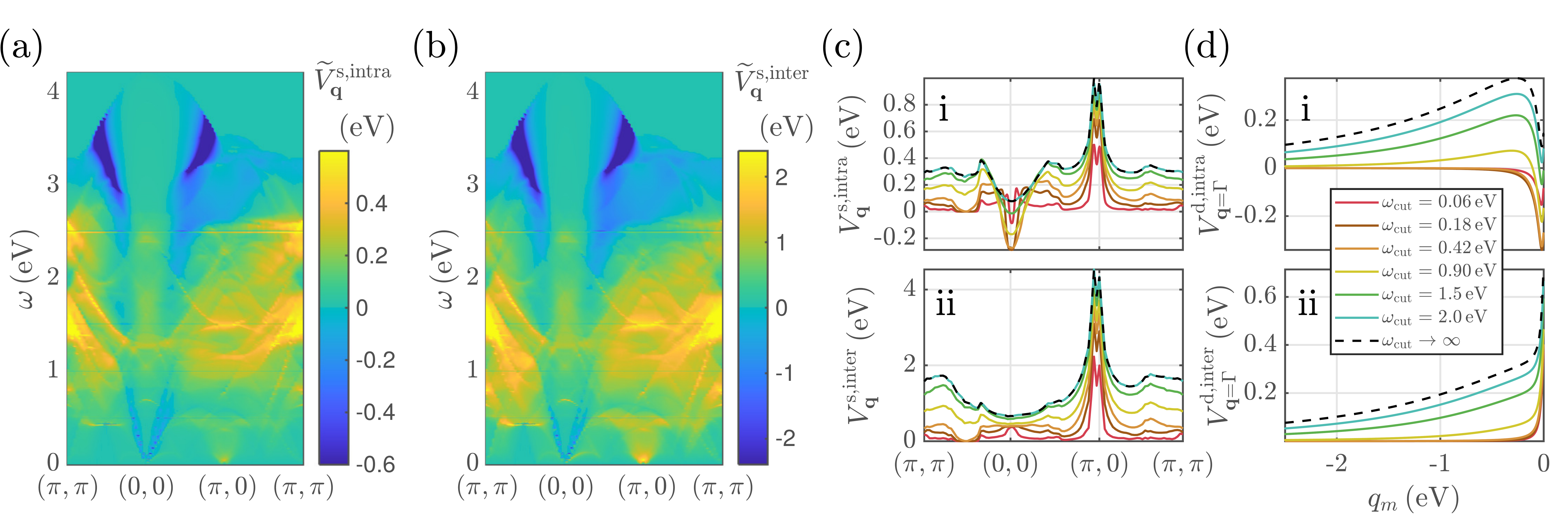}
	\caption{Influence of the truncation parameter $\omega_{\mathrm{cut}}$ used in Eq.\,(\ref{bandkernTrunc}) on intra- and inter-band kernels, with $U=0.8\,\mathrm{eV}$ and $J=U/6$. (a)  Real frequency dependence of $\widetilde{V}^{\mathrm{d,intra}}_{\mathbf{q}}$ as obtained from Eq.\,(\ref{matkernDIntraRealFreq}), along high symmetry points. (b) Same as (a) for $\widetilde{V}^{\mathrm{d,inter}}_{\mathbf{q}}$. (c), (d) Matsubara frequency kernels for various choices of $\omega_{\mathrm{cut}}$, using identical color codes. Rows {\bf i} ({\bf ii}) refer to intra- (inter-) band terms. (c) Static kernels from Eqs.\,(\ref{matkernSIntra}) and (\ref{matkernSInter}). (d) Matsubara frequency dependent dynamic kernels as obtained 
		from Eqs.\,(\ref{matkernDIntra}) and (\ref{matkernDInter}).}
	\label{bulk_cutDependentKern}
\end{figure*}
Next we focus on the Matsubara axis and define frequency dependent counterparts to Eqs.\,(\ref{matkernSIntra}) and (\ref{matkernSInter}) via
\begin{align}
V^{\mathrm{d,intra}}_{\mathbf{q}}(iq_m) &= \frac{1}{2} \sum_n \big[V^{(-)}_{{\bf q}}(iq_m)\big]_{nn} , \label{matkernDIntra} \\
V^{\mathrm{d,inter}}_{\mathbf{q}}(iq_m) &= \frac{1}{2} \sum_{n \neq n'} \! \big[V^{(-)}_{{\bf q}}(iq_m)\big]_{nn'} \label{matkernDInter} .
\end{align}
The results are drawn in Fig.\,\ref{bulk_dynKernel}, where panel (c) and (d) corresponds to the intra- and inter-band terms, respectively. In the first, second and third row of these two panels we show respectively the pairs $(\mathbf{q},J)=(X,U/2)$, $(\Gamma,U/6)$, and $(M,U/10)$. The aforementioned influence of the Stoner continuum is reflected along $q_m$ in all panels. With growing $U$ the kernels are enhanced along the full frequency axis resulting in slowly decaying tails, which in turn increases the computational load of Eliashberg calculations significantly. We further observe that a moderate Hund's rule coupling of $J=U/6$ can lead to attractive intra-band kernels at $\mathbf{q}=\Gamma$, as seen in the central panel of Fig.\,\ref{bulk_dynKernel}(c). This sign change in $V^{\mathrm{d,intra}}_{\mathbf{q}=\Gamma}(iq_m)$ is lost as soon as $U$ approaches the Stoner instability. Such properties, i.e.\ the interplay of repulsive and attractive couplings can have an important impact on Cooper pair formation and the superconducting gap symmetry.

To resolve the aforementioned difficulty concerning the Stoner continuum we introduce a frequency cutoff $\omega_{\mathrm{cut}}>0$ to truncate the integration in Eq.\,(\ref{matsubarakern}). The Matsubara frequency dependent kernels we treat from here on are therefore
\begin{align}
\big[V^{(\pm)}_{{\bf q}}(iq_m)\big]_{nn'} =\!  \frac{1}{\pi}\mathcal{P}\!  \int_{-\omega_{\mathrm{cut}}}^{\omega_{\mathrm{cut}}} \frac{\mathrm{d}\omega}{\omega-iq_m} {\rm Im} \!  \left( \big[V^{(\pm)}_{{\bf q}}(\omega)\big]_{nn'} \right). \label{bandkernTrunc}
\end{align}
At this stage $\omega_{\mathrm{cut}}$ can be considered a variational parameter. Its main effect is to controllably remove high energy parts of the magnetic excitation spectrum, especially the incoherent part which should be irrelevant to superconductivity. 
The need for this cutoff will become clear later below. To make contact between $\omega_{\mathrm{cut}}$ and the real-frequency dependence of kernels as obtained from Eq.\,(\ref{bandkern}), we define 
\begin{align}
\widetilde{V}^{\mathrm{d,intra}}_{\mathbf{q}}(\omega) &= \frac{1}{2} \sum_n  {\rm Im}\left( \big[V^{(-)}_{{\bf q}}(\omega)\big]_{nn}\right) ,\label{matkernDIntraRealFreq} \\
\widetilde{V}^{\mathrm{d,inter}}_{\mathbf{q}}(\omega) &= \frac{1}{2} \sum_{n\neq n'} \! {\rm Im} \left( \big[V^{(-)}_{{\bf q}}(\omega)\big]_{nn'}\right) \label{matkernDInterRealFreq} .
\end{align}
Note that $\widetilde{V}^{\mathrm{d,\cdot}}_{\mathbf{q}}$ is used for kernels as function of $\omega$, while $V^{\mathrm{d,\cdot}}_{\mathbf{q}}$ in Eqs.\,(\ref{matkernDIntra}) and (\ref{matkernDInter}) are Matsubara frequency dependent. 

To show possible consequences arising from the cutoff in Eq.\,(\ref{bandkernTrunc}) we choose $(U,J)=(0.8\,\mathrm{eV},U/6)$ and plot Eq.\,(\ref{matkernDIntraRealFreq}) in Fig.\,\ref{bulk_cutDependentKern}(a). At momenta/frequencies where $\widetilde{V}^{\mathrm{d,intra}}_{\mathbf{q}}(\omega)<0$ the charge contributions dominate, since they enter with negative sign in Eq.\,(\ref{orbitalkernM}), and hence the kernel becomes attractive in the superconducting channel. Wherever the spin dominates over the charge content,
the coupling is repulsive. For $\omega\leq0.5\,\mathrm{eV}$ the spectrum is rather discrete, making it possible to identify clear features at $\Gamma$ (attractive) and $X$ (repulsive). At larger frequencies substantial increases of $\widetilde{V}^{\mathrm{d,intra}}_{\mathbf{q}}(\omega)$ are observed throughout the full BZ, a similar feature as discussed in connection with Fig.\,\ref{bulk_dynKernel}. Going to $\omega\sim3\,\mathrm{eV}$, we see an enhanced influence of charge fluctuations which make the kernel attractive. A qualitatively similar picture is found in Fig.\,\ref{bulk_cutDependentKern}(b) when considering the dynamic inter-band kernel. These graphs indicate that taking the full Stoner continuum into account is not favorable for unconventional superconductivity. Intuitively this becomes clear in the light of FS nesting, which becomes combined with an incoherent background at all $\mathbf{q}$.

In Fig.\,\ref{bulk_cutDependentKern}(c) we use different cutoffs for Eq.\,(\ref{bandkernTrunc}) and show in panel {\bf i} ({\bf ii}) the resulting static intra- (inter-) band kernels. From the lower graph we learn that the incoherent background can be directly controlled via decreasing $\omega_{\mathrm{cut}}$, which makes sense having in mind the discrete nature at low frequencies of $\widetilde{V}^{\mathrm{d,intra}}_{\mathbf{q}}(\omega)$ and $\widetilde{V}^{\mathrm{d,inter}}_{\mathbf{q}}(\omega)$ in panels (a) and (b). Similarly interesting, the kernels in {\bf i} show a sign change at $\Gamma$ with increasing cutoff. Concerning the superconducting gap, see Section \ref{scSuperconducting}, this can lead to different  tendencies concerning the favored gap symmetry since small $\omega_{\mathrm{cut}}$ leads to attractive intra-band coupling on the FS pockets; the limit $\omega_{\mathrm{cut}}\rightarrow\infty$ on the other hand induces a sign change on individual FS sheets, which overall can lead to a different momentum structure of the order parameter. We consider the former situation as more physical, since we employ a low-energy theory, hence any large frequency effects should not drastically change the qualitative picture of our results.

As another direct consequence of $\omega_{\mathrm{cut}}$ we note the changes along Matsubara frequencies in Fig.\,\ref{bulk_cutDependentKern}(d), where we show the dynamic intra- ({\bf i}) and inter-band ({\bf ii}) kernels found from Eqs.\,(\ref{matkernDIntra}) and (\ref{matkernDInter}). Plotting our results at $\mathbf{q}=\Gamma$ we observe in panel (c){\bf i} again the sign change with increasing cutoff, as already discussed. Additionally the tails for small $\omega_{\mathrm{cut}}$ are decaying much faster with $q_m$, which makes the computations more efficient.

\subsection{The superconducting state}\label{scSuperconducting}

We are now in the position to address the selfconsistent Eliashberg problem for spin-fluctuation mediated pairing. The interaction kernels introduced in the previous section are used to solve the following set of coupled and selfconsistent equations:
\begin{align}
Z_{{\bf k}n}(i\omega_m) =& 1 + \frac{T}{\omega_m}\sum_{{\bf k}'m'n'} \big[V^{(+)}_{{\bf q}}(iq_{m-m'})\big]_{nn'} \nonumber\\
&\times \frac{\omega_{m'}Z_{{\bf k}'n'}(i\omega_{m'})}{\Theta_{{\bf k}'n'}(i\omega_{m'})} ~,~~~ \label{z} \\
\Gamma_{{\bf k}n}(i\omega_m) =& -T\sum_{{\bf k}'m'n'} \big[V^{(+)}_{{\bf q}}(iq_{m-m'})\big]_{nn'} \nonumber\\
& \times \frac{\xi_{{\bf k}'n'} + \Gamma_{{\bf k}'n'}(i\omega_{m'})}{\Theta_{{\bf k}'n'}(i\omega_{m'})} ~, \label{gamma}  
\end{align}
\begin{align}
\phi_{{\bf k}n}(i\omega_m) =& -T\sum_{{\bf k}'m'n'} \big[V^{(-)}_{{\bf q}}(iq_{m-m'})\big]_{nn'} \nonumber\\
&\times \frac{\phi_{{\bf k}'n'}(i\omega_{m'})}{\Theta_{{\bf k}'n'}(i\omega_{m'})} ~, \label{phi} \\
\Theta_{{\bf k}n}(i\omega_m) =& \omega_m^2Z^2_{{\bf k}n}(i\omega_m) + [\xi_{{\bf k}n}+\Gamma_{{\bf k}n}(i\omega_m)]^2 \nonumber\\
& + \phi^2_{{\bf k}n}(i\omega_m)  ~. \label{theta}
\end{align}
Here we use $Z_{\mathbf{k}n}(i\omega_m)$ as the electronic mass renormalization with fermionic frequencies $\omega_m=\pi T(2m+1)$, $\Gamma_{\mathbf{k}n}(i\omega_m)$ is the chemical potential renormalization and $\phi_{\mathbf{k}n}(i\omega_m)$ the superconducting order parameter. The gap function is found via $\Delta_{\mathbf{k}n}(i\omega_m)=\phi_{\mathbf{k}n}(i\omega_m)/Z_{\mathbf{k}n}(i\omega_m)$, the zero-frequency component of which is accessible in experiment. Note that all functions in Eqs.\,(\ref{z})-(\ref{theta}) are explicitly momentum, Matsubara frequency, and band dependent. Within our five-band model, this gives rise to 15 coupled selfconsistent Eliashberg equations in total.

The above-presented Eliashberg equations are solved selfconsistently without any further approximation, see Appendix \ref{appMethodDetails} for details. The full mathematical modeling presented in this work has been implemented in the Uppsala Superconductivity code (UppSC) \cite{UppSC,Aperis2015,Aperis2018,Bekaert2018,Schrodi2018,Schrodi2018_2}.

\begin{figure}[t!]
	\centering
	\includegraphics[width=1\columnwidth]{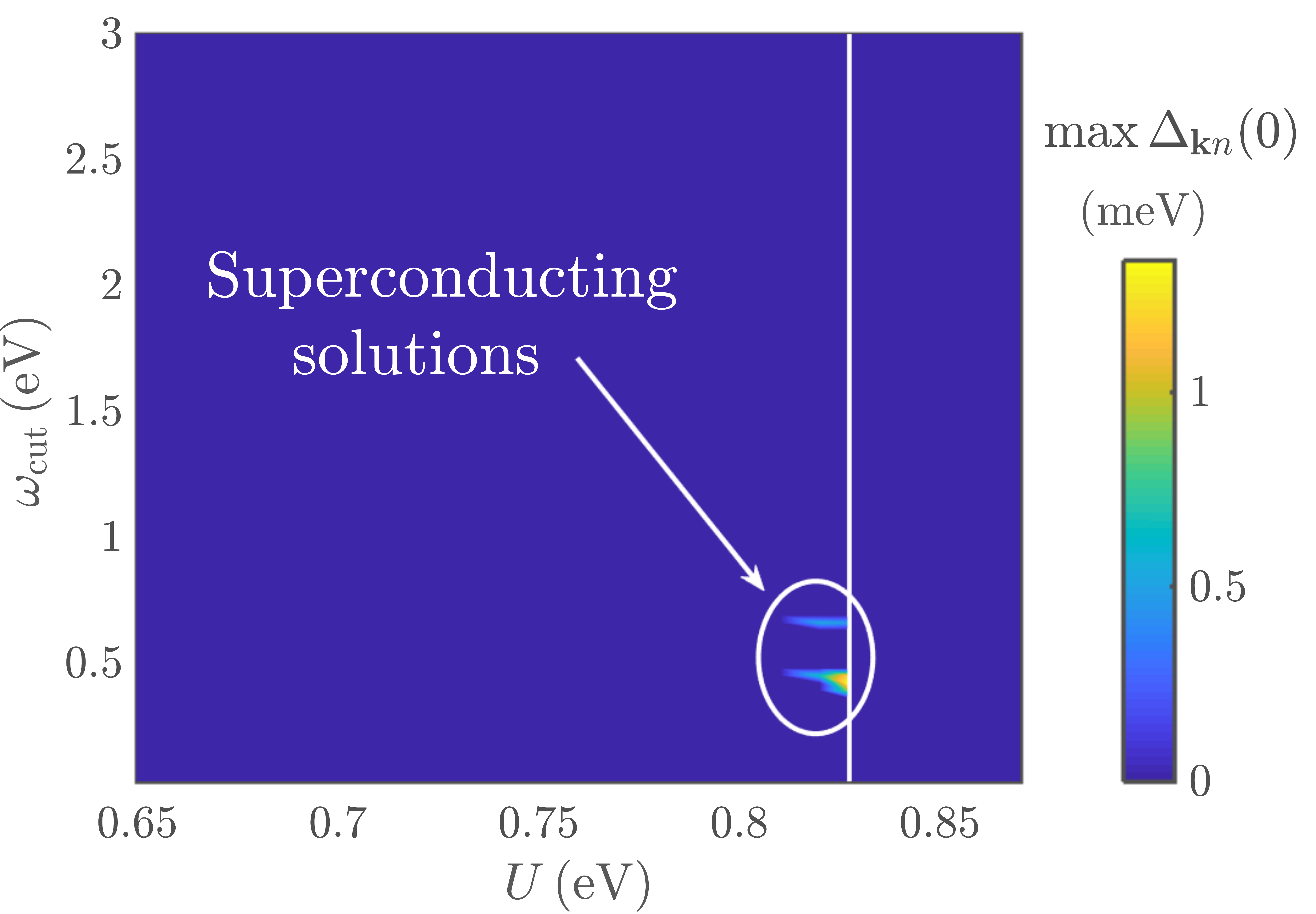}
	\caption{Selfconsistently calculated superconducting gap at zero frequency with $J=U/2$. We show the maximum among all bands and momenta as function of $U$ and $\omega_{\mathrm{cut}}$. The border drawn in white indicates the onset of magnetic order. The nonzero gap solutions are highlighted by the white ellipse.}
	\label{bulk_phasediagramUW}
\end{figure}

We now consider all three $U/J$ ratios highlighted in Fig.\,\ref{bulk_statRPA}(a) and perform a variation in $U$ and $\omega_{\mathrm{cut}}$. For each configuration we test several symmetries for initializing the order parameters to make sure that we capture all possible solutions. Interestingly, our calculations show that a sufficiently large Hund's rule coupling is needed for finding $\phi\neq0$, i.e.\ not a single nonzero gap is found for $J=U/10$ and $J=U/6$. Contrarily, we find that selfconsistent solutions are possible when choosing $J=U/2$. In Fig.\,\ref{bulk_phasediagramUW} we plot the maximum zero-frequency gap in $(U,\omega_{\mathrm{cut}})$ space, keeping $J=U/2$ fixed.
The vertical border drawn in white represents the first Stoner instability. Self-consistent solutions are found only in a very confined region of the phase space, pointing towards two characteristic cutoffs that we identify as $0.42\,\mathrm{eV}$ and $0.66\,\mathrm{eV}$. This corresponds to an energy range where nothing ($0.42\,\mathrm{eV}$) or only a very small fraction (at $0.66\,\mathrm{eV}$) of the Stoner continuum is included when calculating the Matsubara frequency dependent kernels. We therefore see here explicitly what is already discussed in Sec.\ \ref{scCoupling}, namely, that including the Stoner continuum does not allow for a selfconsistently obtained superconducting state. This is due to frustration effects \cite{Kemper2010}, caused by an incoherent distribution of the kernel among nearly all possible exchange wave vectors in the BZ. It is also worth noting that lowering  $\omega_{\mathrm{cut}}$ too much again leads to disappearance of superconducting solutions.
\begin{figure}[t!]
	\centering
	\includegraphics[width=1\columnwidth]{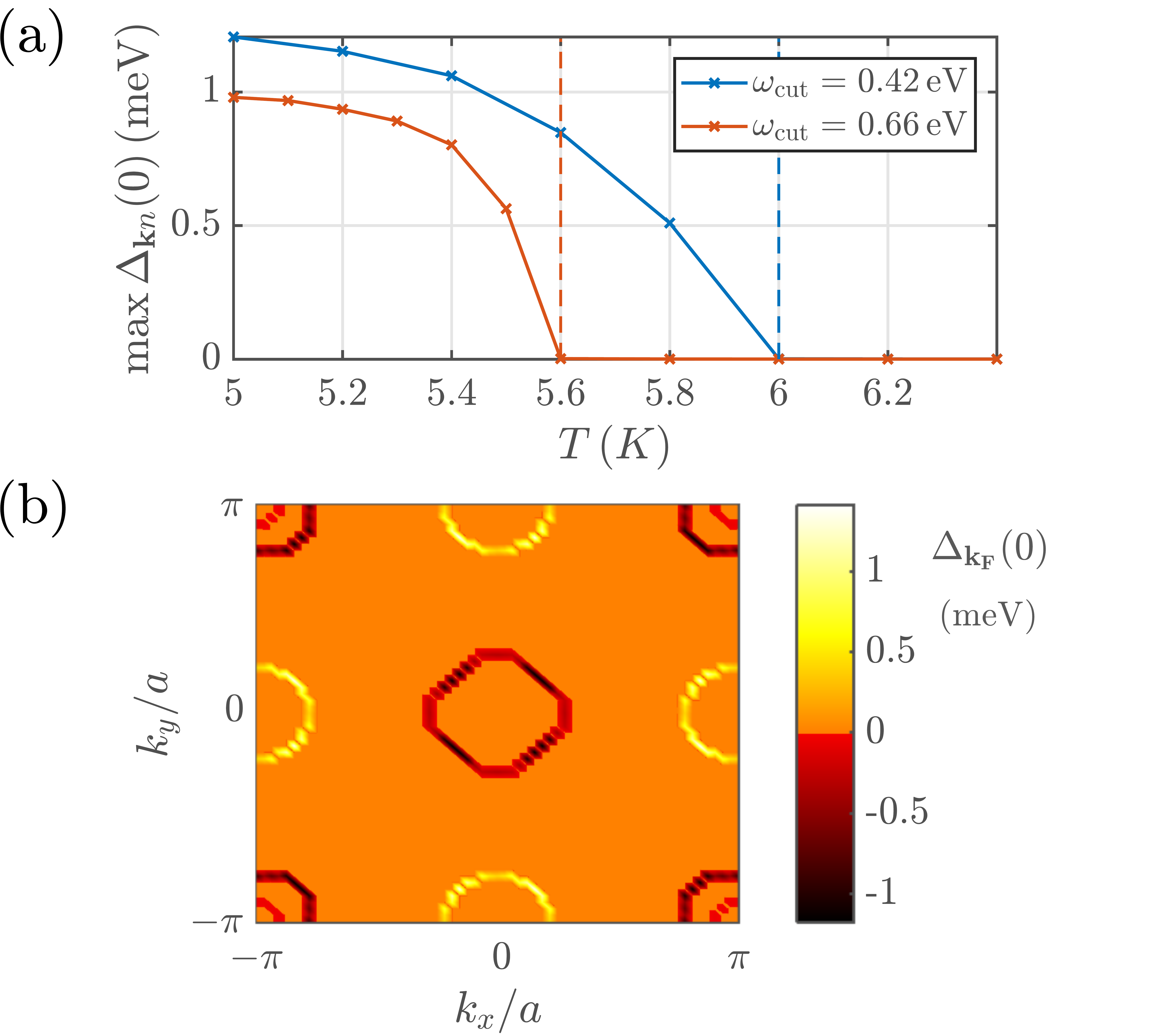}
	\caption{(a) Selfconsistently computed maximum superconducting gap as function of temperature, shown for cutoffs $\omega_{\mathrm{cut}}=0.42\,\mathrm{eV}$ (blue) and $\omega_{\mathrm{cut}}=0.66\,(\mathrm{eV})$ (red), both calculated for $J=U/2$ and $U=0.827\,\mathrm{eV}$. The respective critical temperatures are indicated by dashed lines. (b) Result for the superconducting order parameter projected on the FS, obtained for $U=0.827\,\mathrm{eV}$, $J=U/2$, $\omega_{\mathrm{cut}}=0.42\,\mathrm{eV}$ and $T=5\,\mathrm{K}$.}
	\label{bulk_tempdepGap}
\end{figure}

Next we perform a temperature variation for the aforementioned two cutoffs to obtain the corresponding transition temperatures. Following the evolution of maximal zero frequency gaps in Fig.\,\ref{bulk_tempdepGap}{(a)} we find $T_c\simeq5.6\,\mathrm{K}$ for the cutoff $\omega_{\mathrm{cut}}=0.66\,\mathrm{eV}$ (drawn in red), that is closer to the onset of the Stoner continuum. A slightly larger value of $T_c\simeq6\,\mathrm{K}$ is possible for $\omega_{\mathrm{cut}}=0.42\,\mathrm{eV}$, represented by the blue solid curve in the same graph. As guide for the eye the onset of superconductivity is marked in both cases by dashed lines with respective color code. These results resemble the experimental value of $\sim8\,\mathrm{K}$ remarkably well \cite{Hsu2008}.

We conclude the discussion of bulk FeSe by turning to the gap symmetry, choosing $(U,J)$ as before, $\omega_{\mathrm{cut}}=0.42\,\mathrm{eV}$, and $T=5\,\mathrm{K}$. Having access to the fully momentum dependent zero-frequency component $\Delta_{\mathbf{k}n}(0)$, we project our results on the FS, drawn in Fig.\,\ref{bulk_tempdepGap}(b). As directly evidenced, there is a sign change between electron and hole pockets without any FS nodes. Since the sign on each individual pocket is constant we obtain 
a global $s_{\pm}$ symmetry. Further we note that $\underset{\mathbf{k}_{\rm F}}{\mathrm{max}}\,\Delta_{\mathbf{k}_{\rm F}}(0)\neq |\underset{\mathbf{k}_{\rm F}}{\mathrm{min}}\,\Delta_{\mathbf{k}_{\rm F}}(0)|$, which suggests an additional pure $s$-wave component. Hence, our result for the gap symmetry of superconducting bulk FeSe is $s_{\pm}+s$, matching experimental findings \cite{Jiao2017}. Concerning the magnitude of our calculated superconducting gap we deviate only very slightly from measured values of $\Delta\sim1.67\,\mathrm{meV}$ \cite{Jiao2017,Sprau2017}. The difference to our result of $\Delta\sim1.4\,\mathrm{meV}$ directly explains the small mismatch in the calculated critical temperature. From our selfconsistent results we furthermore observe that superconducting gap values as found experimentally are not primarily related to nematicity \cite{Boehmer2013,Baek2020}. Although our FS obeys $C_4$ symmetry, the main features measured 
in the nematic (orthorhombic) state are reproduced reliably. 

We conclude this part by referring again to Appendix \ref{appMethodDetails}, where several further aspects of our calculations are discussed. These details concern the mathematical and numerical steps in all subsections presented so far. 

\section{Monolayer F$\bf e$S$\bf e$ on STO}\label{scMonolayer}

Having introduced our method in {\rm Sec.}\  \ref{scMethod} we now want to apply it to FeSe/STO, imposing that spin fluctuations are the only relevant ingredient for the superconducting state in this system. Any influence of the interfacial phonon that presumably plays an important role for superconductivity \cite{Lee2014,Rademaker2016,Aperis2018,Schrodi2018} is hence neglected here. After describing some characteristic properties of FeSe/STO within our framework in Sec.\,\ref{scBasicProps}, we directly go to the discussion of our selfconsistent results for the superconducting state, in Sec.\,\ref{scMonolayerSupercond}. We continue in Sec.\,\ref{scInfluenceTB} by examining the effect of changes in our tight-binding model on the superconducting properties. We compare our results to predictions of BCS theory in Sec.\,\ref{scBCS} and conclude by commenting on how our solutions scale with respect to the proximity to antiferromagnetic criticality in Sec.\,\ref{scUpperBoundTc}.

\subsection{Basic properties of FeSe/STO}\label{scBasicProps}

\begin{figure*}[ht!]
	\centering
	\includegraphics[width=1\textwidth]{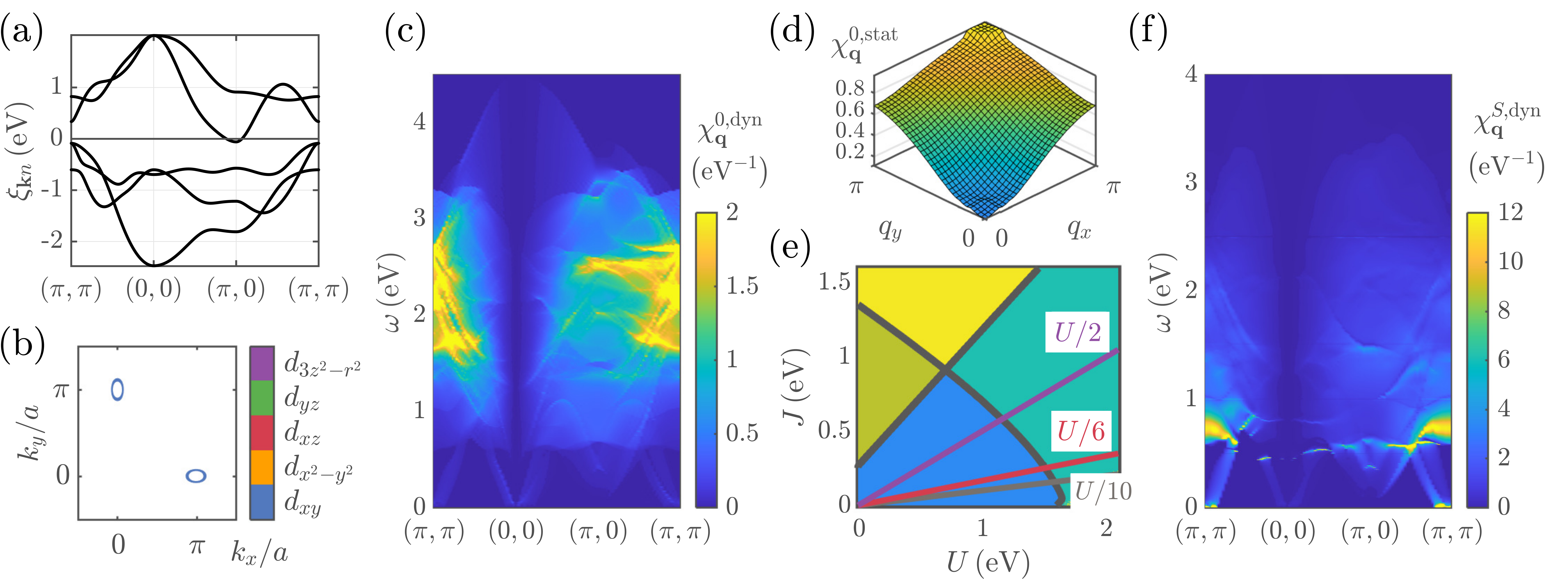}
	\caption{Important characteristics of monolayer FeSe on STO within our tight-binding approach, using the band dispersions derived in Ref.\,\cite{Hao2014}. (a) Electronic dispersions along high-symmetry lines of the BZ. (b) Fermi surface pockets colored by the dominant orbital weights. (c) Dynamic susceptibility as function of frequency, plotted along high-symmetry lines. (d) Static bare susceptibility as calculated from Eq.\,(\ref{statbubble}), shown in the first quadrant. (e) Phase diagram in $(U,J)$ space indicating the onset of charge or magnetic order. The parameter space allowed by the Stoner criterion is drawn in blue. A violation of the Stoner criterion due to charge, spin or charge and spin is indicated in green, cyan, and yellow colors. Three representative $U/J$ ratios are drawn in purple $(J=U/2)$, red $(J=U/6)$ and gray $(J=U/10)$ lines as guide for the eye. (f) Dynamic spin susceptibility calculated from Eq.\,(\ref{dynamicSpin}), using the result of Eq.\,(\ref{rpaMatS}) as input. $U=1.5802\,\mathrm{eV}$ and $J=U/10$ are chosen, which are rather close to the first Stoner instability.}
	\label{ML_hao_setup}
\end{figure*}

The electronic energies are given by a tight-binding model that we take from Ref.\,\cite{Hao2014}.  We show the corresponding energy bands along high-symmetry lines in Fig.\,\ref{ML_hao_setup}(a).
The lattice distortion arising from the deposition of monolayer FeSe on the substrate is explicitly taken into account \cite{Hao2014}. In addition, the hopping parameters are modified in order to move the hole bands present in bulk FeSe to below the Fermi level \cite{Aperis2018}. 
 The FS, plotted with its orbital content in Fig.\,\ref{ML_hao_setup}(b), consists of two electron pockets, which are dominated by $d_{xy}$ character. Compared to experiment the FS sheets are slightly smaller \cite{Zhang2016}, this aspect is further addressed in Sec.\ \ref{scInfluenceTB}.

With these energy dispersions we calculate the real and imaginary parts of the bare susceptibility, Eqs.\,(\ref{bubblereal}) and (\ref{bubbleimag}), which serve as input for obtaining the static and dynamic bare susceptibilities of Eqs.\,(\ref{statbubble}) and (\ref{dynbubble}), respectively. From Fig.\,\ref{ML_hao_setup}(d) we observe that $\chi_{\mathbf{q}}^{0,\mathrm{stat}}$ is peaked near $\mathbf{q}=M$, which is the wave vector connecting the FS pockets. Compared to Fig.\,\ref{bulk_dynbubble}(b) we no longer have pronounced contributions at $X$ since the hole bands can no longer be statically connected to the electron sheets at the FS. A small ring around $\Gamma$ is found due to small wavevectors connecting states within the electron pockets. Turning to the dynamic susceptibility, shown in Fig.\,\ref{ML_hao_setup}(c),
confirms the aforementioned picture clearly, namely, that the leading excitations are located at $M$. Small and distinct branches can be seen at small frequencies as more explicitly shown in Fig.\,\ref{zoomedSuscept}(a) in Appendix \ref{appSuscept}; for $\omega>0.6\,\mathrm{eV}$ a continuum of nonnegligible contributions occurs throughout the BZ. The only exception to this is $\mathbf{q}\sim\Gamma$, where a minimal $\chi_{\mathbf{q}}^{0,\mathrm{dyn}}$ is found for all frequencies. 

Inserting the bare susceptibilities into Eqs.\,(\ref{rpaMatS}) and (\ref{rpaMatC}), we perform a variation in $(U,J)$ to find the onset of charge and magnetic order. The blue region in Fig.\,\ref{ML_hao_setup}(e) represents allowed choices for the Hubbard $U$ and the Hund's rule $J$ coupling. Contrarily, the green, yellow and cyan parts of the diagram correspond to a violation of the Stoner criterion due to 
 $\chi_{\mathbf{q}}^{C,\mathrm{stat}}<0$, $\chi_{\mathbf{q}}^{C,\mathrm{stat}}<0$ and $\chi_{\mathbf{q}}^{S,\mathrm{stat}}<0$ or $\chi_{\mathbf{q}}^{S,\mathrm{stat}}<0$, respectively. We show in the same panel three ratios considered in the following: $J=U/2$ (purple line), $J=U/6$ (red line) and $J=U/10$ (gray line). When comparing this phase diagram with bulk FeSe, Fig.\,\ref{bulk_statRPA}(a), remarkable similarities are observed. Since the bandwidths of both tight-binding models are comparable, the scales of $U$ and $J$ do not differ much. Further, the overall shapes are very alike, except for a small piece of allowed phase space missing in Fig.\,\ref{ML_hao_setup}(e) in the limit $J \ll U$.

From discussions in the previous section we know that a nonzero solution for the superconducting gap is to be expected close to magnetic order. We hence show the dynamic spin susceptibility Eq.\,(\ref{dynamicSpin}) in Fig.\,\ref{ML_hao_setup}(f) for a rather critical pair $(U,J)=(1.5802\,\mathrm{eV},U/10)$. A zoom into the low-energy region is presented in Fig.\,\ref{zoomedSuscept}(b) in Appendix \ref{appSuscept}. At frequencies of approximately $0.6\,\mathrm{eV}$ the Stoner continuum begins, introducing contributions for all momenta sufficiently away from $\Gamma$. For smaller $\omega$ we find enhanced susceptibilities at $X$, which partially represent the connection of FS electron pockets with hole bands below the Fermi level, made possible by a relatively small energy exchange. This aspect is treated in more detail in Sec.\ \ref{scInfluenceTB}. Comparing Fig.\,\ref{ML_hao_setup}(f) with the dynamic bare susceptibility in panel (c) we identify the leading instabilities at $\omega\sim0.8\,\mathrm{eV}$ and $\omega\sim0.1\,\mathrm{eV}$, both around $\mathbf{q}=M$. 

\begin{figure}[t!]
	\centering
	\includegraphics[width=1\columnwidth]{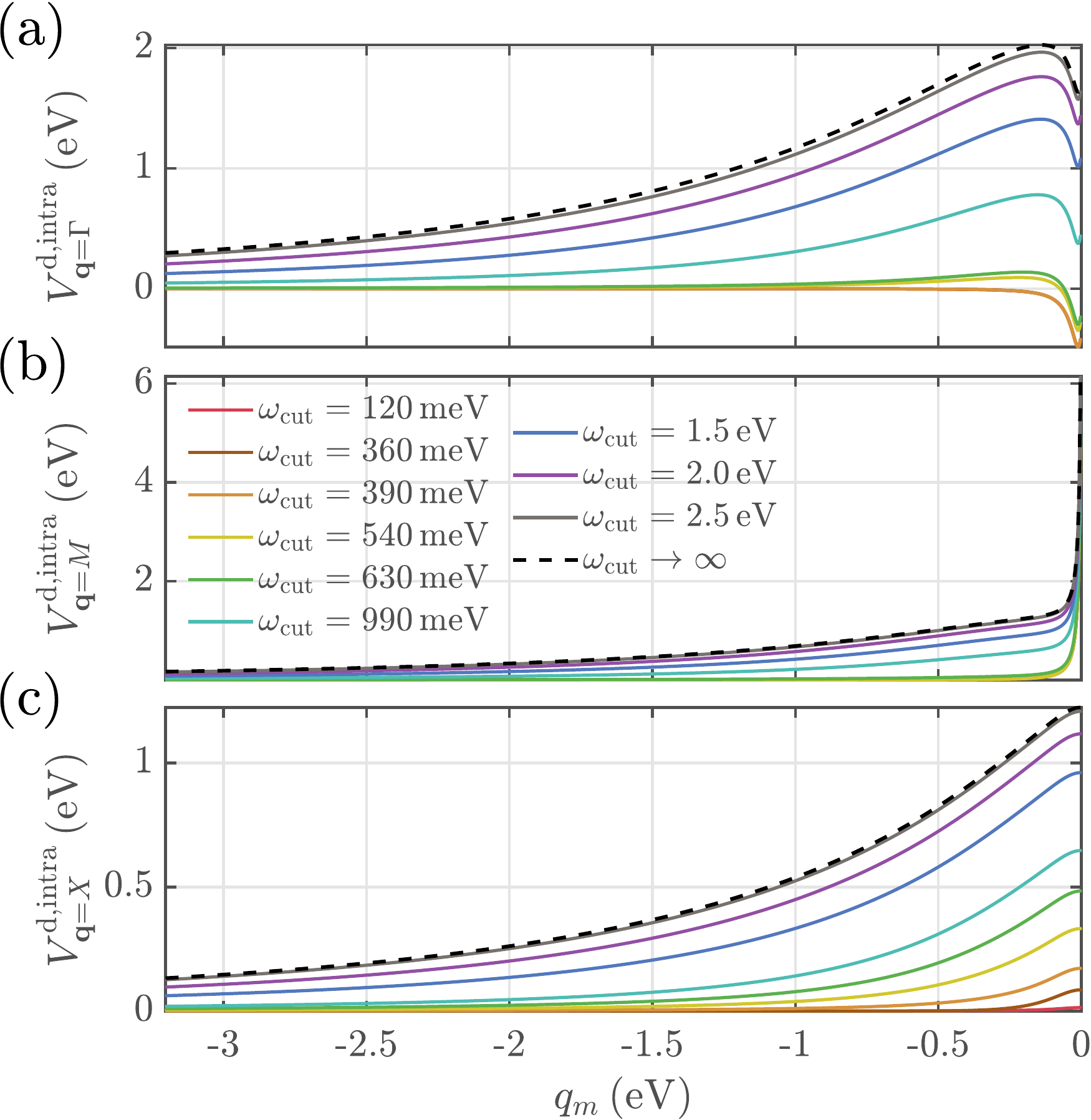}
	\caption{ Matsubara frequency dependent intraband kernels for various $\omega_{\mathrm{cut}}$, each as function of $q_m$, where we choose $U=1.5802\,\mathrm{eV}$ and $J=U/10$. (a), (b) and (c) show the kernels at the high-symmetry momenta $\mathbf{q}=\Gamma$, $M$, and $X$, respectively.}
	\label{ML_hao_kernels}
\end{figure}

The appearance of a Stoner continuum is reflected in interaction kernels similarly as discussed in Sec.\,\ref{scCoupling}. We calculate full frequency, momentum and orbital-dependent couplings from Eqs.\,(\ref{orbitalkernP}) and (\ref{orbitalkernM}), which are then used as input for Eq.\,(\ref{bandkernTrunc}), additionally as function of $\omega_{\mathrm{cut}}$. Keeping parameters $(U,J)=(1.5802\,\mathrm{eV},U/10)$ as in Fig.\,\ref{ML_hao_setup}(f), we show the outcomes at momenta $\Gamma$, $M$, and $X$ in Fig.\,\ref{ML_hao_kernels}(a), (b) and (c) for several cutoffs. From panel (a) we observe a similar behavior as in bulk FeSe: for $\omega_{\mathrm{cut}}$ sufficiently small, i.e.\ below the onset of the Stoner continuum, small-$\mathbf{q}$ couplings are attractive in the superconducting channel for Matsubara frequencies close to zero. As soon as the cutoff gets larger than a threshold of roughly $0.8\,\mathrm{eV}$, $V_{\mathbf{q}=\Gamma}^{\mathrm{d,intra}}$ becomes repulsive for all $q_m$, thus favoring a sign change on the FS pockets.  
Combined with  dominant contributions at $\mathbf{q}=M$, see Fig.\,\ref{ML_hao_kernels}(b), such a configuration could still lead to a nonvanishing gap, possibly with unconventional symmetry,
but we did not find it in any of our calculations. From these considerations one can, even without solving the Eliashberg equations, qualitatively predict that a nonvanishing order parameter is likely to be found only when one excludes the Stoner continuum; as shown in the next subsection this is indeed what we observe. In the above discussion we omit showing explicitly the interband kernels, since these do not provide further insights.

\subsection{Spin-fluctuations mediated pairing} \label{scMonolayerSupercond}

\begin{figure}[t!]
	\centering
	\includegraphics[width=1\columnwidth]{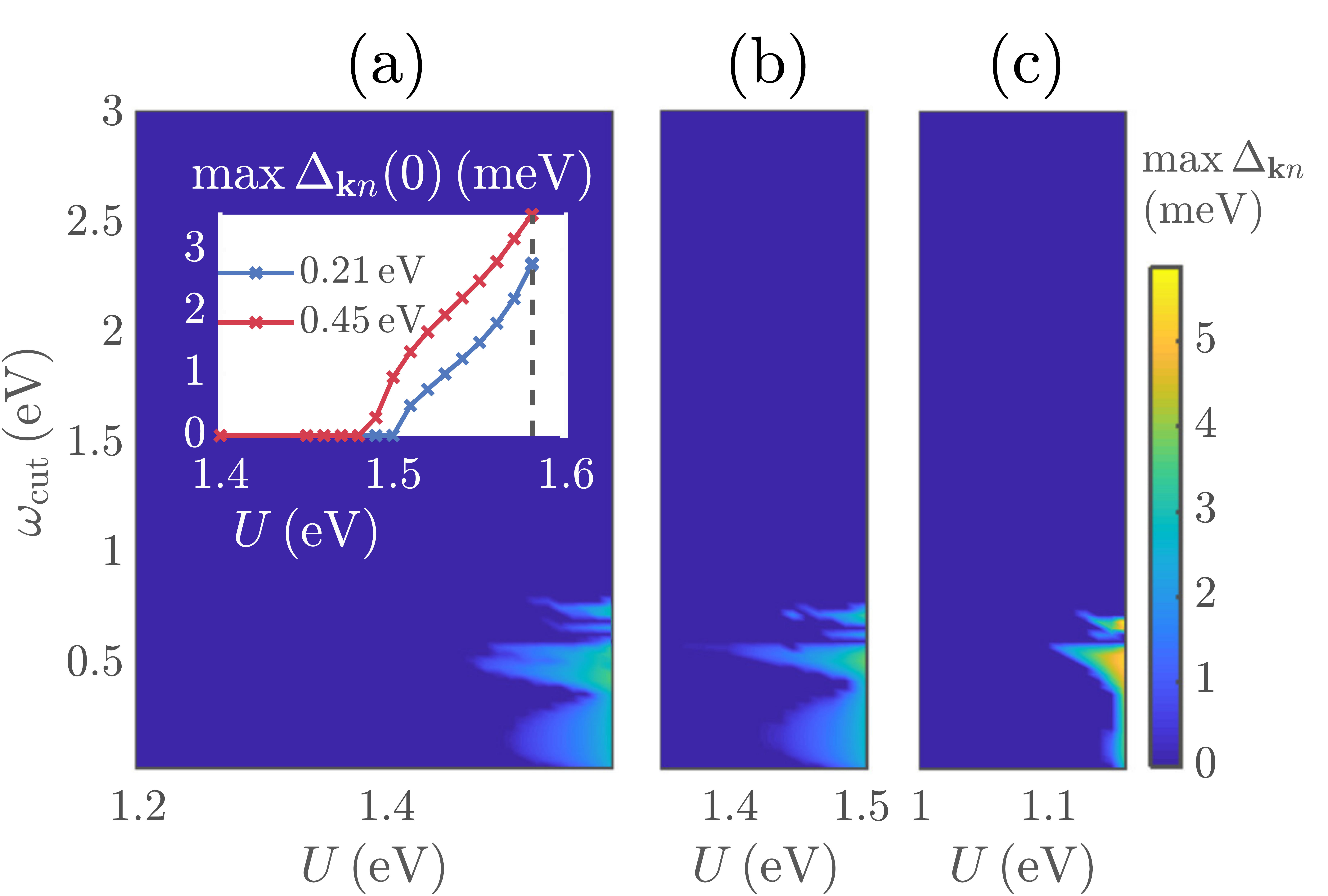}
	\caption{Computed maximum superconducting gap as function of $U$ and $\omega_{\mathrm{cut}}$. Results are shown for (a) $J=U/10$, (b) $J=U/6$, and (c) $J=U/2$. The inset shows the maximum gap as function of $U$ for two specific frequencies, $\omega_{\mathrm{cut}}=0.21\,\mathrm{eV}$ (blue line) and $\omega_{\mathrm{cut}}=0.45\,\mathrm{eV}$  (red line), computed for $J=U/10$.}
	\label{ML_hao_phasediagramUw}
\end{figure}

We solve the coupled set of Eliashberg equations (\ref{z})-(\ref{phi}) for the three ratios $U/J$ as indicated in Fig.\,\ref{ML_hao_setup}(e).  Varying $U$ and $\omega_{\mathrm{cut}}$, compare Sec.\ \ref{scSuperconducting}, we are able to find the available phase space for a nonvanishing order parameter. The selfconsistently calculated maximum superconducting gap is shown in Fig.\,\ref{ML_hao_phasediagramUw}(a), (b), and (c) for $J=U/10$, $J=U/6$, and $J=U/2$, respectively. For two cutoff frequencies $0.21\,\mathrm{eV}$ (blue) and $0.45\,\mathrm{eV}$ (red) we plot $\mathrm{max}\,\Delta_{\mathbf{k}n}(0)$ as function of $U$ in the inset of panel (a) using $J=U/10$. Our results show that the maximal gap possible at $T=5\,\mathrm{K}$ is $\sim5.9\,\mathrm{meV}$ for $\omega_{\mathrm{cut}}=0.69\,\mathrm{eV}$ and $J=U/2$. Close to the Stoner instability we find superconductivity in all three panels of Fig.\,\ref{ML_hao_phasediagramUw}, where maximal values of $\Delta_{\mathbf{k}n}(0)$ at critical $U$ are of the order of $3-6\,\mathrm{meV}$.

As function of $\omega_{\mathrm{cut}}$ there are three dome-like regions allowing for a finite gap. Such behavior points towards distinct branches in $\big[V_{\mathbf{q}}^{(\pm)}(\omega)\big]_{nn'}$, which can be constructive or destructive for Cooper pairing. Without the need of plotting these kernels we can understand the underlying mechanism already on the level of the dynamic spin susceptibility, see Fig.\,\ref{ML_hao_setup}(f), using $U=1.5802$\,eV and $J=U/10$ as example. Note that a direct comparison of frequency values is not appropriate, since the $\omega$-scale in general is shifted to smaller frequencies when going from susceptibilities to interaction kernels. Clearly, the small-$\omega$ contribution at $M$ in Fig.\,\ref{ML_hao_setup}(f) is responsible for the first dome of panel \ref{ML_hao_phasediagramUw}(a). For growing cutoff the pairing is suppressed due to substantial contributions that extends to large parts of the BZ, see Fig.\,\ref{ML_hao_setup}(f) at $\omega\sim0.6\,\mathrm{eV}$. At the high-frequency end we find that the leading instability at $M$, in competition with the Stoner continuum, is responsible for the last region (at largest $\omega_{\mathrm{cut}}$) of nonzero gap. The intermediate frequency regime is less easily understood, since it lies directly within the onset of the Stoner continuum.

When we compare the three panels of Fig.\,\ref{ML_hao_phasediagramUw} we discern that an increase in Hund's rule coupling leads to a smaller phase space, as well as enhanced gap values. For $J=U/2$ we find that the phase diagram is similar to that observed in bulk FeSe, see Fig.\,\ref{bulk_phasediagramUW}. Although the gap sizes are different, the characteristic frequencies at which a nonzero solution is possible, seem to be almost the same. Besides a small  region with $\Delta\neq0$ near $\omega_{\rm cut}\sim0.6\,\mathrm{eV}$, we find the most relevant frequency cutoff at around $0.45\,\mathrm{eV}$. This agreement might be explained by similar choices of $(U,J)$ and the fact that both tight-binding models are derived from Ref.\,\cite{Eschrig2009}. 

\begin{figure}[t!]
	\centering
	\includegraphics[width=1\columnwidth]{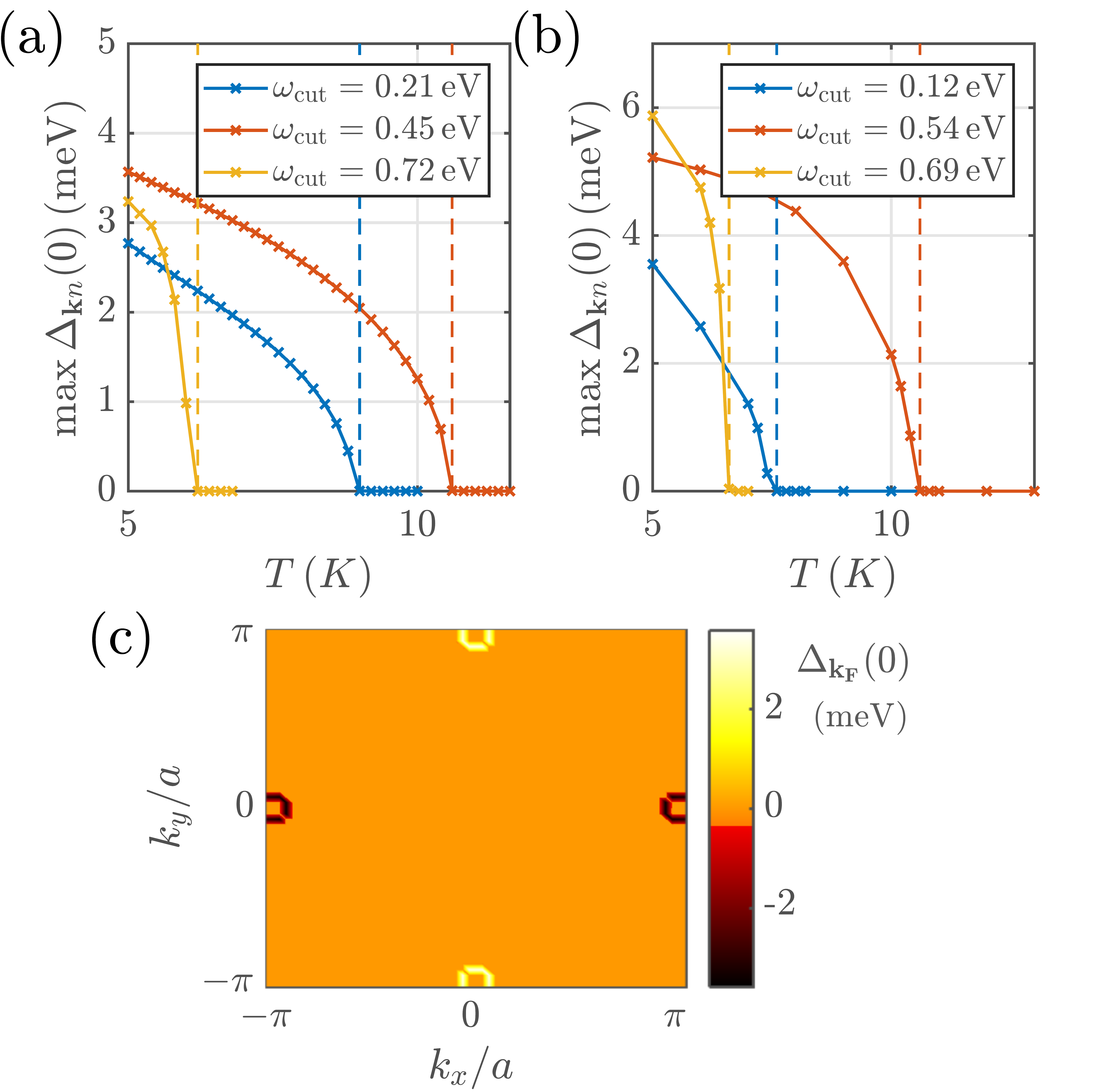}
	\caption{(a),(b) Computed maximum superconducting gap as function of temperature. Values of $T_c$ are indicated by dashed lines. (a)  Taking $(U,J)=(1.5802\,\mathrm{eV},U/10)$ we show results for $\omega_{\mathrm{cut}}=0.21\,\mathrm{eV}$ (blue), $0.45\,\mathrm{eV}$ (red) and $0.72\,\mathrm{eV}$ (yellow line). (b) With $(U,J)=(1.16\,\mathrm{eV},U/2)$ we plot the maximum gap for $\omega_{\mathrm{cut}}=0.12\,\mathrm{eV}$ in blue, $0.54\,\mathrm{eV}$ in red and $0.69\,\mathrm{eV}$ in yellow color. (c) Superconducting gap projected on the renormalized FS for $J=U/10$, $U=1.5802\,\mathrm{eV}$ and $\omega_{\mathrm{cut}}=0.45\,\mathrm{eV}$ at $T=5\,\mathrm{K}$.}
	\label{ML_hao_gapsym}
\end{figure}

As we show in the inset of Fig.\,\ref{ML_hao_phasediagramUw}(a) the gap size increases approximately linear with $U$ going towards its critical value. We observe similar trends for the other choices of Hund's rule coupling (not shown). The behavior of $\Delta$ close to criticality is further discussed in Sec.\ \ref{scBCS}. Focusing now on panels (a) and (c) of Fig.\,\ref{ML_hao_phasediagramUw}, i.e.\ on choices $J=U/10$ and $J=U/2$, we choose three representative cutoffs for both. 
Choosing the respective value of $U$ very close to the Stoner instability,
we calculate the temperature dependence of the maximum gap and show the results in Fig.\,\ref{ML_hao_gapsym}(a) and (b). 

The largest critical temperature for $J=U/10$, panel (a), is found for $\omega_{\mathrm{cut}}=0.45\,\mathrm{eV}$ as $T_c=10.6\,\mathrm{K}$ {($\Delta/T_c\approx 4$)}. Both other cutoffs lead to smaller values of $T_c=9\,\mathrm{K}$ ($\Delta/T_c\approx 3.6$) and $6.2\,\mathrm{K}$ {($\Delta/T_c\approx 6$)}, as seen from the blue and yellow curves. Results for $J=U/2$ do not change much in this respect: The maximal critical temperature is found for the intermediate $\omega_{\mathrm{cut}}=0.54\,\mathrm{eV}$, and is again $T_c=10.6\,\mathrm{K}$ {($\Delta/T_c\approx 5.6$)}. From Fig.\,\ref{ML_hao_gapsym} we also learn that a change in cutoff can lead to 
changes in the ratio of $\Delta=\underset{T\rightarrow0}{\mathrm{lim}}\mathrm{max}\,\Delta_{\mathbf{k}n}(0)$ over $T_c$ which is usually taken as a measure of how strongly coupled superconductivity is in a system.
This can for example be seen in panel (b) when comparing the yellow and the red lines. At $T=5\,\mathrm{K}$ the former shows a larger gap size, although the $T_c$ is higher for the latter, hence the gap over $T_c$ ratio at $T\rightarrow0\,\mathrm{K}$ can not be expected to be the same. We note that a precise limit of zero temperature can not be calculated here due to the associated computational costs.

Next we examine the {FS} momentum {dependence and hence the} symmetry of $\Delta_{\mathbf{k}n}(0)$, which is a direct result from our selfconsistent calculations. To achieve this we calculate the renormalized FS as
\begin{eqnarray}
\tilde{\xi}_{\mathbf{k}n} = \frac{\xi_{\mathbf{k}n} + \Gamma_{\mathbf{k}n}(0)}{Z_{\mathbf{k}n}(0)} =0 ~,
\end{eqnarray}
which does not noticeably deviate from the bare electron pockets. The band-dependent superconducting gap is projected onto $\tilde{\xi}_{\mathbf{k}n}$ in Fig.\,\ref{ML_hao_gapsym}(c), where we choose $U=1.5802\,\mathrm{eV}$ and $J=U/10$ at $T=5\,\mathrm{K}$. As is easily observed the gap follows $d$-wave symmetry, which is similarly true for all solutions presented for FeSe/STO in this work. Although this particular symmetry has been proposed for monolayer FeSe on STO \cite{Ge2019}, the magnitude of $\Delta_{\mathbf{k}n}$ at the Fermi level as found here is not sufficiently large to account for 
experimental findings \cite{Zhang2016,Tang2016}.

From the specifics of the interaction kernels discussed in connection to Fig.\,\ref{ML_hao_kernels} it is worth doing a simplified treatment to explain our calculated gap symmetry. For simplicity we might consider $Z_{\mathbf{k}n}(i\omega_m)=1$ and $\Gamma_{\mathbf{k}n}(i\omega_m)=0$ and focus on the order parameter only. $\phi_{\mathbf{k}n}$ and $\big[V_{\mathbf{q}}^{(-)}\big]_{nn'}$ are both largest at the zeroth frequency, hence we omit the dependence on the Matsubara axis. In this sense we can write Eq.\,(\ref{phi}) as
\begin{align}
\phi_{{\bf k}n} \sim& -T\sum_{{\bf k}'n'} \big[V^{(-)}_{{\bf q}}\big]_{nn'}  \frac{\phi_{{\bf k}'n'}}{\Theta_{{\bf k}'n'}} ~.
\end{align}
Since we are interested in FS properties, and only a single band in our dispersion crosses the Fermi level, we can remove the band index and the associated sum on the right-hand side. As a drastic simplification we may write the potential as sum of delta peaks at the high-symmetry points, such that
\begin{align}
\phi_{{\bf k}} \sim& -T\sum_{{\bf k}'} \left(V^{(\Gamma)}\delta(\mathbf{q}) + V^{(M)}\delta(\mathbf{q}-M) \right. \nonumber \\
&\left.  +V^{(X)}\delta(\mathbf{q}-X)\right)  \frac{\phi_{{\bf k}'}}{\Theta_{{\bf k}'}}  ~,
\end{align}
where $V^{(\Gamma,M,X)}$ is the approximate kernel size at the respective momentum. We directly can neglect the contribution at $X$, since no FS points can be connected by this reciprocal space vector.
Evaluating the sum over $\mathbf{k}'$ leads to
\begin{align}
\phi_{{\bf k}_{\rm F}} \sim & T |V^{(\Gamma)}|\frac{\phi_{{\bf k}_{\rm F}}}{\Theta_{{\bf k}_{\rm F}}}  -T V^{(M)} \frac{\phi_{{\bf k}_{\rm F}-M}}{\Theta_{{\bf k}_{\rm F}-M}}  ~,
\end{align}
where we use the sign change of $V_{\mathbf{q}=\Gamma}^{\mathrm{d,intra}}$, as discussed in Sec.\,\ref{scBasicProps}, and set $V^{(\Gamma)}=-|V^{(\Gamma)}|$. From here it is easy to see that $\phi_{\mathbf{k}_{\rm F}}$ is maximal if $\phi_{\mathbf{k}_{\rm F}-M}=-\phi_{\mathbf{k}_{\rm F}}$, which is just the symmetry observed in Fig.\,\ref{ML_hao_gapsym}(c).

From the results presented in this section one can conclude that spin fluctuation mediated pairing cannot be the prevailing mechanism in monolayer FeSe on STO. The onset of superconductivity as found here does not compare well with experimentally obtained values of $T_c\sim60-100\,\mathrm{K}$ \cite{Qing-Yan2012,Liu2012,Ge2015}, and the maximum calculated superconducting gap is also too small \cite{Zhang2016,Tang2016}. A possible counter argument could be, however, that the specifics of our tight-binding model are not accurately resembling the experimentally observed situation. We therefore check below the influence of the electronic dispersions on our results. 

\subsection{Influence of the tight-binding model and role of incipient pairing}\label{scInfluenceTB}

Now we want to examine the influence of two specific aspects within the tight-binding model used for the calculation. On the one hand we dope the system by means of a global chemical potential, $\xi_{\mathbf{k}n}\rightarrow\xi_{\mathbf{k}n}-\mu$ with $\mu=140\,\mathrm{meV}$. This significantly increases the size of our electronic FS pockets, while placing the hole bands at rather deep energies, which are not confirmed by experiment, see Ref.\,\cite{Lee2014} and Appendix \ref{appModifyTB}. In addition, we want to explore the role that the distance between electron bands at $X$ and hole bands at $M$  plays in the superconducting state. This is achieved by an artificial nonrigid shift of only the hole bands by $\delta\mu=-48\,\mathrm{meV}$. For further details on the modified energies see Appendix \ref{appModifyTB}. In Fig.\,\ref{ML_rigid_phasediag} we show the maximum superconducting gap as function of $U$ and $\omega_{\mathrm{cut}}$. The upper row panels (a), (b), and (c) are obtained by using $\xi_{\mathbf{k}n}-\mu$, while the lower ones (d), (e), and (f) represent
$\xi_{\mathbf{k}n}-\delta\mu$. The columns correspond to different choices of the Hund's rule coupling: (a) and (d) are for  $J=U/10$, (b) and (e) for $J=U/6$, and (c) and (f) for $J=U/2$. Note that we do not show the full frequency range of Fig.\,\ref{ML_hao_phasediagramUw}, since in the current simulations we always find a vanishing gap for $\omega_{\mathrm{cut}}>1\,\mathrm{eV}$.

\begin{figure}[b!]
	\centering
	\includegraphics[width=1\columnwidth]{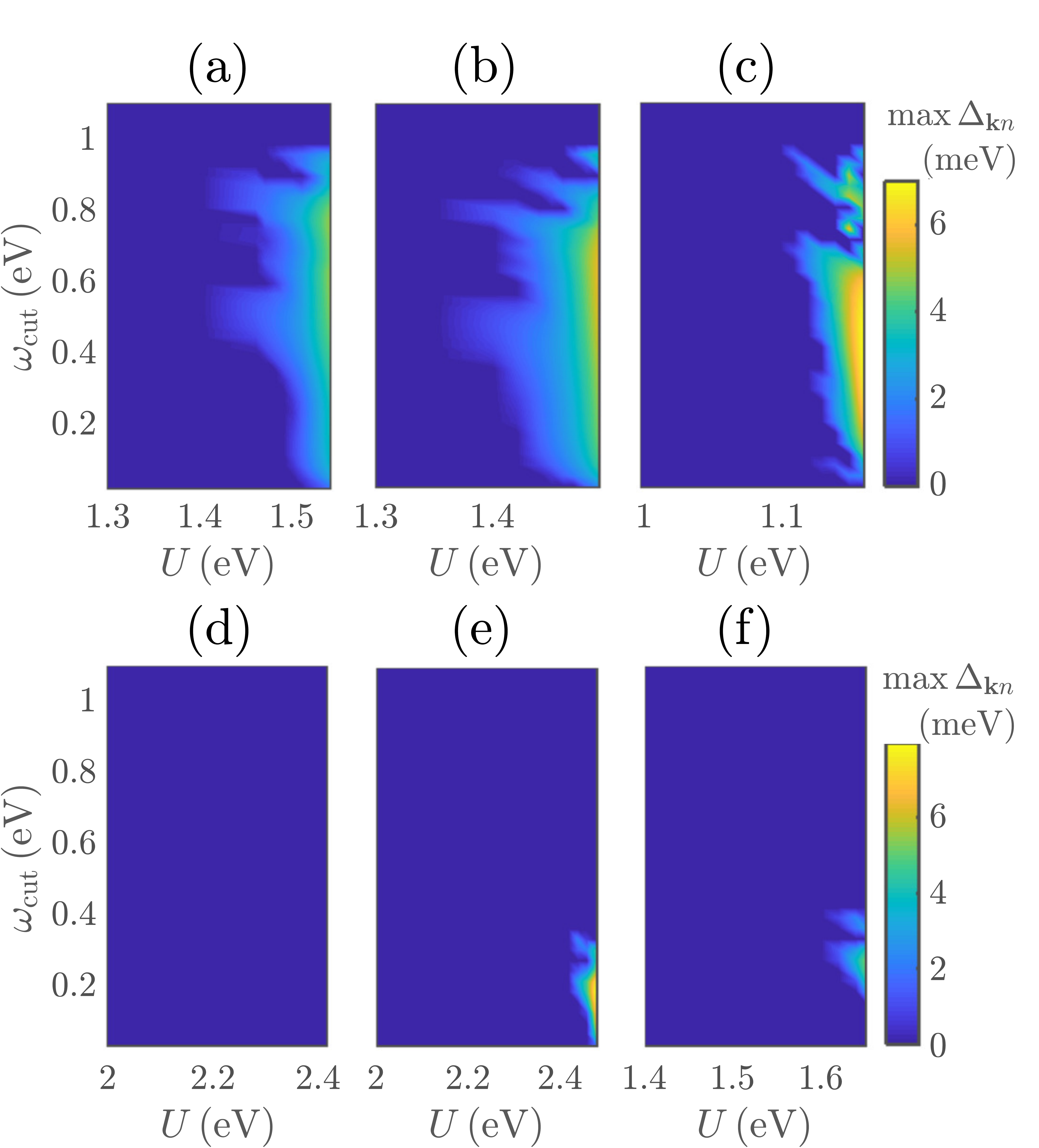}
	\caption{Maximum superconducting gap as function of intraorbital coupling and truncation cutoff $\omega_{\rm cut}$. Upper row: Rigidly shifted dispersions via $\mu=140\,\mathrm{meV}$. Lower row: Shifts of only the hole bands via $\delta\mu=-48\,\mathrm{meV}$. Our results are shown for $J=U/10$ in panels (a) and (d), $J=U/6$ in (b) and (e), and $J=U/2$ in (c) and (f).}
	\label{ML_rigid_phasediag}
\end{figure}
Starting from results for $\xi_{\mathbf{k}n}-\mu$, we see a slight enhancement of allowed phase space for a selfconsistent solution in the superconducting state, compare Fig.\,\ref{ML_hao_phasediagramUw}. The scales along $U$ are slightly different now due to a change in boundaries regarding the first Stoner instability. This aspect is explicitly discussed in Appendix \ref{appModifyTB}. As in the case of our original dispersion, an increase in Hund's rule coupling leads to larger superconducting gaps near the onset of magnetic order, see especially Fig.\,\ref{ML_rigid_phasediag}(c). The largest values possible for a rigidly shifted dispersion with increased FS pockets is about $7\,\mathrm{meV}$. This particular configuration $(U,J,\omega_{\mathrm{cut}})=(1.16\,\mathrm{eV},U/2,0.42\,\mathrm{eV})$ leads to a critical temperature of $T_c\sim8.3\,\mathrm{K}$, which is the maximal value found in all the parameter space explored. Compared to Fig.\,\ref{ML_hao_phasediagramUw} these observations point clearly towards an enhanced  ratio $\Delta/T_c\approx 9.8$, which can not be expected to be compatible with experiment. Generally we observe that the enhancement in the FS size leads to increased superconducting gaps, which can be explained by the larger DOS at the Fermi level, but not to higher critical temperatures. The FS pocket sizes are very well comparable to the experimentally observed situation, although the resulting upper (lower) position of the hole (electron) bands are not \cite{Lee2014}. However, we still observe that spin fluctuations are not sufficient to give the correct characteristics of FeSe/STO.

When examining a decrease in distance between the bottom of the electron band at $X$ and the top of hole bands at $M$ we do not find any superconductivity for $J=U/10$ in Fig.\,\ref{ML_rigid_phasediag}(d). Further, the available phase space for $J=U/6$ (panel (e)) and $J=U/2$ (panel (f)) is significantly smaller than in corresponding graphs of Fig.\,\ref{ML_hao_phasediagramUw}. The maximum gap possible for using a nonrigid shift of $\delta\mu$ occurs at $J=U/6$ and is around $8\,\mathrm{meV}$. We can explain these results by a qualitative comparison to bulk FeSe. Although the hole bands at $M$ do not cross the Fermi level for $\xi_{\mathbf{k}n}-\delta\mu$, the associated coupling to electron pockets is significantly enhanced through an exchange momentum $\mathbf{q}=X$. For the bulk material we find a hole band also at $\Gamma$, which in the current case is still too far away from the Fermi level to contribute significantly, compare Fig.\,\ref{ML_hao_setup}(a). From above arguments it follows that the leading instability for the nonrigidly shifted dispersion is still at $\mathbf{q}=M$, favoring a sign change between FS pockets, and hence the $d$-wave solution as observed in Fig.\,\ref{ML_hao_gapsym}(c). Combining this with an enhanced coupling at $X$ there might be a need of changing sign on the incipient hole band pockets, introducing a node. As apparent, such a solution is hard to accomplish, which means we need to choose $U$ very close to its critical value, such that the relative significance of couplings at $X$ is suppressed. This qualitative argument explains why selfconsistent solutions for the superconducting state are found only very close to magnetic order in panels Fig.\,\ref{ML_rigid_phasediag}(e) and (f), while the gap size for these confined regions is enhanced, in comparison to Fig.\,\ref{ML_hao_phasediagramUw}, to a maximum of almost $8\,\mathrm{meV}$. For dispersion $\xi_{\mathbf{k}n}-\delta\mu$ and $J=U/6$ we calculate the largest critical temperature as $T_c\sim11.4\,\mathrm{K}$ by choosing $(U,\omega_{\mathrm{cut}})=(2.48,0.18)\,\mathrm{eV}$, which is a slight enhancement when compared to results obtained for our actual dispersion, though by far not large enough to account for experimental values. We note that this particular value is rather artificial in any case, since corresponding positions of the hole bands are far from angular resolved photoemission spectroscopy (ARPES) measurements \cite{Lee2014}, see also Appendix \ref{appModifyTB}.

It has been proposed that the just-discussed incipient band coupling can provide an explanation for the high critical temperature in FeSe/STO, imposing an unconventional pairing mechanism \cite{Linscheid2016}. Our calculations presented in this work point, however, towards the opposite direction, i.e.\ that incipient interaction leads to a decrease in the phase space available for superconductivity, while mainly increasing the gap size and hence the $\Delta/T_c$ ratio. This discrepancy can be attributed to neglected intra-band interactions in Ref.\,\cite{Linscheid2016}, which in our framework  provide the dominant contributions, compare Fig.\,\ref{ML_hao_kernels}, since FS points can be connected mainly via $\mathbf{q}\sim\Gamma$ and $\mathbf{q}\sim M$. Moreover, here we do not constrain the symmetry of the superconducting order parameter \cite{Linscheid2016}, instead we obtain it from the selfconsistent calculation.
 
\begin{figure}[b!]
	\centering
	\includegraphics[width=1\columnwidth]{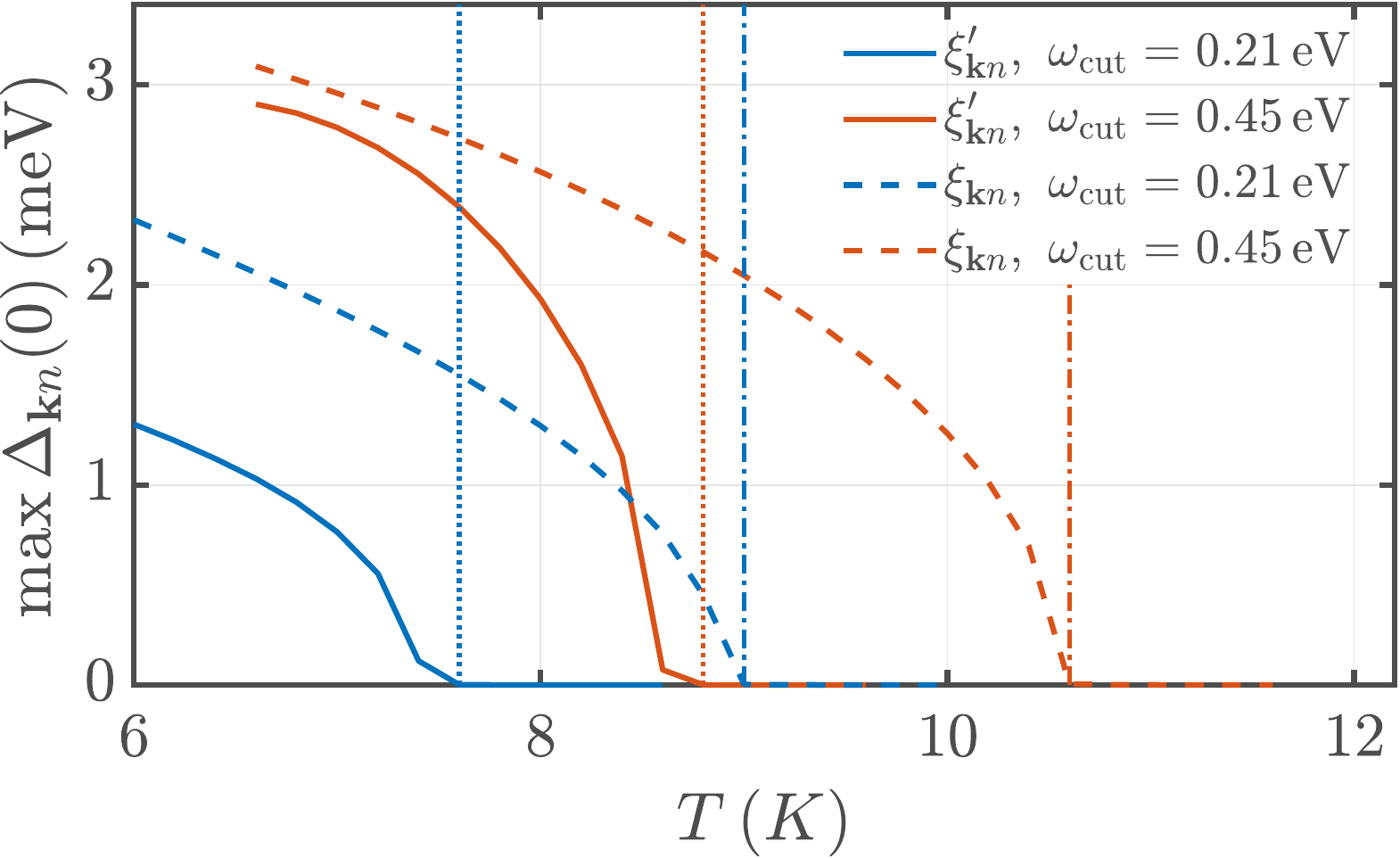}
	\caption{Computed closing of the maximum superconducting gap with temperature for $J=U/10$ and $U=1.5802\,\mathrm{eV}$. Solid (dashed) lines are found from $\xi_{\mathbf{k}n}$ ($\xi_{\mathbf{k}n}'$) dispersions. For the blue and red curves we use cutoffs $\omega_{\mathrm{cut}}=0.21\,\mathrm{eV}$ and $\omega_{\mathrm{cut}}=0.45\,\mathrm{eV}$, respectively.}
	\label{ML_hao_incipientgap}
\end{figure}

To further examine the relevance of incipient pairing within our theory, we now remove all hole bands in the electron dispersions and couplings artificially on the level of Matsubara kernels from Eq.\,(\ref{bandkernTrunc}). With only the electron bands (for simplicity referred to as $\xi_{\mathbf{k}n}'$) left we perform our selfconsistent Eliashberg calculations and follow the maximum superconducting gap with temperature, see Fig.\,\ref{ML_hao_incipientgap}. 
We use $(U,J)=(1.5802\,\mathrm{eV},U/10)$ and run the simulations for our dispersion $\xi_{\mathbf{k}n}$ (dashed) and the reduced $\xi_{\mathbf{k}n}'$ (solid). Two characteristic cutoff frequencies $\omega_{\mathrm{cut}}=0.21\,\mathrm{eV}$ and $\omega_{\mathrm{cut}}=0.45\,\mathrm{eV}$ are shown in blue and red colors, respectively. The onset of superconductivity is marked via dotted (dashed-dotted) straight lines for $\xi_{\mathbf{k}n}$ ($\xi_{\mathbf{k}n}'$).

For cutoff $\omega_{\mathrm{cut}}=0.21\,\mathrm{eV}$ we find a decrease of $T_c$ from $9.0\,\mathrm{K}$ to $7.6\,\mathrm{K}$ due to neglecting the hole bands, while the critical temperature goes from $10.6\,\mathrm{K}$ to $8.8\,\mathrm{K}$ for frequency $\omega_{\mathrm{cut}}=0.45\,\mathrm{eV}$. Hence we find that in both cases almost 85\% of the pairing stems from the electron bands only. This in turn reveals that the role of hole bands is rather minor in monolayer FeSe, compared to its bulk parent compound. From the results in Fig.\,\ref{ML_hao_incipientgap} and the discussion about changing the dispersion via a nonrigid $\delta\mu$ we can safely conclude that the incipient band scenario put forward in Ref.\,\cite{Linscheid2016} does not lead to a satisfactory enhancement of $T_c$ to values near the experimentally observed ones in a full bandwidth Eliashberg theory for spin fluctuations.

\subsection{Simplified calculations}\label{scBCS}

We now want to compare our results to computationally less demanding and more effective theories, as they are rather commonly employed \cite{Scalapino1986,Graser2009,Kemper2010}. For this purpose we use the zero-frequency kernel $V_{\mathbf{q}}^{nn'}=\big[V_{\mathbf{q}}^{(-)}(0)\big]_{nn'}$ and start with a linearized BCS equation at the Fermi surface. By decoupling $\Delta_{\mathbf{k}n}=\Delta g_{\mathbf{k}}$, with $g_{\mathbf{k}}$ a global symmetry form factor, we solve for the coupling via \cite{Scalapino1986,Graser2009}
\begin{align}
\lambda[g_{\mathbf{k}}] = - \frac{\sum_{nn'} \oint_{C_n} \frac{d\mathbf{k}}{v_{\mathbf{k}}}   \oint_{C_{n'}}\frac{d\mathbf{k}'}{v_{\mathbf{k}'}} g_{\mathbf{k}} V_{\mathbf{q}}^{nn'} g_{\mathbf{k}'}}{(2\pi)^2 \sum_n \oint_{C_n} \frac{d\mathbf{k}}{v_{\mathbf{k}}} g^2_{\mathbf{k}}}  . \label{linearBCS}
\end{align}
In Eq.\,(\ref{linearBCS}) we integrate over FS sheets $C_n$ and test form factors $g_{\mathbf{k}}\in\{\hat{1},\cos(k_x)+\cos(k_y),\cos(k_x)-\cos(k_y),$ $\cos(k_x)\cos(k_y),\sin(k_x)\sin(k_y)\}$. Among the $g_{\mathbf{k}}$, the prevailing symmetry is determined by the largest coupling. We test this setup for all cutoffs  shown in Fig.\,\ref{ML_hao_phasediagramUw} with an  enlarged range for $U$ towards smaller values, and with all three ratios of $U/J$ as in panels (a), (b) and (c) of this figure. Our results reveal that the leading $\lambda$ has exclusively $d$-wave symmetry, which  implies  that our selfconsistent result in Fig.\,\ref{ML_hao_gapsym}(c) is governed by FS contributions. Since {on this level of approximation} any value $\lambda>0$ leads to a finite $T_c$ \cite{Carbotte2003}, solutions are found for all $U$ and $\omega_{\mathrm{cut}}$, which generally overestimates largely the available phase space allowing for superconductivity. In addition, we find hardly any dependence on $\omega_{\mathrm{cut}}$, which is easily explained by the fact that this approximation neglects the frequencies and all momenta away from the Fermi level.

\begin{figure}[t!]
	\centering
	\includegraphics[width=1\columnwidth]{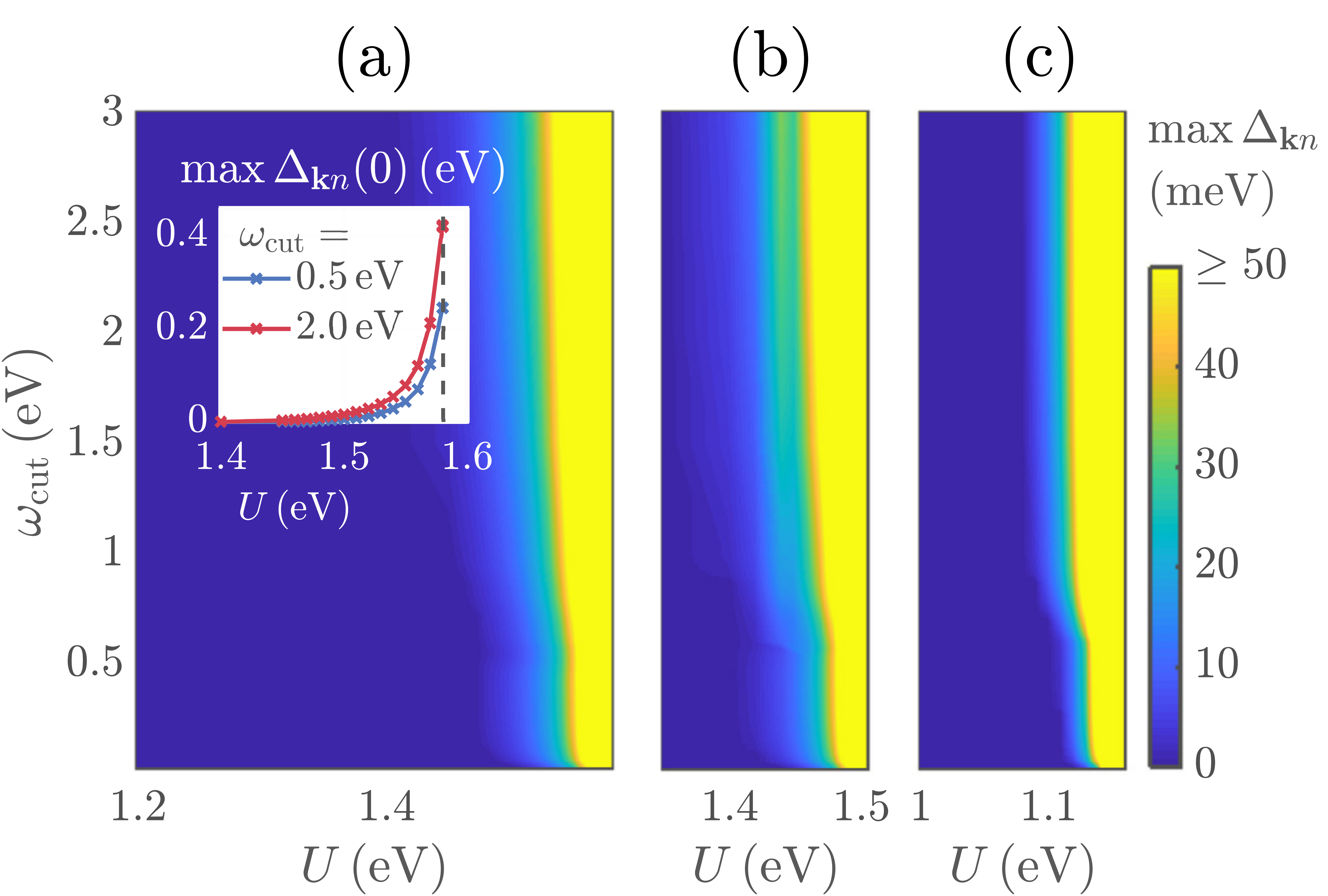}
	\caption{Solution for the maximum superconducting gap from BCS Eq.\,(\ref{bcsDelta}), calculated at $T=5\,\mathrm{K}$. (a) $J=U/10$, (b) $J=U/6$, (c) $J=U/2$. Inset: Maximum gap as function of $U$ for two cutoff frequencies, $\omega_{\mathrm{cut}}=0.5\,\mathrm{eV}$ (blue curve) and $2.0\,\mathrm{eV}$ (red curve), using $J=U/10$.}
	\label{ML_hao_bcs_phasediagramUw}
\end{figure}

A less drastic approximation can be made by neglecting only the Matsubara frequency components but keeping the full momentum dependence. The selfconsistent BCS equation for the superconducting gap is then found as
\begin{align}
&\Delta_{\mathbf{k}n} = -\sum_{\mathbf{k}'n'}V_{\mathbf{q}}^{nn'}\frac{\Delta_{\mathbf{k}'n'}}{2E_{\mathbf{k}'n'}} \tanh\left(\frac{E_{\mathbf{k}'n'}}{2T}\right) , \label{bcsDelta}
\end{align}
with $E_{\mathbf{k}n} = \sqrt{\Delta_{\mathbf{k}n}^2 + \xi_{\mathbf{k}n}^2}$ \cite{Carbotte2003}. We solve the above equation at $T=5\,\mathrm{K}$ as function of $U$ and $\omega_{\mathrm{cut}}$. Results for the maximum superconducting gap in case of $J=U/10$, $J=U/6$, and $J=U/2$ are shown in panels (a), (b) and (c) of Fig.\,\ref{ML_hao_bcs_phasediagramUw}. The inset of (a) shows the dependence of $\mathrm{max}\,\Delta_{\mathbf{k}n}$ on $U=10J$ for two frequency cutoffs $\omega_{\mathrm{cut}}=0.5\,\mathrm{eV}$ in blue and $\omega_{\mathrm{cut}}=2.0\,\mathrm{eV}$ in red.
Comparing these results to Fig.\,\ref{ML_hao_phasediagramUw} immediately reveals a large increase of the superconducting gap sizes in all three panels. Close to the Stoner instability $\Delta$ reaches values as large as few hundred $\mathrm{meV}$'s, which is by no means compatible with experiment.  Such a large enhancement is not particularly surprising given that BCS theory has no quantitative use beyond the very weak-coupling limit where the mass renormalization is negligible. In addition the onset of superconductivity occurs at smaller $U$, while the dependence on $\omega_{\mathrm{cut}}$, as observed in Fig.\,\ref{ML_hao_phasediagramUw}, is almost completely lost. As the inset of Fig.\,\ref{ML_hao_bcs_phasediagramUw}(a) clearly shows, the gap diverges for $U$ close to the Stoner instability. This behavior translates into almost arbitrary gap magnitudes and corresponding critical temperatures, which is clearly a deficiency of BCS theory. In the following Sec.\,\ref{scUpperBoundTc} we discuss the aspects of criticality in $U$ within both, BCS and Eliashberg treatments, and explicitly show how the aforementioned problem is cured when solving the full set of Eqs.\,(\ref{z})-(\ref{phi}).

Our calculations in the current section reveal that spin-fluctuation mediated pairing is drastically overestimated within theories as BCS that neglect any frequency dependence \cite{Gao2017,Linscheid2016,Kreisel2017}. This holds true for the gap magnitude (and hence $T_c$) and the available phase space with respect to the choice of $(U,J)$.

\subsection{An upper bound for the gap magnitude}\label{scUpperBoundTc}

The inset of Fig.\,\ref{ML_hao_phasediagramUw} shows an approximately linear decrease of the superconducting gap as function of $U$, obtained from our Eliashberg formalism. Contrarily we find approximately a $1/U$ trend when solving the BCS equation in Sec.\,\ref{scBCS}{, compare with the inset of Fig.\,\ref{ML_hao_bcs_phasediagramUw}(a)}. Below we use some {qualitative arguments} 
to reproduce these trends, which in case of the Eliashberg theory allows us to give a realistic estimation of the maximally possible gap size in FeSe/STO within our unconventional theory presented here.

\begin{figure}[h!]
	\centering
	\includegraphics[width=1\columnwidth]{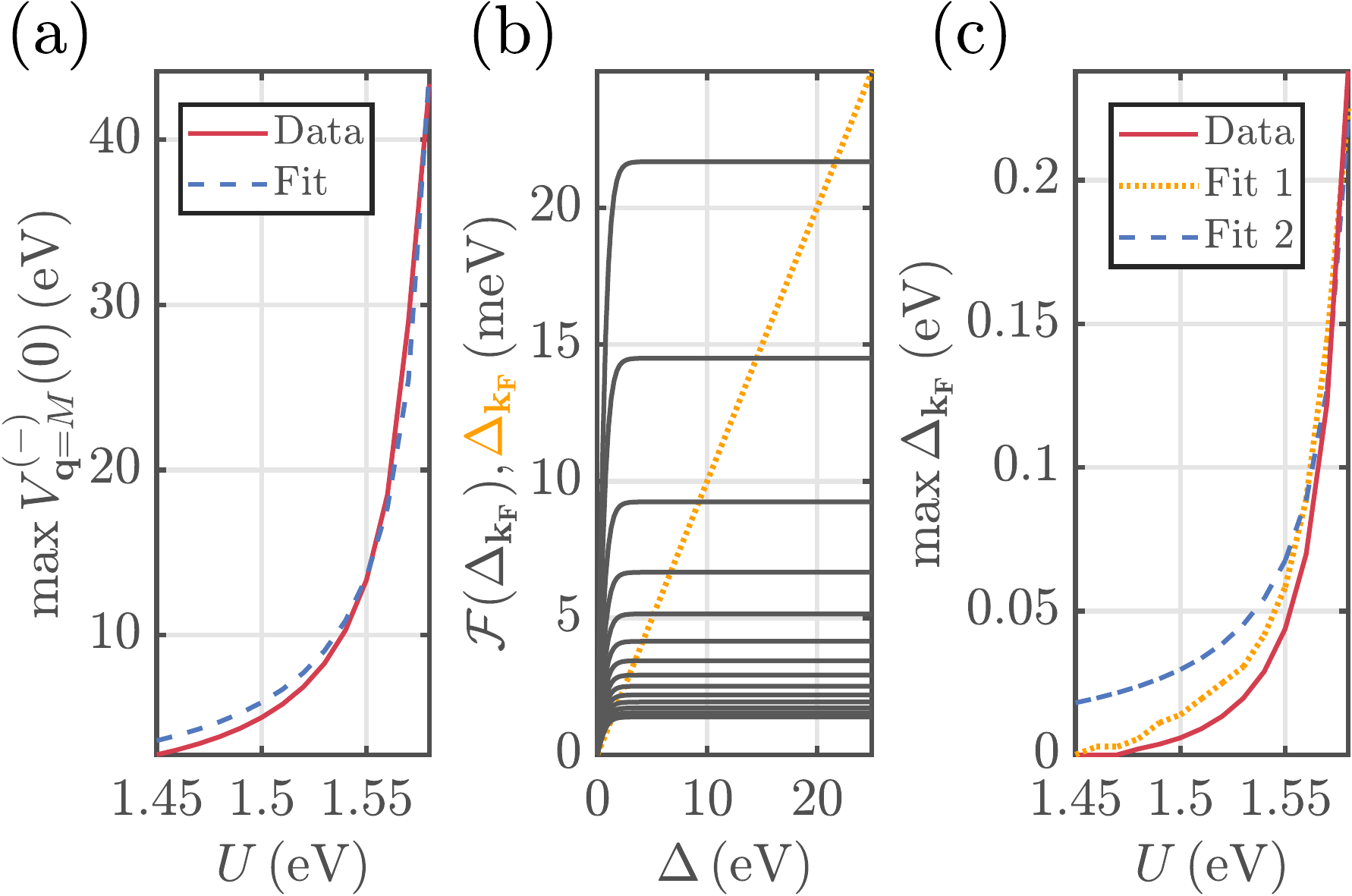}
	\caption{Calculated and fitted scalings for $(U,J)=(1.5802\,\mathrm{eV},U/10)$ and $\omega_{\mathrm{cut}}=0.45\,\mathrm{eV}$. (a) Maximum static interaction kernel at $\mathbf{q}=M$ as function of $U$. The actual values are shown by the solid red curve, the fit by the blue dashed line using Eq.\,(\ref{kernelFit}). (b) Orange dotted (black solid) curve shows the left (right) hand side of Eq.\,(\ref{simplifiedGap}). (c) Increase of the superconducting gap as function of $U$ in the BCS framework. Red solid line: selfconsistently calculated results. Blue dashed line: Fit obtained by using (a) and the scaling $\Delta_{\mathbf{k}_{\rm F}}\propto V$. Orange dotted curve: graphical solution of (b) scaled appropriately.}
	\label{ML_bcs_scaling}
\end{figure}

Let us start from the BCS Eq.\,(\ref{bcsDelta}). If we confine ourselves to the FS we can set $\xi_{\mathbf{k}n}=0$, hence $E_{\mathbf{k}n}=\Delta_{\mathbf{k}n}$. Since in the unfolded BZ only a single band is crossing the Fermi level we omit the band index $n$ and get
\begin{align}
\Delta_{\mathbf{k}_{\rm F}} \sim -\frac{1}{2}\sum_{\mathbf{k}'_{\rm F}} V_{\mathbf{q}} \tanh\left(\frac{\Delta_{\mathbf{k}'_{\rm F}}}{2T}\right) .
\end{align}
The leading instability occurs at $\mathbf{q}=M$, so we write $V_{\mathbf{q}}=V\delta(\mathbf{q}-M)$ with $V$ a function of the intraorbital coupling only. Using the sign change among FS pockets $\Delta_{\mathbf{k}_{\rm F}-M}=-\Delta_{\mathbf{k}n}$ yields
\begin{align}
\Delta_{\mathbf{k}_{\rm F} }\sim -\frac{V}{2}\tanh\left(\frac{\Delta_{\mathbf{k}_{\rm F}}}{2T}\right) 
 \equiv  \mathcal{F}(\Delta_{\mathbf{k}_{\rm F}}), \label{simplifiedGap}
\end{align}
which can be solved graphically as function of $V=V(U)$, see Fig.\,\ref{ML_bcs_scaling}(b). There we show the left and right hand side of Eq.\,(\ref{simplifiedGap}) in dotted orange and solid black, respectively, for $U=1.5802$\,eV, $J=U/10$, and $\omega_{\mathrm{cut}}=0.45\,\mathrm{eV}$, focusing on a low-energy region of $\Delta$. The crossing points indicate valid solutions, which are shown in panel (c) as the dotted orange curve. The actual values drawn in solid red are reproduced quite accurately, indicating that we have the correct functional dependencies.

As an alternative one can start from an estimate of the scaling for $V(U)$ by considering the spin susceptibility in Eq.\,(\ref{rpaMatS}). Close to the Stoner instability $\hat{\chi}^{S}_{\mathbf{q}}(0)$ grows as $1/(U^{\mathrm{crit}}-U)$, with leading contributions at $\mathbf{q}=M$. Inserting this in Eq.\,(\ref{orbitalkernM}) and neglecting all {but the first contribution, which represents the coupling via the spin degree of freedom,} 
gives a scaling $V\propto U^2/(U^{\mathrm{crit}}-U)$. We fit the zero-frequency maximum kernel for the superconducting channel by
\begin{align}
V \sim \frac{U^2}{U^{\mathrm{crit}}-U} \label{kernelFit}
\end{align}
and find $U^{\mathrm{crit}}=1.594\,\mathrm{eV}$. If $\Delta_{\mathbf{k}_{\rm F}}$ is sufficiently large we can neglect the hyperbolic tangent in Eq.\,(\ref{simplifiedGap}), and hence the gap grows as $\Delta_{\mathbf{k}_{\rm F}}\propto V\propto U^2/(U^{\mathrm{crit}}-U)$. This approach is depicted by blue dashed lines in Fig.\,\ref{ML_bcs_scaling}(a) and (c). In panel (a) we compare our actual data (red, solid) for the maximum static Matsubara kernels with the fitted form of Eq.\,(\ref{kernelFit}) in blue and dashed, leading to an accurate agreement. This particular functional form serves as input for the comparison in panel (c), using similar color code. As expected the scaling is very accurate for critical $U\rightarrow U^{\mathrm{crit}}$, while less precise for smaller $U$.

From the above discussions we learn that the scaling of the interaction kernel is directly translated to the order parameter, leading to a $1/U$ divergence of the superconducting gap as $U\rightarrow U^{\mathrm{crit}}$ in the BCS approximation. Such a behavior does not correspond to the actual physical situation and is an artifact of a too simplistic modeling. Below we show that such a divergence does not occur in our Eliashberg formalism due to the explicit inclusion of the electronic mass renormalization.

We use similar assumptions as before, i.e.\ we confine Matsubara frequency indices to $m=m'=0$ and $\mathbf{k}$ at the Fermi level. The latter condition within Eliashberg theory translates as $\xi_{\mathbf{k}n}+\Gamma_{\mathbf{k}n}(0)=0$, hence we only consider Eqs.\,(\ref{z}) and (\ref{phi}). Let us assume for the moment that $T>T_c$, then $\phi_{\mathbf{k}_{\rm F}}=0$ and
\begin{align}
Z_{\mathbf{k}_{\rm F}} \sim 1 + \frac{1}{\pi^2 T}\sum_{\mathbf{k}'_{\rm F}} V_{\mathbf{q}}\frac{1}{Z_{\mathbf{k}'_{\rm F}}} ~.
\end{align}
Taking the kernel again as $V_{\mathbf{q}}=V\delta(\mathbf{q}-M)$, together with $Z_{\mathbf{k}_{\rm F}-M}=Z_{\mathbf{k}_{\rm F}}$ leads to $Z_{\mathbf{k}_{\rm F}} \sim 1 + V/\pi^2 TZ_{\mathbf{k}_{\rm F}}$. The solution to this second order polynomial is given by
\begin{align}
Z_{\mathbf{k}_{\rm F}} \sim \frac{1}{2} + \sqrt{\frac{1}{4} + \frac{V}{\pi^2 T}} ~. \label{Zfit}
\end{align}
Next, we assume that the mass renormalization does not change significantly when going to the superconducting state. The order parameter can then be simplified as $\phi_{\mathbf{k}_{\rm F}}=\sqrt{TV-\pi^2T^2Z_{\mathbf{k}_{\rm F}}^2}$, where we use similar arguments that led to Eqs.\,(\ref{simplifiedGap}) and (\ref{Zfit}). Finally, the superconducting gap function then reads
\begin{align}
\Delta_{\mathbf{k}_{\rm F}} \sim \sqrt{\frac{TV}{Z_{\mathbf{k}_{\rm F}}^2} - \pi^2T^2} ~. \label{DeltaFit}
\end{align}

\begin{figure}[t!]
	\centering
	\includegraphics[width=1\columnwidth]{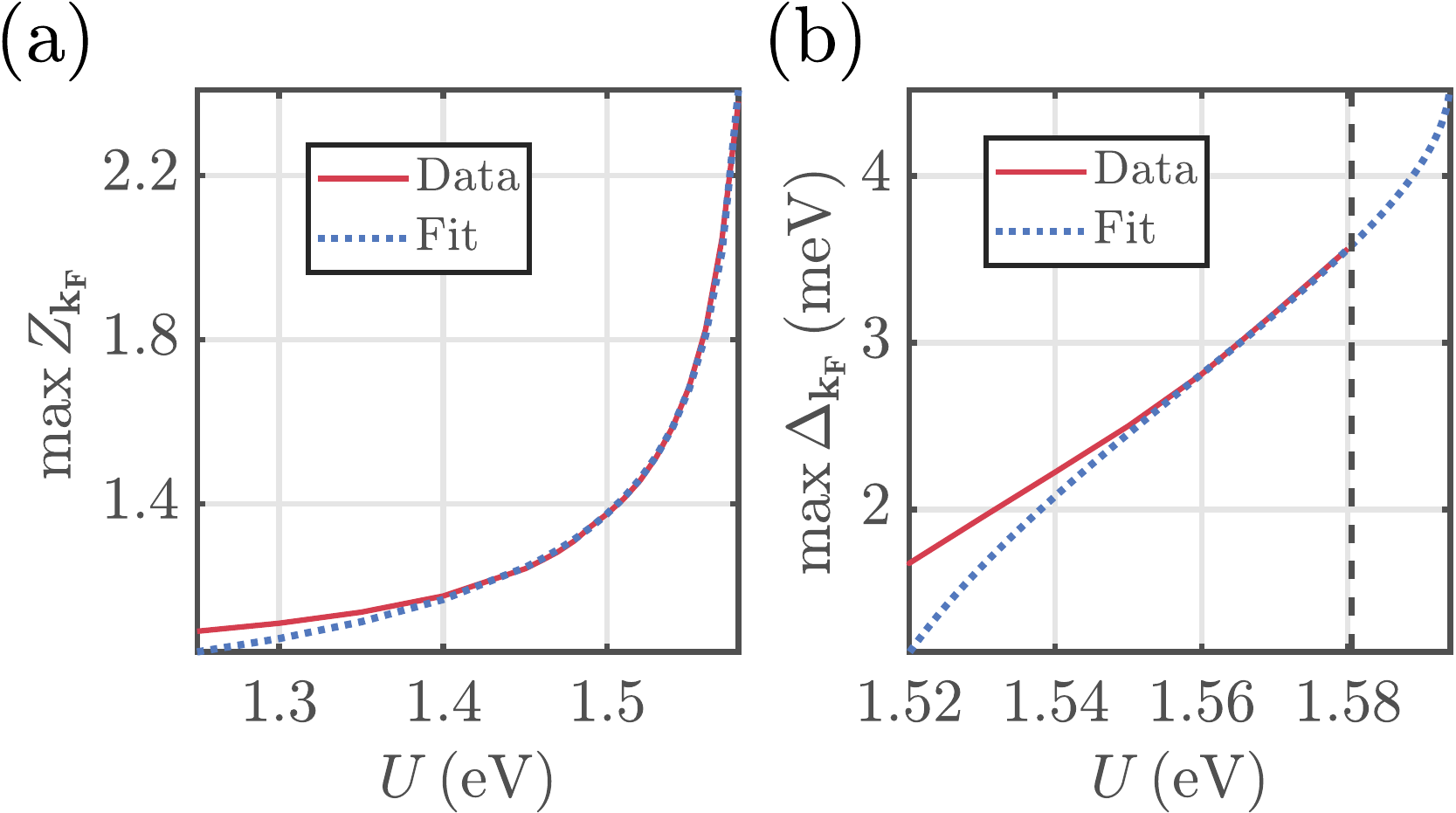}
	\caption{Extremal values of $Z_{\mathbf{k}_{\rm F}}$ and $\Delta_{\mathbf{k}_{\rm F}}$ close to the Stoner instability, obtained at $T=5\,\mathrm{K}$, $J=U/10$ and $\omega_{\mathrm{cut}}=0.45\,\mathrm{eV}$. (a) Maximum mass renormalization as function of $U$. The red solid line shows the actual data, the blue dotted curve is our fitting result from Eq.\,(\ref{Zfit}). (b) Maximum superconducting gap with the same color code as (a). The fitting is done via Eq.\,(\ref{DeltaFit})}
	\label{ML_eliash_scaling}
\end{figure}

In Fig.\,\ref{ML_eliash_scaling} we present the solutions to fitting our actual data to the simplified dependencies of Eqs.\,(\ref{Zfit}) and (\ref{DeltaFit}). Panel (a) shows the maximum mass renormalization on the FS as function of intraorbital coupling, where selfconsistently obtained results from the Eliashberg equations are drawn in solid red lines and the fit via Eq.\,(\ref{Zfit}) in dotted blue lines. Close to the instability we retrieve the actual data very accurately, confirming the proposed fitting function. Note that $Z_{\mathbf{k}_{\rm F}}$ diverges as $1/\sqrt{U^{\mathrm{crit}}-U}$. In graph (b) of the same figure we model the maximum superconducting gap on the Fermi level via Eq.\,(\ref{DeltaFit}) by the dotted blue curve, while our actual results are again given by solid red lines. The onset of magnetic order is indicated as gray dashed border, where the corresponding $U$ is found from Eq.\,(\ref{rpaMatS}), yielding a value slightly smaller than the fitted $U^{\mathrm{crit}}$. Since, in contrast to the BCS situation, $\Delta_{\mathbf{k}_{\rm F}}$ does not diverge close to the border we can extract the maximally possible gap by imposing $U=1.5805\,\mathrm{eV}$ precisely at the instability. The upper bound obtained in this way is $\underset{U}{\mathrm{max}}\,\Delta_{\mathbf{k}_{\rm F}}=3.585\,\mathrm{meV}$, which is close to results already found in Fig.\,\ref{ML_hao_gapsym}(a) due to a rather critical choice of $U$ therein.

From these results we learn two important aspects: First, the Eliashberg formalism employed in this work removes the unphysical divergence of the superconducting gap (and hence $T_c$) as function of $U$. This is due to a mild divergence of the mass renormalization, which counteracts the scaling of the interaction kernel, such that $\Delta$ grows only linear with $U$ in the instability region close to magnetic order. On the other hand, we can consider our results for the superconducting gap, and hence the transition temperatures, of Fig.\,\ref{ML_hao_gapsym} as upper bounds already, since all values of $U$ employed for associated calculations have been chosen very close to the Stoner instability.

\section{Discussion and Conclusions}\label{scDiscussion}


Before formulating our conclusions it is appropriate to discuss possible limitations of the here-developed formalism.

First, to start with, there is the influence that the employed tight-binding energy bands could have.  For example, the tight-binding models used here do not account for spin-orbit coupling. Nonetheless, we have taken care to simulate the influences due to changes in the near-Fermi energy bands {of FeSe/STO} in Sec.\,\ref{scInfluenceTB} and did not find an appreciable change in the superconductivity characteristics. We have furthermore performed our simulations in the unfolded, instead of the folded BZ due to computational performance, but changes due to this aspect are presumably small as well.

Second, in the theory presented in Sec.\,\ref{scMethod} we introduce a frequency cutoff to truncate the real-frequency dependent kernels when transforming into Matsubara space. This particular cutoff is one of the key ingredients to find selfconsistent solutions in the superconducting state. 
With this procedure, we are able to controllably remove the Stoner continuum-like incoherent  end of the magnetic spectrum and the concomitant high energy diverging tendencies which are not relevant
to the low-energy superconducting phenomena. Ideally, these high-energy degrees of freedom should be integrated out and used to renormalize the remaining interactions, in a manner similar to the well-known treatment of the Coulomb interaction in the electron-phonon problem that leads to the low-energy Coulomb pseudopotential $\mu^*$ \cite{Carbotte2003}. Such a procedure would however increase considerably the already high complexity and computational cost of the here- proposed framework and is therefore out of the scope of the present work. 
The influence of such corrections to our method is expected to be rather minor since our results for bulk FeSe are very accurate already. By enabling the systematic numerical solution of the Eliashberg equations over a broad energy range, our method provides a way of mapping out the relevant parts of the spin-fluctuation spectrum that are important to the pairing and therefore provides new insights to electronic mechanisms of superconductivity.

{Third}, as a further step closer to the definite answer to the superconductivity mediating mechanisms in bulk FeSe and FeSe/STO, Eliashberg theory calculations, wherein both spin fluctuations and electron-phonon coupling are treated on the same footing are required. Nonetheless, a clear hint towards the outcome of such an investigation  is provided in this work, together with studies \cite{Lee2014,Rademaker2016,Aperis2018,Schrodi2018} 
of phonon-mediated superconductivity in FeSe/STO. The evidence collected in the latter studies speaks clearly in favor of interfacial electron-phonon coupling as dominant contribution to the superconducting gap, and hence its high $T_c$.

Summarizing,  we have presented a full Eliashberg treatment of spin-fluctuations mediated superconductivity with broad applicability. While our results for bulk and monolayer FeSe are representative for a wide spectrum in the class of iron-based superconductors, the here-presented Eliashberg formalism provides a way of examining the influence of magnetic fluctuations in arbitrary materials, with and without superconductivity. With a faithful tight-binding model for bulk FeSe as the only input needed for our treatment, we achieve for FeSe good agreement with known experimental quantities in the superconducting state: our maximum gap is $1.4\,\mathrm{meV}$, compared to measurements that yield $1.67\,\mathrm{meV}$ \cite{Jiao2017}. The selfconsistently computed gap symmetry of $s_{\pm}+s$ type is also obtained correctly. Further, we find a critical temperature of $6\,\mathrm{K}$, which compares well to the measured $T_c\sim8\,\mathrm{K}$ \cite{Hsu2008}. 
As the outcomes presented are consistent with the main characteristics of superconductivity in this material, we are confident that our microscopic treatment captures the important physics and thus provides strong support for spin-fluctuations mediated superconductivity in bulk FeSe.

Applying the same methodology to monolayer FeSe on STO we find a clear discrepancy between computed results for the superconducting state and well established experimental facts as e.g.\ the critical temperature. 
Although the tight-binding model we use for our main findings might deviate to {some degree} from reality, we have explicitly  tested the influence of changing the FS size {and} distance between electron and hole bands. For the former situation, i.e.\ a rigid shift in the electronic dispersions, the maximum gap size is increased to $\sim7\,\mathrm{meV}$. Changing the electron-hole distance on the other hand leads to even $\sim8\,\mathrm{meV}$, which is not far from experimental values. The critical temperature, however, does not noticeably increase and stays at the order of $10\,\mathrm{K}$. Our results further indicate that band incipiency has little effect on $T_c$ and therefore,
solely spin fluctuation in combination with band incipiency
cannot explain superconductivity in FeSe/STO.
{As we showed, the  $T_c$ value computed with full-bandwidth Eliashberg theory is notably lower than that predicted by BCS theories based on spin fluctuations, 
due to approximations involved in the latter. 
We have explicitly shown that within Eliashberg theory there exists an upper limit to the obtained $T_c$ as $U \rightarrow U^{\rm crit}$, in contrast to BCS theory calculations where the gap values and critical temperatures can be almost arbitrarily increased by choosing
$U$ close to $U^{\rm crit}$ of the 
antiferromagnetic transition. This emphasizes an important quantitative limitation of the RPA-BCS approach one should be aware of when comparing with experiments.
 Overall, our results lead us to surmise that superconductivity due to spin and charge fluctuations could be possible in FeSe/STO, but only at temperatures slightly larger than the $T_c\sim8\,\mathrm{K}$ of bulk FeSe.  Consequently, we conclude that a spin-fluctuations mechanism alone cannot explain the observed high $T_c$ and that another, dominant pairing mechanism such as interfacial electron-phonon coupling must be responsible.

\begin{acknowledgments}
	This work has been supported by the Swedish Research Council (VR), the R{\"o}ntgen-{\AA}ngstr{\"o}m Cluster, the K.\ and A.\ Wallenberg Foundation (grant No.\ 2015.0060), and the Swedish National Infrastructure for Computing (SNIC).	
\end{acknowledgments}


\begin{thebibliography}{83}%
	\makeatletter
	\providecommand \@ifxundefined [1]{%
		\@ifx{#1\undefined}
	}%
	\providecommand \@ifnum [1]{%
		\ifnum #1\expandafter \@firstoftwo
		\else \expandafter \@secondoftwo
		\fi
	}%
	\providecommand \@ifx [1]{%
		\ifx #1\expandafter \@firstoftwo
		\else \expandafter \@secondoftwo
		\fi
	}%
	\providecommand \natexlab [1]{#1}%
	\providecommand \enquote  [1]{``#1''}%
	\providecommand \bibnamefont  [1]{#1}%
	\providecommand \bibfnamefont [1]{#1}%
	\providecommand \citenamefont [1]{#1}%
	\providecommand \href@noop [0]{\@secondoftwo}%
	\providecommand \href [0]{\begingroup \@sanitize@url \@href}%
	\providecommand \@href[1]{\@@startlink{#1}\@@href}%
	\providecommand \@@href[1]{\endgroup#1\@@endlink}%
	\providecommand \@sanitize@url [0]{\catcode `\\12\catcode `\$12\catcode
		`\&12\catcode `\#12\catcode `\^12\catcode `\_12\catcode `\%12\relax}%
	\providecommand \@@startlink[1]{}%
	\providecommand \@@endlink[0]{}%
	\providecommand \url  [0]{\begingroup\@sanitize@url \@url }%
	\providecommand \@url [1]{\endgroup\@href {#1}{\urlprefix }}%
	\providecommand \urlprefix  [0]{URL }%
	\providecommand \Eprint [0]{\href }%
	\providecommand \doibase [0]{http://dx.doi.org/}%
	\providecommand \selectlanguage [0]{\@gobble}%
	\providecommand \bibinfo  [0]{\@secondoftwo}%
	\providecommand \bibfield  [0]{\@secondoftwo}%
	\providecommand \translation [1]{[#1]}%
	\providecommand \BibitemOpen [0]{}%
	\providecommand \bibitemStop [0]{}%
	\providecommand \bibitemNoStop [0]{.\EOS\space}%
	\providecommand \EOS [0]{\spacefactor3000\relax}%
	\providecommand \BibitemShut  [1]{\csname bibitem#1\endcsname}%
	\let\auto@bib@innerbib\@empty
	\bibitem [{\citenamefont {Kamihara}\ \emph {et~al.}(2008)\citenamefont
		{Kamihara}, \citenamefont {Watanabe}, \citenamefont {Hirano},\ and\
		\citenamefont {Hosono}}]{Kamihara2008}%
	\BibitemOpen
	\bibfield  {author} {\bibinfo {author} {\bibfnamefont {Y.}~\bibnamefont
			{Kamihara}}, \bibinfo {author} {\bibfnamefont {T.}~\bibnamefont {Watanabe}},
		\bibinfo {author} {\bibfnamefont {M.}~\bibnamefont {Hirano}}, \ and\ \bibinfo
		{author} {\bibfnamefont {H.}~\bibnamefont {Hosono}},\ }\href@noop {}
	{\bibfield  {journal} {\bibinfo  {journal} {J. Am. Chem. Soc.}\ }\textbf
		{\bibinfo {volume} {130}},\ \bibinfo {pages} {3296} (\bibinfo {year}
		{2008})}\BibitemShut {NoStop}%
	\bibitem [{\citenamefont {Hsu}\ \emph {et~al.}(2008)\citenamefont {Hsu},
		\citenamefont {Luo}, \citenamefont {Yeh}, \citenamefont {Chen}, \citenamefont
		{Huang}, \citenamefont {Wu}, \citenamefont {Lee}, \citenamefont {Huang},
		\citenamefont {Chu}, \citenamefont {Yan},\ and\ \citenamefont
		{Wu}}]{Hsu2008}%
	\BibitemOpen
	\bibfield  {author} {\bibinfo {author} {\bibfnamefont {F.-C.}\ \bibnamefont
			{Hsu}}, \bibinfo {author} {\bibfnamefont {J.-Y.}\ \bibnamefont {Luo}},
		\bibinfo {author} {\bibfnamefont {K.-W.}\ \bibnamefont {Yeh}}, \bibinfo
		{author} {\bibfnamefont {T.-K.}\ \bibnamefont {Chen}}, \bibinfo {author}
		{\bibfnamefont {T.-W.}\ \bibnamefont {Huang}}, \bibinfo {author}
		{\bibfnamefont {P.~M.}\ \bibnamefont {Wu}}, \bibinfo {author} {\bibfnamefont
			{Y.-C.}\ \bibnamefont {Lee}}, \bibinfo {author} {\bibfnamefont {Y.-L.}\
			\bibnamefont {Huang}}, \bibinfo {author} {\bibfnamefont {Y.-Y.}\ \bibnamefont
			{Chu}}, \bibinfo {author} {\bibfnamefont {D.-C.}\ \bibnamefont {Yan}}, \ and\
		\bibinfo {author} {\bibfnamefont {M.-K.}\ \bibnamefont {Wu}},\ }\href
	{\doibase 10.1073/pnas.0807325105} {\bibfield  {journal} {\bibinfo  {journal}
			{Proc. Natl. Acad. Sci. USA}\ }\textbf {\bibinfo {volume} {105}},\ \bibinfo
		{pages} {14262 } (\bibinfo {year} {2008})}\BibitemShut {NoStop}%
	\bibitem [{\citenamefont {Rotter}\ \emph {et~al.}(2008)\citenamefont {Rotter},
		\citenamefont {Tegel},\ and\ \citenamefont {Johrendt}}]{Rotter2008}%
	\BibitemOpen
	\bibfield  {author} {\bibinfo {author} {\bibfnamefont {M.}~\bibnamefont
			{Rotter}}, \bibinfo {author} {\bibfnamefont {M.}~\bibnamefont {Tegel}}, \
		and\ \bibinfo {author} {\bibfnamefont {D.}~\bibnamefont {Johrendt}},\
	}\href@noop {} {\bibfield  {journal} {\bibinfo  {journal} {Phys. Rev. Lett.}\
	}\textbf {\bibinfo {volume} {101}},\ \bibinfo {pages} {107006} (\bibinfo
	{year} {2008})}\BibitemShut {NoStop}%
\bibitem [{\citenamefont {Medvedev}\ \emph {et~al.}(2009)\citenamefont
	{Medvedev}, \citenamefont {McQueen}, \citenamefont {Troyan}, \citenamefont
	{Palasyuk}, \citenamefont {Eremets}, \citenamefont {Cava}, \citenamefont
	{Naghavi}, \citenamefont {Casper}, \citenamefont {Ksenofontov}, \citenamefont
	{Wortmann},\ and\ \citenamefont {Felser}}]{Medvedev2009}%
\BibitemOpen
\bibfield  {author} {\bibinfo {author} {\bibfnamefont {S.}~\bibnamefont
		{Medvedev}}, \bibinfo {author} {\bibfnamefont {T.~M.}\ \bibnamefont
		{McQueen}}, \bibinfo {author} {\bibfnamefont {I.~A.}\ \bibnamefont {Troyan}},
	\bibinfo {author} {\bibfnamefont {T.}~\bibnamefont {Palasyuk}}, \bibinfo
	{author} {\bibfnamefont {M.~I.}\ \bibnamefont {Eremets}}, \bibinfo {author}
	{\bibfnamefont {R.~J.}\ \bibnamefont {Cava}}, \bibinfo {author}
	{\bibfnamefont {S.}~\bibnamefont {Naghavi}}, \bibinfo {author} {\bibfnamefont
		{F.}~\bibnamefont {Casper}}, \bibinfo {author} {\bibfnamefont
		{V.}~\bibnamefont {Ksenofontov}}, \bibinfo {author} {\bibfnamefont
		{G.}~\bibnamefont {Wortmann}}, \ and\ \bibinfo {author} {\bibfnamefont
		{C.}~\bibnamefont {Felser}},\ }\href@noop {} {\bibfield  {journal} {\bibinfo
		{journal} {Nat. Mater.}\ }\textbf {\bibinfo {volume} {8}},\ \bibinfo {pages}
	{630} (\bibinfo {year} {2009})}\BibitemShut {NoStop}%
\bibitem [{\citenamefont {Mazin}\ and\ \citenamefont
	{Schmalian}(2009)}]{Mazin2009}%
\BibitemOpen
\bibfield  {author} {\bibinfo {author} {\bibfnamefont {I.~I.}\ \bibnamefont
		{Mazin}}\ and\ \bibinfo {author} {\bibfnamefont {J.}~\bibnamefont
		{Schmalian}},\ }\href {\doibase
	http://dx.doi.org/10.1016/j.physc.2009.03.019} {\bibfield  {journal}
	{\bibinfo  {journal} {Physica C}\ }\textbf {\bibinfo {volume} {469}},\
	\bibinfo {pages} {614 } (\bibinfo {year} {2009})}\BibitemShut {NoStop}%
\bibitem [{\citenamefont {Johnston}(2010)}]{Johnston2010}%
\BibitemOpen
\bibfield  {author} {\bibinfo {author} {\bibfnamefont {D.~C.}\ \bibnamefont
		{Johnston}},\ }\href@noop {} {\bibfield  {journal} {\bibinfo  {journal} {Adv.
			Phys.}\ }\textbf {\bibinfo {volume} {59}},\ \bibinfo {pages} {803 } (\bibinfo
	{year} {2010})}\BibitemShut {NoStop}%
\bibitem [{\citenamefont {Stewart}(2011)}]{Stewart2011}%
\BibitemOpen
\bibfield  {author} {\bibinfo {author} {\bibfnamefont {G.~R.}\ \bibnamefont
		{Stewart}},\ }\href {\doibase 10.1103/RevModPhys.83.1589} {\bibfield
	{journal} {\bibinfo  {journal} {Rev. Mod. Phys.}\ }\textbf {\bibinfo {volume}
		{83}},\ \bibinfo {pages} {1589} (\bibinfo {year} {2011})}\BibitemShut
{NoStop}%
\bibitem [{\citenamefont {Wang}\ and\ \citenamefont {Lee}(2011)}]{Wang2011}%
\BibitemOpen
\bibfield  {author} {\bibinfo {author} {\bibfnamefont {F.}~\bibnamefont
		{Wang}}\ and\ \bibinfo {author} {\bibfnamefont {D.-H.}\ \bibnamefont {Lee}},\
}\href@noop {} {\bibfield  {journal} {\bibinfo  {journal} {Science}\ }\textbf
{\bibinfo {volume} {332}},\ \bibinfo {pages} {200 } (\bibinfo {year}
{2011})}\BibitemShut {NoStop}%
\bibitem [{\citenamefont {Dai}(2015)}]{Dai2015}%
\BibitemOpen
\bibfield  {author} {\bibinfo {author} {\bibfnamefont {P.}~\bibnamefont
		{Dai}},\ }\href@noop {} {\bibfield  {journal} {\bibinfo  {journal} {Rev. Mod.
			Phys.}\ }\textbf {\bibinfo {volume} {87}},\ \bibinfo {pages} {855 } (\bibinfo
	{year} {2015})}\BibitemShut {NoStop}%
\bibitem [{\citenamefont {Si}\ \emph {et~al.}(2016)\citenamefont {Si},
	\citenamefont {Yu},\ and\ \citenamefont {Abrahams}}]{Si2016}%
\BibitemOpen
\bibfield  {author} {\bibinfo {author} {\bibfnamefont {Q.}~\bibnamefont
		{Si}}, \bibinfo {author} {\bibfnamefont {R.}~\bibnamefont {Yu}}, \ and\
	\bibinfo {author} {\bibfnamefont {E.}~\bibnamefont {Abrahams}},\ }\href@noop
{} {\bibfield  {journal} {\bibinfo  {journal} {Nat. Rev. Mater.}\ }\textbf
	{\bibinfo {volume} {1}},\ \bibinfo {pages} {16017} (\bibinfo {year}
	{2016})}\BibitemShut {NoStop}%
\bibitem [{\citenamefont {Qing-Yan}\ \emph {et~al.}(2012)\citenamefont
	{Qing-Yan}, \citenamefont {Zhi}, \citenamefont {Wen-Hao}, \citenamefont
	{Zuo-Cheng}, \citenamefont {Jin-Song}, \citenamefont {Wei}, \citenamefont
	{Hao}, \citenamefont {Yun-Bo}, \citenamefont {Peng}, \citenamefont {Kai},
	\citenamefont {Jing}, \citenamefont {Can-Li}, \citenamefont {Ke},
	\citenamefont {Jin-Feng}, \citenamefont {Shuai-Hua}, \citenamefont {Ya-Yu},
	\citenamefont {Li-Li}, \citenamefont {Xi}, \citenamefont {Xu-Cun},\ and\
	\citenamefont {Qi-Kun}}]{Qing-Yan2012}%
\BibitemOpen
\bibfield  {author} {\bibinfo {author} {\bibfnamefont {W.}~\bibnamefont
		{Qing-Yan}}, \bibinfo {author} {\bibfnamefont {L.}~\bibnamefont {Zhi}},
	\bibinfo {author} {\bibfnamefont {Z.}~\bibnamefont {Wen-Hao}}, \bibinfo
	{author} {\bibfnamefont {Z.}~\bibnamefont {Zuo-Cheng}}, \bibinfo {author}
	{\bibfnamefont {Z.}~\bibnamefont {Jin-Song}}, \bibinfo {author}
	{\bibfnamefont {L.}~\bibnamefont {Wei}}, \bibinfo {author} {\bibfnamefont
		{D.}~\bibnamefont {Hao}}, \bibinfo {author} {\bibfnamefont {O.}~\bibnamefont
		{Yun-Bo}}, \bibinfo {author} {\bibfnamefont {D.}~\bibnamefont {Peng}},
	\bibinfo {author} {\bibfnamefont {C.}~\bibnamefont {Kai}}, \bibinfo {author}
	{\bibfnamefont {W.}~\bibnamefont {Jing}}, \bibinfo {author} {\bibfnamefont
		{S.}~\bibnamefont {Can-Li}}, \bibinfo {author} {\bibfnamefont
		{H.}~\bibnamefont {Ke}}, \bibinfo {author} {\bibfnamefont {J.}~\bibnamefont
		{Jin-Feng}}, \bibinfo {author} {\bibfnamefont {J.}~\bibnamefont {Shuai-Hua}},
	\bibinfo {author} {\bibfnamefont {W.}~\bibnamefont {Ya-Yu}}, \bibinfo
	{author} {\bibfnamefont {W.}~\bibnamefont {Li-Li}}, \bibinfo {author}
	{\bibfnamefont {C.}~\bibnamefont {Xi}}, \bibinfo {author} {\bibfnamefont
		{M.}~\bibnamefont {Xu-Cun}}, \ and\ \bibinfo {author} {\bibfnamefont
		{X.}~\bibnamefont {Qi-Kun}},\ }\href
{http://stacks.iop.org/0256-307X/29/i=3/a=037402} {\bibfield  {journal}
	{\bibinfo  {journal} {Chin. Phys. Lett.}\ }\textbf {\bibinfo {volume} {29}},\
	\bibinfo {pages} {037402} (\bibinfo {year} {2012})}\BibitemShut {NoStop}%
\bibitem [{\citenamefont {Liu}\ \emph {et~al.}(2012)\citenamefont {Liu},
	\citenamefont {Zhang}, \citenamefont {Mou}, \citenamefont {He}, \citenamefont
	{Ou}, \citenamefont {Wang}, \citenamefont {Li}, \citenamefont {Wang},
	\citenamefont {Zhao}, \citenamefont {He}, \citenamefont {Peng}, \citenamefont
	{Liu}, \citenamefont {Chen}, \citenamefont {Yu}, \citenamefont {Liu},
	\citenamefont {Dong}, \citenamefont {Zhang}, \citenamefont {Chen},
	\citenamefont {Xu}, \citenamefont {Hu}, \citenamefont {Chen}, \citenamefont
	{Ma}, \citenamefont {Xue},\ and\ \citenamefont {Zhou}}]{Liu2012}%
\BibitemOpen
\bibfield  {author} {\bibinfo {author} {\bibfnamefont {D.}~\bibnamefont
		{Liu}}, \bibinfo {author} {\bibfnamefont {W.}~\bibnamefont {Zhang}}, \bibinfo
	{author} {\bibfnamefont {D.}~\bibnamefont {Mou}}, \bibinfo {author}
	{\bibfnamefont {J.}~\bibnamefont {He}}, \bibinfo {author} {\bibfnamefont
		{Y.-B.}\ \bibnamefont {Ou}}, \bibinfo {author} {\bibfnamefont {Q.-Y.}\
		\bibnamefont {Wang}}, \bibinfo {author} {\bibfnamefont {Z.}~\bibnamefont
		{Li}}, \bibinfo {author} {\bibfnamefont {L.}~\bibnamefont {Wang}}, \bibinfo
	{author} {\bibfnamefont {L.}~\bibnamefont {Zhao}}, \bibinfo {author}
	{\bibfnamefont {S.}~\bibnamefont {He}}, \bibinfo {author} {\bibfnamefont
		{Y.}~\bibnamefont {Peng}}, \bibinfo {author} {\bibfnamefont {X.}~\bibnamefont
		{Liu}}, \bibinfo {author} {\bibfnamefont {C.}~\bibnamefont {Chen}}, \bibinfo
	{author} {\bibfnamefont {L.}~\bibnamefont {Yu}}, \bibinfo {author}
	{\bibfnamefont {G.}~\bibnamefont {Liu}}, \bibinfo {author} {\bibfnamefont
		{X.}~\bibnamefont {Dong}}, \bibinfo {author} {\bibfnamefont {J.}~\bibnamefont
		{Zhang}}, \bibinfo {author} {\bibfnamefont {C.}~\bibnamefont {Chen}},
	\bibinfo {author} {\bibfnamefont {Z.}~\bibnamefont {Xu}}, \bibinfo {author}
	{\bibfnamefont {J.}~\bibnamefont {Hu}}, \bibinfo {author} {\bibfnamefont
		{X.}~\bibnamefont {Chen}}, \bibinfo {author} {\bibfnamefont {X.}~\bibnamefont
		{Ma}}, \bibinfo {author} {\bibfnamefont {Q.}~\bibnamefont {Xue}}, \ and\
	\bibinfo {author} {\bibfnamefont {X.}~\bibnamefont {Zhou}},\ }\href
{http://dx.doi.org/10.1038/ncomms1946} {\bibfield  {journal} {\bibinfo
		{journal} {Nat. Commun.}\ }\textbf {\bibinfo {volume} {3}},\ \bibinfo {pages}
	{931} (\bibinfo {year} {2012})}\BibitemShut {NoStop}%
\bibitem [{\citenamefont {He}\ \emph {et~al.}(2013)\citenamefont {He},
	\citenamefont {He}, \citenamefont {Zhang}, \citenamefont {Zhao},
	\citenamefont {Liu}, \citenamefont {Liu}, \citenamefont {Mou}, \citenamefont
	{Ou}, \citenamefont {Wang}, \citenamefont {Li}, \citenamefont {Wang},
	\citenamefont {Peng}, \citenamefont {Liu}, \citenamefont {Chen},
	\citenamefont {Yu}, \citenamefont {Liu}, \citenamefont {Dong}, \citenamefont
	{Zhang}, \citenamefont {Chen}, \citenamefont {Xu}, \citenamefont {Chen},
	\citenamefont {Ma}, \citenamefont {Xue},\ and\ \citenamefont
	{Zhou}}]{He2013}%
\BibitemOpen
\bibfield  {author} {\bibinfo {author} {\bibfnamefont {S.}~\bibnamefont
		{He}}, \bibinfo {author} {\bibfnamefont {J.}~\bibnamefont {He}}, \bibinfo
	{author} {\bibfnamefont {W.}~\bibnamefont {Zhang}}, \bibinfo {author}
	{\bibfnamefont {L.}~\bibnamefont {Zhao}}, \bibinfo {author} {\bibfnamefont
		{D.}~\bibnamefont {Liu}}, \bibinfo {author} {\bibfnamefont {X.}~\bibnamefont
		{Liu}}, \bibinfo {author} {\bibfnamefont {D.}~\bibnamefont {Mou}}, \bibinfo
	{author} {\bibfnamefont {Y.-B.}\ \bibnamefont {Ou}}, \bibinfo {author}
	{\bibfnamefont {Q.-Y.}\ \bibnamefont {Wang}}, \bibinfo {author}
	{\bibfnamefont {Z.}~\bibnamefont {Li}}, \bibinfo {author} {\bibfnamefont
		{L.}~\bibnamefont {Wang}}, \bibinfo {author} {\bibfnamefont {Y.}~\bibnamefont
		{Peng}}, \bibinfo {author} {\bibfnamefont {Y.}~\bibnamefont {Liu}}, \bibinfo
	{author} {\bibfnamefont {C.}~\bibnamefont {Chen}}, \bibinfo {author}
	{\bibfnamefont {L.}~\bibnamefont {Yu}}, \bibinfo {author} {\bibfnamefont
		{G.}~\bibnamefont {Liu}}, \bibinfo {author} {\bibfnamefont {X.}~\bibnamefont
		{Dong}}, \bibinfo {author} {\bibfnamefont {J.}~\bibnamefont {Zhang}},
	\bibinfo {author} {\bibfnamefont {C.}~\bibnamefont {Chen}}, \bibinfo {author}
	{\bibfnamefont {Z.}~\bibnamefont {Xu}}, \bibinfo {author} {\bibfnamefont
		{X.}~\bibnamefont {Chen}}, \bibinfo {author} {\bibfnamefont {X.}~\bibnamefont
		{Ma}}, \bibinfo {author} {\bibfnamefont {Q.}~\bibnamefont {Xue}}, \ and\
	\bibinfo {author} {\bibfnamefont {X.~J.}\ \bibnamefont {Zhou}},\ }\href
{http://dx.doi.org/10.1038/nmat3648} {\bibfield  {journal} {\bibinfo
		{journal} {Nat. Mater.}\ }\textbf {\bibinfo {volume} {12}},\ \bibinfo {pages}
	{605 } (\bibinfo {year} {2013})}\BibitemShut {NoStop}%
\bibitem [{\citenamefont {Tan}\ \emph {et~al.}(2013)\citenamefont {Tan},
	\citenamefont {Zhang}, \citenamefont {Xia}, \citenamefont {Ye}, \citenamefont
	{Chen}, \citenamefont {Xie}, \citenamefont {Peng}, \citenamefont {Xu},
	\citenamefont {Fan}, \citenamefont {Xu}, \citenamefont {Jiang}, \citenamefont
	{Zhang}, \citenamefont {Lai}, \citenamefont {Xiang}, \citenamefont {Hu},
	\citenamefont {Xie},\ and\ \citenamefont {Feng}}]{Tan2013}%
\BibitemOpen
\bibfield  {author} {\bibinfo {author} {\bibfnamefont {S.}~\bibnamefont
		{Tan}}, \bibinfo {author} {\bibfnamefont {Y.}~\bibnamefont {Zhang}}, \bibinfo
	{author} {\bibfnamefont {M.}~\bibnamefont {Xia}}, \bibinfo {author}
	{\bibfnamefont {Z.}~\bibnamefont {Ye}}, \bibinfo {author} {\bibfnamefont
		{F.}~\bibnamefont {Chen}}, \bibinfo {author} {\bibfnamefont {X.}~\bibnamefont
		{Xie}}, \bibinfo {author} {\bibfnamefont {R.}~\bibnamefont {Peng}}, \bibinfo
	{author} {\bibfnamefont {D.}~\bibnamefont {Xu}}, \bibinfo {author}
	{\bibfnamefont {Q.}~\bibnamefont {Fan}}, \bibinfo {author} {\bibfnamefont
		{H.}~\bibnamefont {Xu}}, \bibinfo {author} {\bibfnamefont {J.}~\bibnamefont
		{Jiang}}, \bibinfo {author} {\bibfnamefont {T.}~\bibnamefont {Zhang}},
	\bibinfo {author} {\bibfnamefont {X.}~\bibnamefont {Lai}}, \bibinfo {author}
	{\bibfnamefont {T.}~\bibnamefont {Xiang}}, \bibinfo {author} {\bibfnamefont
		{J.}~\bibnamefont {Hu}}, \bibinfo {author} {\bibfnamefont {B.}~\bibnamefont
		{Xie}}, \ and\ \bibinfo {author} {\bibfnamefont {D.}~\bibnamefont {Feng}},\
}\href {http://dx.doi.org/10.1038/nmat3654} {\bibfield  {journal} {\bibinfo
	{journal} {Nat. Mater.}\ }\textbf {\bibinfo {volume} {12}},\ \bibinfo {pages}
{634 } (\bibinfo {year} {2013})}\BibitemShut {NoStop}%
\bibitem [{\citenamefont {Peng}\ \emph {et~al.}(2014)\citenamefont {Peng},
	\citenamefont {Shen}, \citenamefont {Xie}, \citenamefont {Xu}, \citenamefont
	{Tan}, \citenamefont {Xia}, \citenamefont {Zhang}, \citenamefont {Cao},
	\citenamefont {Gong}, \citenamefont {Hu}, \citenamefont {Xie},\ and\
	\citenamefont {Feng}}]{Peng2014}%
\BibitemOpen
\bibfield  {author} {\bibinfo {author} {\bibfnamefont {R.}~\bibnamefont
		{Peng}}, \bibinfo {author} {\bibfnamefont {X.~P.}\ \bibnamefont {Shen}},
	\bibinfo {author} {\bibfnamefont {X.}~\bibnamefont {Xie}}, \bibinfo {author}
	{\bibfnamefont {H.~C.}\ \bibnamefont {Xu}}, \bibinfo {author} {\bibfnamefont
		{S.~Y.}\ \bibnamefont {Tan}}, \bibinfo {author} {\bibfnamefont
		{M.}~\bibnamefont {Xia}}, \bibinfo {author} {\bibfnamefont {T.}~\bibnamefont
		{Zhang}}, \bibinfo {author} {\bibfnamefont {H.~Y.}\ \bibnamefont {Cao}},
	\bibinfo {author} {\bibfnamefont {X.~G.}\ \bibnamefont {Gong}}, \bibinfo
	{author} {\bibfnamefont {J.~P.}\ \bibnamefont {Hu}}, \bibinfo {author}
	{\bibfnamefont {B.~P.}\ \bibnamefont {Xie}}, \ and\ \bibinfo {author}
	{\bibfnamefont {D.~L.}\ \bibnamefont {Feng}},\ }\href {\doibase
	10.1103/PhysRevLett.112.107001} {\bibfield  {journal} {\bibinfo  {journal}
		{Phys. Rev. Lett.}\ }\textbf {\bibinfo {volume} {112}},\ \bibinfo {pages}
	{107001} (\bibinfo {year} {2014})}\BibitemShut {NoStop}%
\bibitem [{\citenamefont {Lee}\ \emph {et~al.}(2014)\citenamefont {Lee},
	\citenamefont {Schmitt}, \citenamefont {Moore}, \citenamefont {Johnston},
	\citenamefont {Cui}, \citenamefont {Li}, \citenamefont {Yi}, \citenamefont
	{Liu}, \citenamefont {Hashimoto}, \citenamefont {Zhang}, \citenamefont {Lu},
	\citenamefont {Devereaux}, \citenamefont {Lee},\ and\ \citenamefont
	{Shen}}]{Lee2014}%
\BibitemOpen
\bibfield  {author} {\bibinfo {author} {\bibfnamefont {J.~J.}\ \bibnamefont
		{Lee}}, \bibinfo {author} {\bibfnamefont {F.~T.}\ \bibnamefont {Schmitt}},
	\bibinfo {author} {\bibfnamefont {R.~G.}\ \bibnamefont {Moore}}, \bibinfo
	{author} {\bibfnamefont {S.}~\bibnamefont {Johnston}}, \bibinfo {author}
	{\bibfnamefont {Y.-T.}\ \bibnamefont {Cui}}, \bibinfo {author} {\bibfnamefont
		{W.}~\bibnamefont {Li}}, \bibinfo {author} {\bibfnamefont {M.}~\bibnamefont
		{Yi}}, \bibinfo {author} {\bibfnamefont {Z.~K.}\ \bibnamefont {Liu}},
	\bibinfo {author} {\bibfnamefont {M.}~\bibnamefont {Hashimoto}}, \bibinfo
	{author} {\bibfnamefont {Y.}~\bibnamefont {Zhang}}, \bibinfo {author}
	{\bibfnamefont {D.~H.}\ \bibnamefont {Lu}}, \bibinfo {author} {\bibfnamefont
		{T.~P.}\ \bibnamefont {Devereaux}}, \bibinfo {author} {\bibfnamefont {D.-H.}\
		\bibnamefont {Lee}}, \ and\ \bibinfo {author} {\bibfnamefont {Z.-X.}\
		\bibnamefont {Shen}},\ }\href {http://dx.doi.org/10.1038/nature13894}
{\bibfield  {journal} {\bibinfo  {journal} {Nature}\ }\textbf {\bibinfo
		{volume} {515}},\ \bibinfo {pages} {245 } (\bibinfo {year}
	{2014})}\BibitemShut {NoStop}%
\bibitem [{\citenamefont {Ge}\ \emph {et~al.}(2015)\citenamefont {Ge},
	\citenamefont {Liu}, \citenamefont {Liu}, \citenamefont {Gao}, \citenamefont
	{Qian}, \citenamefont {Xue}, \citenamefont {Liu},\ and\ \citenamefont
	{Jia}}]{Ge2015}%
\BibitemOpen
\bibfield  {author} {\bibinfo {author} {\bibfnamefont {J.-F.}\ \bibnamefont
		{Ge}}, \bibinfo {author} {\bibfnamefont {Z.-L.}\ \bibnamefont {Liu}},
	\bibinfo {author} {\bibfnamefont {C.}~\bibnamefont {Liu}}, \bibinfo {author}
	{\bibfnamefont {C.-L.}\ \bibnamefont {Gao}}, \bibinfo {author} {\bibfnamefont
		{D.}~\bibnamefont {Qian}}, \bibinfo {author} {\bibfnamefont {Q.-K.}\
		\bibnamefont {Xue}}, \bibinfo {author} {\bibfnamefont {Y.}~\bibnamefont
		{Liu}}, \ and\ \bibinfo {author} {\bibfnamefont {J.-F.}\ \bibnamefont
		{Jia}},\ }\href {http://dx.doi.org/10.1038/nmat4153} {\bibfield  {journal}
	{\bibinfo  {journal} {Nat. Mater.}\ }\textbf {\bibinfo {volume} {14}},\
	\bibinfo {pages} {285} (\bibinfo {year} {2015})}\BibitemShut {NoStop}%
\bibitem [{\citenamefont {Jiao}\ \emph {et~al.}(2017)\citenamefont {Jiao},
	\citenamefont {Huang}, \citenamefont {R{\"o}{\ss}ler}, \citenamefont {Koz},
	\citenamefont {R{\"o}{\ss}ler}, \citenamefont {Schwarz},\ and\ \citenamefont
	{Wirth}}]{Jiao2017}%
\BibitemOpen
\bibfield  {author} {\bibinfo {author} {\bibfnamefont {L.}~\bibnamefont
		{Jiao}}, \bibinfo {author} {\bibfnamefont {C.-L.}\ \bibnamefont {Huang}},
	\bibinfo {author} {\bibfnamefont {S.}~\bibnamefont {R{\"o}{\ss}ler}},
	\bibinfo {author} {\bibfnamefont {C.}~\bibnamefont {Koz}}, \bibinfo {author}
	{\bibfnamefont {U.~K.}\ \bibnamefont {R{\"o}{\ss}ler}}, \bibinfo {author}
	{\bibfnamefont {U.}~\bibnamefont {Schwarz}}, \ and\ \bibinfo {author}
	{\bibfnamefont {S.}~\bibnamefont {Wirth}},\ }\href
{https://doi.org/10.1038/srep44024} {\bibfield  {journal} {\bibinfo
		{journal} {Sci. Rep.}\ }\textbf {\bibinfo {volume} {7}},\ \bibinfo {pages}
	{44024} (\bibinfo {year} {2017})}\BibitemShut {NoStop}%
\bibitem [{\citenamefont {Scalapino}\ \emph {et~al.}(1986)\citenamefont
	{Scalapino}, \citenamefont {Loh},\ and\ \citenamefont
	{Hirsch}}]{Scalapino1986}%
\BibitemOpen
\bibfield  {author} {\bibinfo {author} {\bibfnamefont {D.~J.}\ \bibnamefont
		{Scalapino}}, \bibinfo {author} {\bibfnamefont {E.}~\bibnamefont {Loh}}, \
	and\ \bibinfo {author} {\bibfnamefont {J.~E.}\ \bibnamefont {Hirsch}},\
}\href {\doibase 10.1103/PhysRevB.34.8190} {\bibfield  {journal} {\bibinfo
	{journal} {Phys. Rev. B}\ }\textbf {\bibinfo {volume} {34}},\ \bibinfo
{pages} {8190} (\bibinfo {year} {1986})}\BibitemShut {NoStop}%
\bibitem [{\citenamefont {Carbotte}\ and\ \citenamefont
	{Marsiglio}(2003)}]{Carbotte2003}%
\BibitemOpen
\bibfield  {author} {\bibinfo {author} {\bibfnamefont {J.~P.}\ \bibnamefont
		{Carbotte}}\ and\ \bibinfo {author} {\bibfnamefont {F.}~\bibnamefont
		{Marsiglio}},\ }\enquote {\bibinfo {title} {Electron-phonon
		superconductivity},}\ in\ \href {\doibase 10.1007/978-3-642-55675-3_4} {\emph
	{\bibinfo {booktitle} {The Physics of Superconductors: Vol. I. Conventional
			and High-Tc Superconductors}}},\ \bibinfo {editor} {edited by\ \bibinfo
	{editor} {\bibfnamefont {K.~H.}\ \bibnamefont {Bennemann}}\ and\ \bibinfo
	{editor} {\bibfnamefont {J.~B.}\ \bibnamefont {Ketterson}}}\ (\bibinfo
{publisher} {Springer},\ \bibinfo {address} {Berlin, Heidelberg},\ \bibinfo
{year} {2003})\ pp.\ \bibinfo {pages} {233--345}\BibitemShut {NoStop}%
\bibitem [{\citenamefont {Kubo}(2007)}]{Kubo2007}%
\BibitemOpen
\bibfield  {author} {\bibinfo {author} {\bibfnamefont {K.}~\bibnamefont
		{Kubo}},\ }\href {\doibase 10.1103/PhysRevB.75.224509} {\bibfield  {journal}
	{\bibinfo  {journal} {Phys. Rev. B}\ }\textbf {\bibinfo {volume} {75}},\
	\bibinfo {pages} {224509} (\bibinfo {year} {2007})}\BibitemShut {NoStop}%
\bibitem [{\citenamefont {Graser}\ \emph {et~al.}(2009)\citenamefont {Graser},
	\citenamefont {Maier}, \citenamefont {Hirschfeld},\ and\ \citenamefont
	{Scalapino}}]{Graser2009}%
\BibitemOpen
\bibfield  {author} {\bibinfo {author} {\bibfnamefont {S.}~\bibnamefont
		{Graser}}, \bibinfo {author} {\bibfnamefont {T.~A.}\ \bibnamefont {Maier}},
	\bibinfo {author} {\bibfnamefont {P.~J.}\ \bibnamefont {Hirschfeld}}, \ and\
	\bibinfo {author} {\bibfnamefont {D.~J.}\ \bibnamefont {Scalapino}},\ }\href
{\doibase 10.1088/1367-2630/11/2/025016} {\bibfield  {journal} {\bibinfo
		{journal} {New J. Phys.}\ }\textbf {\bibinfo {volume} {11}},\ \bibinfo
	{pages} {025016} (\bibinfo {year} {2009})}\BibitemShut {NoStop}%
\bibitem [{\citenamefont {Kemper}\ \emph {et~al.}(2010)\citenamefont {Kemper},
	\citenamefont {Maier}, \citenamefont {Graser}, \citenamefont {Cheng},
	\citenamefont {Hirschfeld},\ and\ \citenamefont {Scalapino}}]{Kemper2010}%
\BibitemOpen
\bibfield  {author} {\bibinfo {author} {\bibfnamefont {A.~F.}\ \bibnamefont
		{Kemper}}, \bibinfo {author} {\bibfnamefont {T.~A.}\ \bibnamefont {Maier}},
	\bibinfo {author} {\bibfnamefont {S.}~\bibnamefont {Graser}}, \bibinfo
	{author} {\bibfnamefont {H.-P.}\ \bibnamefont {Cheng}}, \bibinfo {author}
	{\bibfnamefont {P.~J.}\ \bibnamefont {Hirschfeld}}, \ and\ \bibinfo {author}
	{\bibfnamefont {D.~J.}\ \bibnamefont {Scalapino}},\ }\href {\doibase
	10.1088/1367-2630/12/7/073030} {\bibfield  {journal} {\bibinfo  {journal}
		{New J. Phys.}\ }\textbf {\bibinfo {volume} {12}},\ \bibinfo {pages} {073030}
	(\bibinfo {year} {2010})}\BibitemShut {NoStop}%
\bibitem [{\citenamefont {Margadonna}\ \emph {et~al.}(2008)\citenamefont
	{Margadonna}, \citenamefont {Takabayashi}, \citenamefont {McDonald},
	\citenamefont {Kasperkiewicz}, \citenamefont {Mizuguchi}, \citenamefont
	{Takano}, \citenamefont {Fitch}, \citenamefont {Suarde},\ and\ \citenamefont
	{Prassides}}]{Margadonna2008}%
\BibitemOpen
\bibfield  {author} {\bibinfo {author} {\bibfnamefont {S.}~\bibnamefont
		{Margadonna}}, \bibinfo {author} {\bibfnamefont {Y.}~\bibnamefont
		{Takabayashi}}, \bibinfo {author} {\bibfnamefont {M.~T.}\ \bibnamefont
		{McDonald}}, \bibinfo {author} {\bibfnamefont {K.}~\bibnamefont
		{Kasperkiewicz}}, \bibinfo {author} {\bibfnamefont {Y.}~\bibnamefont
		{Mizuguchi}}, \bibinfo {author} {\bibfnamefont {Y.}~\bibnamefont {Takano}},
	\bibinfo {author} {\bibfnamefont {A.~N.}\ \bibnamefont {Fitch}}, \bibinfo
	{author} {\bibfnamefont {E.}~\bibnamefont {Suarde}}, \ and\ \bibinfo {author}
	{\bibfnamefont {K.}~\bibnamefont {Prassides}},\ }\href@noop {} {\bibfield
	{journal} {\bibinfo  {journal} {Chem. Comm.}\ }\textbf {\bibinfo {volume}
		{43}},\ \bibinfo {pages} {5607 } (\bibinfo {year} {2008})}\BibitemShut
{NoStop}%
\bibitem [{\citenamefont {Fernandes}\ \emph {et~al.}(2012)\citenamefont
	{Fernandes}, \citenamefont {Chubukov}, \citenamefont {Knolle}, \citenamefont
	{Eremin},\ and\ \citenamefont {Schmalian}}]{Fernandes2012}%
\BibitemOpen
\bibfield  {author} {\bibinfo {author} {\bibfnamefont {R.~M.}\ \bibnamefont
		{Fernandes}}, \bibinfo {author} {\bibfnamefont {A.~V.}\ \bibnamefont
		{Chubukov}}, \bibinfo {author} {\bibfnamefont {J.}~\bibnamefont {Knolle}},
	\bibinfo {author} {\bibfnamefont {I.}~\bibnamefont {Eremin}}, \ and\ \bibinfo
	{author} {\bibfnamefont {J.}~\bibnamefont {Schmalian}},\ }\href {\doibase
	10.1103/PhysRevB.85.024534} {\bibfield  {journal} {\bibinfo  {journal} {Phys.
			Rev. B}\ }\textbf {\bibinfo {volume} {85}},\ \bibinfo {pages} {024534}
	(\bibinfo {year} {2012})}\BibitemShut {NoStop}%
\bibitem [{\citenamefont {B\"ohmer}\ \emph {et~al.}(2013)\citenamefont
	{B\"ohmer}, \citenamefont {Hardy}, \citenamefont {Eilers}, \citenamefont
	{Ernst}, \citenamefont {Adelmann}, \citenamefont {Schweiss}, \citenamefont
	{Wolf},\ and\ \citenamefont {Meingast}}]{Boehmer2013}%
\BibitemOpen
\bibfield  {author} {\bibinfo {author} {\bibfnamefont {A.~E.}\ \bibnamefont
		{B\"ohmer}}, \bibinfo {author} {\bibfnamefont {F.}~\bibnamefont {Hardy}},
	\bibinfo {author} {\bibfnamefont {F.}~\bibnamefont {Eilers}}, \bibinfo
	{author} {\bibfnamefont {D.}~\bibnamefont {Ernst}}, \bibinfo {author}
	{\bibfnamefont {P.}~\bibnamefont {Adelmann}}, \bibinfo {author}
	{\bibfnamefont {P.}~\bibnamefont {Schweiss}}, \bibinfo {author}
	{\bibfnamefont {T.}~\bibnamefont {Wolf}}, \ and\ \bibinfo {author}
	{\bibfnamefont {C.}~\bibnamefont {Meingast}},\ }\href {\doibase
	10.1103/PhysRevB.87.180505} {\bibfield  {journal} {\bibinfo  {journal} {Phys.
			Rev. B}\ }\textbf {\bibinfo {volume} {87}},\ \bibinfo {pages} {180505}
	(\bibinfo {year} {2013})}\BibitemShut {NoStop}%
\bibitem [{\citenamefont {Fernandes}\ \emph {et~al.}(2014)\citenamefont
	{Fernandes}, \citenamefont {Chubukov},\ and\ \citenamefont
	{Schmalian}}]{Fernandes2014}%
\BibitemOpen
\bibfield  {author} {\bibinfo {author} {\bibfnamefont {R.~M.}\ \bibnamefont
		{Fernandes}}, \bibinfo {author} {\bibfnamefont {A.~V.}\ \bibnamefont
		{Chubukov}}, \ and\ \bibinfo {author} {\bibfnamefont {J.}~\bibnamefont
		{Schmalian}},\ }\href@noop {} {\bibfield  {journal} {\bibinfo  {journal}
		{Nature Phys.}\ }\textbf {\bibinfo {volume} {10}},\ \bibinfo {pages} {97}
	(\bibinfo {year} {2014})}\BibitemShut {NoStop}%
\bibitem [{\citenamefont {Kuo}\ \emph {et~al.}(2016)\citenamefont {Kuo},
	\citenamefont {Chu}, \citenamefont {Palmstrom}, \citenamefont {Kivelson},\
	and\ \citenamefont {Fisher}}]{Kuo2016}%
\BibitemOpen
\bibfield  {author} {\bibinfo {author} {\bibfnamefont {H.-H.}\ \bibnamefont
		{Kuo}}, \bibinfo {author} {\bibfnamefont {J.-H.}\ \bibnamefont {Chu}},
	\bibinfo {author} {\bibfnamefont {J.~C.}\ \bibnamefont {Palmstrom}}, \bibinfo
	{author} {\bibfnamefont {S.~A.}\ \bibnamefont {Kivelson}}, \ and\ \bibinfo
	{author} {\bibfnamefont {I.~R.}\ \bibnamefont {Fisher}},\ }\href@noop {}
{\bibfield  {journal} {\bibinfo  {journal} {Science}\ }\textbf {\bibinfo
		{volume} {352}},\ \bibinfo {pages} {958 } (\bibinfo {year}
	{2016})}\BibitemShut {NoStop}%
\bibitem [{\citenamefont {Chubukov}\ \emph {et~al.}(2016)\citenamefont
	{Chubukov}, \citenamefont {Khodas},\ and\ \citenamefont
	{Fernandes}}]{Chubukov2016}%
\BibitemOpen
\bibfield  {author} {\bibinfo {author} {\bibfnamefont {A.~V.}\ \bibnamefont
		{Chubukov}}, \bibinfo {author} {\bibfnamefont {M.}~\bibnamefont {Khodas}}, \
	and\ \bibinfo {author} {\bibfnamefont {R.~M.}\ \bibnamefont {Fernandes}},\
}\href@noop {} {\bibfield  {journal} {\bibinfo  {journal} {Phys. Rev. X}\
}\textbf {\bibinfo {volume} {6}},\ \bibinfo {pages} {041045} (\bibinfo {year}
{2016})}\BibitemShut {NoStop}%
\bibitem [{\citenamefont {B{\"o}hmer}\ and\ \citenamefont
	{Kreisel}(2018)}]{Boehmer2018}%
\BibitemOpen
\bibfield  {author} {\bibinfo {author} {\bibfnamefont {A.~E.}\ \bibnamefont
		{B{\"o}hmer}}\ and\ \bibinfo {author} {\bibfnamefont {A.}~\bibnamefont
		{Kreisel}},\ }\href {\doibase 10.1088/1361-648x/aa9caa} {\bibfield  {journal}
	{\bibinfo  {journal} {J. Phys.: Condens. Matter}\ }\textbf {\bibinfo {volume}
		{30}},\ \bibinfo {pages} {023001} (\bibinfo {year} {2018})}\BibitemShut
{NoStop}%
\bibitem [{\citenamefont {Baek}\ \emph {et~al.}(2020)\citenamefont {Baek},
	\citenamefont {Mok}, \citenamefont {Kim}, \citenamefont {Aswartham},
	\citenamefont {Morozov}, \citenamefont {Chareev}, \citenamefont {Urata},
	\citenamefont {Tanigaki}, \citenamefont {Tanabe}, \citenamefont
	{B{\"u}chner},\ and\ \citenamefont {Efremov}}]{Baek2020}%
\BibitemOpen
\bibfield  {author} {\bibinfo {author} {\bibfnamefont {S.-H.}\ \bibnamefont
		{Baek}}, \bibinfo {author} {\bibfnamefont {J.}~\bibnamefont {Mok}}, \bibinfo
	{author} {\bibfnamefont {J.~S.}\ \bibnamefont {Kim}}, \bibinfo {author}
	{\bibfnamefont {S.}~\bibnamefont {Aswartham}}, \bibinfo {author}
	{\bibfnamefont {I.}~\bibnamefont {Morozov}}, \bibinfo {author} {\bibfnamefont
		{D.}~\bibnamefont {Chareev}}, \bibinfo {author} {\bibfnamefont
		{T.}~\bibnamefont {Urata}}, \bibinfo {author} {\bibfnamefont
		{K.}~\bibnamefont {Tanigaki}}, \bibinfo {author} {\bibfnamefont
		{Y.}~\bibnamefont {Tanabe}}, \bibinfo {author} {\bibfnamefont
		{B.}~\bibnamefont {B{\"u}chner}}, \ and\ \bibinfo {author} {\bibfnamefont
		{D.~V.}\ \bibnamefont {Efremov}},\ }\href@noop {} {\bibfield  {journal}
	{\bibinfo  {journal} {njp Quant. Mater.}\ }\textbf {\bibinfo {volume} {5}},\
	\bibinfo {pages} {8} (\bibinfo {year} {2020})}\BibitemShut {NoStop}%
\bibitem [{\citenamefont {Imai}\ \emph {et~al.}(2009)\citenamefont {Imai},
	\citenamefont {Ahilan}, \citenamefont {Ning}, \citenamefont {McQueen},\ and\
	\citenamefont {Cava}}]{Imai2009}%
\BibitemOpen
\bibfield  {author} {\bibinfo {author} {\bibfnamefont {T.}~\bibnamefont
		{Imai}}, \bibinfo {author} {\bibfnamefont {K.}~\bibnamefont {Ahilan}},
	\bibinfo {author} {\bibfnamefont {F.~L.}\ \bibnamefont {Ning}}, \bibinfo
	{author} {\bibfnamefont {T.~M.}\ \bibnamefont {McQueen}}, \ and\ \bibinfo
	{author} {\bibfnamefont {R.~J.}\ \bibnamefont {Cava}},\ }\href@noop {}
{\bibfield  {journal} {\bibinfo  {journal} {Phys. Rev. Lett.}\ }\textbf
	{\bibinfo {volume} {102}},\ \bibinfo {pages} {177005} (\bibinfo {year}
	{2009})}\BibitemShut {NoStop}%
\bibitem [{\citenamefont {Rahn}\ \emph {et~al.}(2015)\citenamefont {Rahn},
	\citenamefont {Ewings}, \citenamefont {Sedlmaier}, \citenamefont {Clarke},\
	and\ \citenamefont {Boothroyd}}]{Rahn2015}%
\BibitemOpen
\bibfield  {author} {\bibinfo {author} {\bibfnamefont {M.~C.}\ \bibnamefont
		{Rahn}}, \bibinfo {author} {\bibfnamefont {R.~A.}\ \bibnamefont {Ewings}},
	\bibinfo {author} {\bibfnamefont {S.~J.}\ \bibnamefont {Sedlmaier}}, \bibinfo
	{author} {\bibfnamefont {S.~J.}\ \bibnamefont {Clarke}}, \ and\ \bibinfo
	{author} {\bibfnamefont {A.~T.}\ \bibnamefont {Boothroyd}},\ }\href@noop {}
{\bibfield  {journal} {\bibinfo  {journal} {Phys. Rev. B}\ }\textbf {\bibinfo
		{volume} {91}},\ \bibinfo {pages} {180501} (\bibinfo {year}
	{2015})}\BibitemShut {NoStop}%
\bibitem [{\citenamefont {Wang}\ \emph
	{et~al.}(2016{\natexlab{a}})\citenamefont {Wang}, \citenamefont {Shen},
	\citenamefont {Pan}, \citenamefont {Hao}, \citenamefont {Ma}, \citenamefont
	{Zhou}, \citenamefont {Steffens}, \citenamefont {Schmalzl}, \citenamefont
	{Forrest}, \citenamefont {Abdel-Hafiez}, \citenamefont {Chareev},
	\citenamefont {Vasiliev}, \citenamefont {Bourges}, \citenamefont {Sidis},
	\citenamefont {Cao},\ and\ \citenamefont {Zhao}}]{Wang2016NatMat}%
\BibitemOpen
\bibfield  {author} {\bibinfo {author} {\bibfnamefont {Q.}~\bibnamefont
		{Wang}}, \bibinfo {author} {\bibfnamefont {Y.}~\bibnamefont {Shen}}, \bibinfo
	{author} {\bibfnamefont {B.}~\bibnamefont {Pan}}, \bibinfo {author}
	{\bibfnamefont {Y.}~\bibnamefont {Hao}}, \bibinfo {author} {\bibfnamefont
		{M.}~\bibnamefont {Ma}}, \bibinfo {author} {\bibfnamefont {F.}~\bibnamefont
		{Zhou}}, \bibinfo {author} {\bibfnamefont {P.}~\bibnamefont {Steffens}},
	\bibinfo {author} {\bibfnamefont {K.}~\bibnamefont {Schmalzl}}, \bibinfo
	{author} {\bibfnamefont {T.~R.}\ \bibnamefont {Forrest}}, \bibinfo {author}
	{\bibfnamefont {M.}~\bibnamefont {Abdel-Hafiez}}, \bibinfo {author}
	{\bibfnamefont {D.~A.}\ \bibnamefont {Chareev}}, \bibinfo {author}
	{\bibfnamefont {A.~N.}\ \bibnamefont {Vasiliev}}, \bibinfo {author}
	{\bibfnamefont {P.}~\bibnamefont {Bourges}}, \bibinfo {author} {\bibfnamefont
		{Y.}~\bibnamefont {Sidis}}, \bibinfo {author} {\bibfnamefont
		{H.}~\bibnamefont {Cao}}, \ and\ \bibinfo {author} {\bibfnamefont
		{J.}~\bibnamefont {Zhao}},\ }\href@noop {} {\bibfield  {journal} {\bibinfo
		{journal} {Nat. Mater.}\ }\textbf {\bibinfo {volume} {15}},\ \bibinfo {pages}
	{159 } (\bibinfo {year} {2016}{\natexlab{a}})}\BibitemShut {NoStop}%
\bibitem [{\citenamefont {Wang}\ \emph
	{et~al.}(2016{\natexlab{b}})\citenamefont {Wang}, \citenamefont {Shen},
	\citenamefont {Pan}, \citenamefont {Zhang}, \citenamefont {Ikeuchi},
	\citenamefont {Iida}, \citenamefont {Christianson}, \citenamefont {Walker},
	\citenamefont {Adroja}, \citenamefont {Abdel-Hafiez}, \citenamefont {Chen},
	\citenamefont {Chareev}, \citenamefont {Vasiliev},\ and\ \citenamefont
	{Zhao}}]{Wang2016NatCom}%
\BibitemOpen
\bibfield  {author} {\bibinfo {author} {\bibfnamefont {Q.}~\bibnamefont
		{Wang}}, \bibinfo {author} {\bibfnamefont {Y.}~\bibnamefont {Shen}}, \bibinfo
	{author} {\bibfnamefont {B.}~\bibnamefont {Pan}}, \bibinfo {author}
	{\bibfnamefont {X.}~\bibnamefont {Zhang}}, \bibinfo {author} {\bibfnamefont
		{K.}~\bibnamefont {Ikeuchi}}, \bibinfo {author} {\bibfnamefont
		{K.}~\bibnamefont {Iida}}, \bibinfo {author} {\bibfnamefont {A.~D.}\
		\bibnamefont {Christianson}}, \bibinfo {author} {\bibfnamefont {H.~C.}\
		\bibnamefont {Walker}}, \bibinfo {author} {\bibfnamefont {D.~T.}\
		\bibnamefont {Adroja}}, \bibinfo {author} {\bibfnamefont {M.}~\bibnamefont
		{Abdel-Hafiez}}, \bibinfo {author} {\bibfnamefont {X.}~\bibnamefont {Chen}},
	\bibinfo {author} {\bibfnamefont {D.~A.}\ \bibnamefont {Chareev}}, \bibinfo
	{author} {\bibfnamefont {A.~N.}\ \bibnamefont {Vasiliev}}, \ and\ \bibinfo
	{author} {\bibfnamefont {J.}~\bibnamefont {Zhao}},\ }\href@noop {} {\bibfield
	{journal} {\bibinfo  {journal} {Nat. Commun.}\ }\textbf {\bibinfo {volume}
		{7}},\ \bibinfo {pages} {12182} (\bibinfo {year}
	{2016}{\natexlab{b}})}\BibitemShut {NoStop}%
\bibitem [{\citenamefont {de~la Cruz}\ \emph {et~al.}(2008)\citenamefont {de~la
		Cruz}, \citenamefont {Huang}, \citenamefont {Lynn}, \citenamefont {Li},
	\citenamefont {II}, \citenamefont {Zarestky}, \citenamefont {Mook},
	\citenamefont {Chen}, \citenamefont {Luo}, \citenamefont {Wang},\ and\
	\citenamefont {Dai}}]{delaCruz2008}%
\BibitemOpen
\bibfield  {author} {\bibinfo {author} {\bibfnamefont {C.}~\bibnamefont
		{de~la Cruz}}, \bibinfo {author} {\bibfnamefont {Q.}~\bibnamefont {Huang}},
	\bibinfo {author} {\bibfnamefont {J.~W.}\ \bibnamefont {Lynn}}, \bibinfo
	{author} {\bibfnamefont {J.}~\bibnamefont {Li}}, \bibinfo {author}
	{\bibfnamefont {W.~R.}\ \bibnamefont {II}}, \bibinfo {author} {\bibfnamefont
		{J.~L.}\ \bibnamefont {Zarestky}}, \bibinfo {author} {\bibfnamefont {H.~A.}\
		\bibnamefont {Mook}}, \bibinfo {author} {\bibfnamefont {G.~F.}\ \bibnamefont
		{Chen}}, \bibinfo {author} {\bibfnamefont {J.~L.}\ \bibnamefont {Luo}},
	\bibinfo {author} {\bibfnamefont {N.~L.}\ \bibnamefont {Wang}}, \ and\
	\bibinfo {author} {\bibfnamefont {P.}~\bibnamefont {Dai}},\ }\href@noop {}
{\bibfield  {journal} {\bibinfo  {journal} {Nature}\ }\textbf {\bibinfo
		{volume} {453}},\ \bibinfo {pages} {899} (\bibinfo {year}
	{2008})}\BibitemShut {NoStop}%
\bibitem [{\citenamefont {Christianson}\ \emph {et~al.}(2009)\citenamefont
	{Christianson}, \citenamefont {Lumsden}, \citenamefont {Nagler},
	\citenamefont {MacDougall}, \citenamefont {McGuire}, \citenamefont {Sefat},
	\citenamefont {Jin}, \citenamefont {Sales},\ and\ \citenamefont
	{Mandrus}}]{Christianson2009}%
\BibitemOpen
\bibfield  {author} {\bibinfo {author} {\bibfnamefont {A.~D.}\ \bibnamefont
		{Christianson}}, \bibinfo {author} {\bibfnamefont {M.~D.}\ \bibnamefont
		{Lumsden}}, \bibinfo {author} {\bibfnamefont {S.~E.}\ \bibnamefont {Nagler}},
	\bibinfo {author} {\bibfnamefont {G.~J.}\ \bibnamefont {MacDougall}},
	\bibinfo {author} {\bibfnamefont {M.~A.}\ \bibnamefont {McGuire}}, \bibinfo
	{author} {\bibfnamefont {A.~S.}\ \bibnamefont {Sefat}}, \bibinfo {author}
	{\bibfnamefont {R.}~\bibnamefont {Jin}}, \bibinfo {author} {\bibfnamefont
		{B.~C.}\ \bibnamefont {Sales}}, \ and\ \bibinfo {author} {\bibfnamefont
		{D.}~\bibnamefont {Mandrus}},\ }\href {\doibase
	10.1103/PhysRevLett.103.087002} {\bibfield  {journal} {\bibinfo  {journal}
		{Phys. Rev. Lett.}\ }\textbf {\bibinfo {volume} {103}},\ \bibinfo {pages}
	{087002} (\bibinfo {year} {2009})}\BibitemShut {NoStop}%
\bibitem [{\citenamefont {Lumsden}\ \emph {et~al.}(2009)\citenamefont
	{Lumsden}, \citenamefont {Christianson}, \citenamefont {Parshall},
	\citenamefont {Stone}, \citenamefont {Nagler}, \citenamefont {MacDougall},
	\citenamefont {Mook}, \citenamefont {Lokshin}, \citenamefont {Egami},
	\citenamefont {Abernathy}, \citenamefont {Goremychkin}, \citenamefont
	{Osborn}, \citenamefont {McGuire}, \citenamefont {Sefat}, \citenamefont
	{Jin}, \citenamefont {Sales},\ and\ \citenamefont {Mandrus}}]{Lumsden2009}%
\BibitemOpen
\bibfield  {author} {\bibinfo {author} {\bibfnamefont {M.~D.}\ \bibnamefont
		{Lumsden}}, \bibinfo {author} {\bibfnamefont {A.~D.}\ \bibnamefont
		{Christianson}}, \bibinfo {author} {\bibfnamefont {D.}~\bibnamefont
		{Parshall}}, \bibinfo {author} {\bibfnamefont {M.~B.}\ \bibnamefont {Stone}},
	\bibinfo {author} {\bibfnamefont {S.~E.}\ \bibnamefont {Nagler}}, \bibinfo
	{author} {\bibfnamefont {G.~J.}\ \bibnamefont {MacDougall}}, \bibinfo
	{author} {\bibfnamefont {H.~A.}\ \bibnamefont {Mook}}, \bibinfo {author}
	{\bibfnamefont {K.}~\bibnamefont {Lokshin}}, \bibinfo {author} {\bibfnamefont
		{T.}~\bibnamefont {Egami}}, \bibinfo {author} {\bibfnamefont {D.~L.}\
		\bibnamefont {Abernathy}}, \bibinfo {author} {\bibfnamefont {E.~A.}\
		\bibnamefont {Goremychkin}}, \bibinfo {author} {\bibfnamefont
		{R.}~\bibnamefont {Osborn}}, \bibinfo {author} {\bibfnamefont {M.~A.}\
		\bibnamefont {McGuire}}, \bibinfo {author} {\bibfnamefont {A.~S.}\
		\bibnamefont {Sefat}}, \bibinfo {author} {\bibfnamefont {R.}~\bibnamefont
		{Jin}}, \bibinfo {author} {\bibfnamefont {B.~C.}\ \bibnamefont {Sales}}, \
	and\ \bibinfo {author} {\bibfnamefont {D.}~\bibnamefont {Mandrus}},\ }\href
{\doibase 10.1103/PhysRevLett.102.107005} {\bibfield  {journal} {\bibinfo
		{journal} {Phys. Rev. Lett.}\ }\textbf {\bibinfo {volume} {102}},\ \bibinfo
	{pages} {107005} (\bibinfo {year} {2009})}\BibitemShut {NoStop}%
\bibitem [{\citenamefont {Yin}\ \emph {et~al.}(2011{\natexlab{a}})\citenamefont
	{Yin}, \citenamefont {Haule},\ and\ \citenamefont
	{Kotliar}}]{Yin2011NatPhys}%
\BibitemOpen
\bibfield  {author} {\bibinfo {author} {\bibfnamefont {Z.~P.}\ \bibnamefont
		{Yin}}, \bibinfo {author} {\bibfnamefont {K.}~\bibnamefont {Haule}}, \ and\
	\bibinfo {author} {\bibfnamefont {G.}~\bibnamefont {Kotliar}},\ }\href@noop
{} {\bibfield  {journal} {\bibinfo  {journal} {Nat. Phys.}\ }\textbf
	{\bibinfo {volume} {7}},\ \bibinfo {pages} {294} (\bibinfo {year}
	{2011}{\natexlab{a}})}\BibitemShut {NoStop}%
\bibitem [{\citenamefont {Yin}\ \emph {et~al.}(2011{\natexlab{b}})\citenamefont
	{Yin}, \citenamefont {Haule},\ and\ \citenamefont {Kotliar}}]{Yin2011NatMat}%
\BibitemOpen
\bibfield  {author} {\bibinfo {author} {\bibfnamefont {Z.~P.}\ \bibnamefont
		{Yin}}, \bibinfo {author} {\bibfnamefont {K.}~\bibnamefont {Haule}}, \ and\
	\bibinfo {author} {\bibfnamefont {G.}~\bibnamefont {Kotliar}},\ }\href@noop
{} {\bibfield  {journal} {\bibinfo  {journal} {Nat. Mater.}\ }\textbf
	{\bibinfo {volume} {10}},\ \bibinfo {pages} {932} (\bibinfo {year}
	{2011}{\natexlab{b}})}\BibitemShut {NoStop}%
\bibitem [{\citenamefont {Yin}\ \emph {et~al.}(2014)\citenamefont {Yin},
	\citenamefont {Haule},\ and\ \citenamefont {Kotliar}}]{Yin2014}%
\BibitemOpen
\bibfield  {author} {\bibinfo {author} {\bibfnamefont {Z.~P.}\ \bibnamefont
		{Yin}}, \bibinfo {author} {\bibfnamefont {K.}~\bibnamefont {Haule}}, \ and\
	\bibinfo {author} {\bibfnamefont {G.}~\bibnamefont {Kotliar}},\ }\href
{https://doi.org/10.1038/nphys3116} {\bibfield  {journal} {\bibinfo
		{journal} {Nat. Phys.}\ }\textbf {\bibinfo {volume} {10}},\ \bibinfo {pages}
	{845} (\bibinfo {year} {2014})}\BibitemShut {NoStop}%
\bibitem [{\citenamefont {Scherer}\ \emph {et~al.}(2017)\citenamefont
	{Scherer}, \citenamefont {Jacko}, \citenamefont {Friedrich}, \citenamefont
	{\ifmmode \mbox{\c{S}}\else \c{S}\fi{}a\ifmmode \mbox{\c{s}}\else
		\c{s}\fi{}\ifmmode \imath \else \i \fi{}o\ifmmode~\breve{g}\else
		\u{g}\fi{}lu}, \citenamefont {Bl\"ugel}, \citenamefont {Valent\'{\i}},\ and\
	\citenamefont {Andersen}}]{Scherer2017}%
\BibitemOpen
\bibfield  {author} {\bibinfo {author} {\bibfnamefont {D.~D.}\ \bibnamefont
		{Scherer}}, \bibinfo {author} {\bibfnamefont {A.~C.}\ \bibnamefont {Jacko}},
	\bibinfo {author} {\bibfnamefont {C.}~\bibnamefont {Friedrich}}, \bibinfo
	{author} {\bibfnamefont {E.}~\bibnamefont {\ifmmode \mbox{\c{S}}\else
			\c{S}\fi{}a\ifmmode \mbox{\c{s}}\else \c{s}\fi{}\ifmmode \imath \else \i
			\fi{}o\ifmmode~\breve{g}\else \u{g}\fi{}lu}}, \bibinfo {author}
	{\bibfnamefont {S.}~\bibnamefont {Bl\"ugel}}, \bibinfo {author}
	{\bibfnamefont {R.}~\bibnamefont {Valent\'{\i}}}, \ and\ \bibinfo {author}
	{\bibfnamefont {B.~M.}\ \bibnamefont {Andersen}},\ }\href {\doibase
	10.1103/PhysRevB.95.094504} {\bibfield  {journal} {\bibinfo  {journal} {Phys.
			Rev. B}\ }\textbf {\bibinfo {volume} {95}},\ \bibinfo {pages} {094504}
	(\bibinfo {year} {2017})}\BibitemShut {NoStop}%
\bibitem [{\citenamefont {Boeri}\ \emph {et~al.}(2008)\citenamefont {Boeri},
	\citenamefont {Dolgov},\ and\ \citenamefont {Golubov}}]{Boeri2008}%
\BibitemOpen
\bibfield  {author} {\bibinfo {author} {\bibfnamefont {L.}~\bibnamefont
		{Boeri}}, \bibinfo {author} {\bibfnamefont {O.~V.}\ \bibnamefont {Dolgov}}, \
	and\ \bibinfo {author} {\bibfnamefont {A.~A.}\ \bibnamefont {Golubov}},\
}\href {\doibase 10.1103/PhysRevLett.101.026403} {\bibfield  {journal}
{\bibinfo  {journal} {Phys. Rev. Lett.}\ }\textbf {\bibinfo {volume} {101}},\
\bibinfo {pages} {026403} (\bibinfo {year} {2008})}\BibitemShut {NoStop}%
\bibitem [{\citenamefont {Nomura}\ \emph {et~al.}(2014)\citenamefont {Nomura},
	\citenamefont {Nakamura},\ and\ \citenamefont {Arita}}]{Nomura2014}%
\BibitemOpen
\bibfield  {author} {\bibinfo {author} {\bibfnamefont {Y.}~\bibnamefont
		{Nomura}}, \bibinfo {author} {\bibfnamefont {K.}~\bibnamefont {Nakamura}}, \
	and\ \bibinfo {author} {\bibfnamefont {R.}~\bibnamefont {Arita}},\ }\href
{\doibase 10.1103/PhysRevLett.112.027002} {\bibfield  {journal} {\bibinfo
		{journal} {Phys. Rev. Lett.}\ }\textbf {\bibinfo {volume} {112}},\ \bibinfo
	{pages} {027002} (\bibinfo {year} {2014})}\BibitemShut {NoStop}%
\bibitem [{\citenamefont {Mazin}\ \emph {et~al.}(2008)\citenamefont {Mazin},
	\citenamefont {Singh}, \citenamefont {Johannes},\ and\ \citenamefont
	{Du}}]{Mazin2008}%
\BibitemOpen
\bibfield  {author} {\bibinfo {author} {\bibfnamefont {I.~I.}\ \bibnamefont
		{Mazin}}, \bibinfo {author} {\bibfnamefont {D.~J.}\ \bibnamefont {Singh}},
	\bibinfo {author} {\bibfnamefont {M.~D.}\ \bibnamefont {Johannes}}, \ and\
	\bibinfo {author} {\bibfnamefont {M.~H.}\ \bibnamefont {Du}},\ }\href
{\doibase 10.1103/PhysRevLett.101.057003} {\bibfield  {journal} {\bibinfo
		{journal} {Phys. Rev. Lett.}\ }\textbf {\bibinfo {volume} {101}},\ \bibinfo
	{pages} {057003} (\bibinfo {year} {2008})}\BibitemShut {NoStop}%
\bibitem [{\citenamefont {Kuroki}\ \emph {et~al.}(2008)\citenamefont {Kuroki},
	\citenamefont {Onari}, \citenamefont {Arita}, \citenamefont {Usui},
	\citenamefont {Tanaka}, \citenamefont {Kontani},\ and\ \citenamefont
	{Aoki}}]{Kuroki2008}%
\BibitemOpen
\bibfield  {author} {\bibinfo {author} {\bibfnamefont {K.}~\bibnamefont
		{Kuroki}}, \bibinfo {author} {\bibfnamefont {S.}~\bibnamefont {Onari}},
	\bibinfo {author} {\bibfnamefont {R.}~\bibnamefont {Arita}}, \bibinfo
	{author} {\bibfnamefont {H.}~\bibnamefont {Usui}}, \bibinfo {author}
	{\bibfnamefont {Y.}~\bibnamefont {Tanaka}}, \bibinfo {author} {\bibfnamefont
		{H.}~\bibnamefont {Kontani}}, \ and\ \bibinfo {author} {\bibfnamefont
		{H.}~\bibnamefont {Aoki}},\ }\href {\doibase 10.1103/PhysRevLett.101.087004}
{\bibfield  {journal} {\bibinfo  {journal} {Phys. Rev. Lett.}\ }\textbf
	{\bibinfo {volume} {101}},\ \bibinfo {pages} {087004} (\bibinfo {year}
	{2008})}\BibitemShut {NoStop}%
\bibitem [{\citenamefont {Platt}\ \emph {et~al.}(2011)\citenamefont {Platt},
	\citenamefont {Thomale},\ and\ \citenamefont {Hanke}}]{Platt2011}%
\BibitemOpen
\bibfield  {author} {\bibinfo {author} {\bibfnamefont {C.}~\bibnamefont
		{Platt}}, \bibinfo {author} {\bibfnamefont {R.}~\bibnamefont {Thomale}}, \
	and\ \bibinfo {author} {\bibfnamefont {W.}~\bibnamefont {Hanke}},\ }\href
{\doibase 10.1103/PhysRevB.84.235121} {\bibfield  {journal} {\bibinfo
		{journal} {Phys. Rev. B}\ }\textbf {\bibinfo {volume} {84}},\ \bibinfo
	{pages} {235121} (\bibinfo {year} {2011})}\BibitemShut {NoStop}%
\bibitem [{\citenamefont {Aperis}\ \emph {et~al.}(2011)\citenamefont {Aperis},
	\citenamefont {Kotetes}, \citenamefont {Varelogiannis},\ and\ \citenamefont
	{Oppeneer}}]{Aperis2011}%
\BibitemOpen
\bibfield  {author} {\bibinfo {author} {\bibfnamefont {A.}~\bibnamefont
		{Aperis}}, \bibinfo {author} {\bibfnamefont {P.}~\bibnamefont {Kotetes}},
	\bibinfo {author} {\bibfnamefont {G.}~\bibnamefont {Varelogiannis}}, \ and\
	\bibinfo {author} {\bibfnamefont {P.~M.}\ \bibnamefont {Oppeneer}},\ }\href
{\doibase 10.1103/PhysRevB.83.092505} {\bibfield  {journal} {\bibinfo
		{journal} {Phys. Rev. B}\ }\textbf {\bibinfo {volume} {83}},\ \bibinfo
	{pages} {092505} (\bibinfo {year} {2011})}\BibitemShut {NoStop}%
\bibitem [{\citenamefont {Ikeda}\ \emph {et~al.}(2010)\citenamefont {Ikeda},
	\citenamefont {Arita},\ and\ \citenamefont {Kune\ifmmode~\check{s}\else
		\v{s}\fi{}}}]{Ikeda2010}%
\BibitemOpen
\bibfield  {author} {\bibinfo {author} {\bibfnamefont {H.}~\bibnamefont
		{Ikeda}}, \bibinfo {author} {\bibfnamefont {R.}~\bibnamefont {Arita}}, \ and\
	\bibinfo {author} {\bibfnamefont {J.}~\bibnamefont
		{Kune\ifmmode~\check{s}\else \v{s}\fi{}}},\ }\href {\doibase
	10.1103/PhysRevB.82.024508} {\bibfield  {journal} {\bibinfo  {journal} {Phys.
			Rev. B}\ }\textbf {\bibinfo {volume} {82}},\ \bibinfo {pages} {024508}
	(\bibinfo {year} {2010})}\BibitemShut {NoStop}%
\bibitem [{\citenamefont {Scalapino}(2012)}]{Scalapino2012}%
\BibitemOpen
\bibfield  {author} {\bibinfo {author} {\bibfnamefont {D.~J.}\ \bibnamefont
		{Scalapino}},\ }\href {\doibase 10.1103/RevModPhys.84.1383} {\bibfield
	{journal} {\bibinfo  {journal} {Rev. Mod. Phys.}\ }\textbf {\bibinfo {volume}
		{84}},\ \bibinfo {pages} {1383} (\bibinfo {year} {2012})}\BibitemShut
{NoStop}%
\bibitem [{\citenamefont {Essenberger}\ \emph {et~al.}(2012)\citenamefont
	{Essenberger}, \citenamefont {Buczek}, \citenamefont {Ernst}, \citenamefont
	{Sandratskii},\ and\ \citenamefont {Gross}}]{Essenberger2012}%
\BibitemOpen
\bibfield  {author} {\bibinfo {author} {\bibfnamefont {F.}~\bibnamefont
		{Essenberger}}, \bibinfo {author} {\bibfnamefont {P.}~\bibnamefont {Buczek}},
	\bibinfo {author} {\bibfnamefont {A.}~\bibnamefont {Ernst}}, \bibinfo
	{author} {\bibfnamefont {L.}~\bibnamefont {Sandratskii}}, \ and\ \bibinfo
	{author} {\bibfnamefont {E.~K.~U.}\ \bibnamefont {Gross}},\ }\href {\doibase
	10.1103/PhysRevB.86.060412} {\bibfield  {journal} {\bibinfo  {journal} {Phys.
			Rev. B}\ }\textbf {\bibinfo {volume} {86}},\ \bibinfo {pages} {060412}
	(\bibinfo {year} {2012})}\BibitemShut {NoStop}%
\bibitem [{\citenamefont {Lischner}\ \emph {et~al.}(2015)\citenamefont
	{Lischner}, \citenamefont {Bazhirov}, \citenamefont {MacDonald},
	\citenamefont {Cohen},\ and\ \citenamefont {Louie}}]{Lischner2015}%
\BibitemOpen
\bibfield  {author} {\bibinfo {author} {\bibfnamefont {J.}~\bibnamefont
		{Lischner}}, \bibinfo {author} {\bibfnamefont {T.}~\bibnamefont {Bazhirov}},
	\bibinfo {author} {\bibfnamefont {A.~H.}\ \bibnamefont {MacDonald}}, \bibinfo
	{author} {\bibfnamefont {M.~L.}\ \bibnamefont {Cohen}}, \ and\ \bibinfo
	{author} {\bibfnamefont {S.~G.}\ \bibnamefont {Louie}},\ }\href {\doibase
	10.1103/PhysRevB.91.020502} {\bibfield  {journal} {\bibinfo  {journal} {Phys.
			Rev. B}\ }\textbf {\bibinfo {volume} {91}},\ \bibinfo {pages} {020502}
	(\bibinfo {year} {2015})}\BibitemShut {NoStop}%
\bibitem [{\citenamefont {Essenberger}\ \emph {et~al.}(2016)\citenamefont
	{Essenberger}, \citenamefont {Sanna}, \citenamefont {Buczek}, \citenamefont
	{Ernst}, \citenamefont {Sandratskii},\ and\ \citenamefont
	{Gross}}]{Essenberger2016}%
\BibitemOpen
\bibfield  {author} {\bibinfo {author} {\bibfnamefont {F.}~\bibnamefont
		{Essenberger}}, \bibinfo {author} {\bibfnamefont {A.}~\bibnamefont {Sanna}},
	\bibinfo {author} {\bibfnamefont {P.}~\bibnamefont {Buczek}}, \bibinfo
	{author} {\bibfnamefont {A.}~\bibnamefont {Ernst}}, \bibinfo {author}
	{\bibfnamefont {L.}~\bibnamefont {Sandratskii}}, \ and\ \bibinfo {author}
	{\bibfnamefont {E.~K.~U.}\ \bibnamefont {Gross}},\ }\href {\doibase
	10.1103/PhysRevB.94.014503} {\bibfield  {journal} {\bibinfo  {journal} {Phys.
			Rev. B}\ }\textbf {\bibinfo {volume} {94}},\ \bibinfo {pages} {014503}
	(\bibinfo {year} {2016})}\BibitemShut {NoStop}%
\bibitem [{\citenamefont {Rebec}\ \emph {et~al.}(2017)\citenamefont {Rebec},
	\citenamefont {Jia}, \citenamefont {Zhang}, \citenamefont {Hashimoto},
	\citenamefont {Lu}, \citenamefont {Moore},\ and\ \citenamefont
	{Shen}}]{Rebec2017}%
\BibitemOpen
\bibfield  {author} {\bibinfo {author} {\bibfnamefont {S.~N.}\ \bibnamefont
		{Rebec}}, \bibinfo {author} {\bibfnamefont {T.}~\bibnamefont {Jia}}, \bibinfo
	{author} {\bibfnamefont {C.}~\bibnamefont {Zhang}}, \bibinfo {author}
	{\bibfnamefont {M.}~\bibnamefont {Hashimoto}}, \bibinfo {author}
	{\bibfnamefont {D.-H.}\ \bibnamefont {Lu}}, \bibinfo {author} {\bibfnamefont
		{R.~G.}\ \bibnamefont {Moore}}, \ and\ \bibinfo {author} {\bibfnamefont
		{Z.-X.}\ \bibnamefont {Shen}},\ }\href {\doibase
	10.1103/PhysRevLett.118.067002} {\bibfield  {journal} {\bibinfo  {journal}
		{Phys. Rev. Lett.}\ }\textbf {\bibinfo {volume} {118}},\ \bibinfo {pages}
	{067002} (\bibinfo {year} {2017})}\BibitemShut {NoStop}%
\bibitem [{\citenamefont {Fan}\ \emph {et~al.}(2015)\citenamefont {Fan},
	\citenamefont {Zhang}, \citenamefont {Liu}, \citenamefont {Yan},
	\citenamefont {Ren}, \citenamefont {Peng}, \citenamefont {Xu}, \citenamefont
	{Xie}, \citenamefont {Hu}, \citenamefont {Zhang},\ and\ \citenamefont
	{Feng}}]{Fan2015}%
\BibitemOpen
\bibfield  {author} {\bibinfo {author} {\bibfnamefont {Q.}~\bibnamefont
		{Fan}}, \bibinfo {author} {\bibfnamefont {W.~H.}\ \bibnamefont {Zhang}},
	\bibinfo {author} {\bibfnamefont {X.}~\bibnamefont {Liu}}, \bibinfo {author}
	{\bibfnamefont {Y.~J.}\ \bibnamefont {Yan}}, \bibinfo {author} {\bibfnamefont
		{M.~Q.}\ \bibnamefont {Ren}}, \bibinfo {author} {\bibfnamefont
		{R.}~\bibnamefont {Peng}}, \bibinfo {author} {\bibfnamefont {H.~C.}\
		\bibnamefont {Xu}}, \bibinfo {author} {\bibfnamefont {B.~P.}\ \bibnamefont
		{Xie}}, \bibinfo {author} {\bibfnamefont {J.~P.}\ \bibnamefont {Hu}},
	\bibinfo {author} {\bibfnamefont {T.}~\bibnamefont {Zhang}}, \ and\ \bibinfo
	{author} {\bibfnamefont {D.~L.}\ \bibnamefont {Feng}},\ }\href
{http://dx.doi.org/10.1038/nphys3450} {\bibfield  {journal} {\bibinfo
		{journal} {Nat. Phys.}\ }\textbf {\bibinfo {volume} {11}},\ \bibinfo {pages}
	{946} (\bibinfo {year} {2015})}\BibitemShut {NoStop}%
\bibitem [{\citenamefont {Huang}\ and\ \citenamefont
	{Hoffman}(2017)}]{Huang2017_2}%
\BibitemOpen
\bibfield  {author} {\bibinfo {author} {\bibfnamefont {D.}~\bibnamefont
		{Huang}}\ and\ \bibinfo {author} {\bibfnamefont {J.~E.}\ \bibnamefont
		{Hoffman}},\ }\href {\doibase 10.1146/annurev-conmatphys-031016-025242}
{\bibfield  {journal} {\bibinfo  {journal} {Ann. Rev. Condens. Matter Phys.}\
	}\textbf {\bibinfo {volume} {8}},\ \bibinfo {pages} {311} (\bibinfo {year}
	{2017})}\BibitemShut {NoStop}%
\bibitem [{\citenamefont {Sadovskii}(2016)}]{Sadovskii2016}%
\BibitemOpen
\bibfield  {author} {\bibinfo {author} {\bibfnamefont {M.~V.}\ \bibnamefont
		{Sadovskii}},\ }\href {\doibase 10.3367/ufne.2016.06.037825} {\bibfield
	{journal} {\bibinfo  {journal} {Physics-Uspekhi}\ }\textbf {\bibinfo {volume}
		{59}},\ \bibinfo {pages} {947} (\bibinfo {year} {2016})}\BibitemShut
{NoStop}%
\bibitem [{\citenamefont {Rademaker}\ \emph {et~al.}(2016)\citenamefont
	{Rademaker}, \citenamefont {Wang}, \citenamefont {Berlijn},\ and\
	\citenamefont {Johnston}}]{Rademaker2016}%
\BibitemOpen
\bibfield  {author} {\bibinfo {author} {\bibfnamefont {L.}~\bibnamefont
		{Rademaker}}, \bibinfo {author} {\bibfnamefont {Y.}~\bibnamefont {Wang}},
	\bibinfo {author} {\bibfnamefont {T.}~\bibnamefont {Berlijn}}, \ and\
	\bibinfo {author} {\bibfnamefont {S.}~\bibnamefont {Johnston}},\ }\href
{http://stacks.iop.org/1367-2630/18/i=2/a=022001} {\bibfield  {journal}
	{\bibinfo  {journal} {New J. Phys.}\ }\textbf {\bibinfo {volume} {18}},\
	\bibinfo {pages} {022001} (\bibinfo {year} {2016})}\BibitemShut {NoStop}%
\bibitem [{\citenamefont {Song}\ \emph {et~al.}(2019)\citenamefont {Song},
	\citenamefont {Yu}, \citenamefont {Lou}, \citenamefont {Xie}, \citenamefont
	{Xu}, \citenamefont {Wen}, \citenamefont {Yao}, \citenamefont {Zhang},
	\citenamefont {Zhu}, \citenamefont {Guo}, \citenamefont {Peng},\ and\
	\citenamefont {Feng}}]{Song2019}%
\BibitemOpen
\bibfield  {author} {\bibinfo {author} {\bibfnamefont {Q.}~\bibnamefont
		{Song}}, \bibinfo {author} {\bibfnamefont {T.~L.}\ \bibnamefont {Yu}},
	\bibinfo {author} {\bibfnamefont {X.}~\bibnamefont {Lou}}, \bibinfo {author}
	{\bibfnamefont {B.~P.}\ \bibnamefont {Xie}}, \bibinfo {author} {\bibfnamefont
		{H.~C.}\ \bibnamefont {Xu}}, \bibinfo {author} {\bibfnamefont {C.~H.~P.}\
		\bibnamefont {Wen}}, \bibinfo {author} {\bibfnamefont {Q.}~\bibnamefont
		{Yao}}, \bibinfo {author} {\bibfnamefont {S.~Y.}\ \bibnamefont {Zhang}},
	\bibinfo {author} {\bibfnamefont {X.~T.}\ \bibnamefont {Zhu}}, \bibinfo
	{author} {\bibfnamefont {J.~D.}\ \bibnamefont {Guo}}, \bibinfo {author}
	{\bibfnamefont {R.}~\bibnamefont {Peng}}, \ and\ \bibinfo {author}
	{\bibfnamefont {D.~L.}\ \bibnamefont {Feng}},\ }\href {\doibase
	10.1038/s41467-019-08560-z} {\bibfield  {journal} {\bibinfo  {journal} {Nat.
			Commun.}\ }\textbf {\bibinfo {volume} {10}},\ \bibinfo {pages} {758}
	(\bibinfo {year} {2019})}\BibitemShut {NoStop}%
\bibitem [{\citenamefont {Aperis}\ and\ \citenamefont
	{Oppeneer}(2018)}]{Aperis2018}%
\BibitemOpen
\bibfield  {author} {\bibinfo {author} {\bibfnamefont {A.}~\bibnamefont
		{Aperis}}\ and\ \bibinfo {author} {\bibfnamefont {P.~M.}\ \bibnamefont
		{Oppeneer}},\ }\href {\doibase 10.1103/PhysRevB.97.060501} {\bibfield
	{journal} {\bibinfo  {journal} {Phys. Rev. B}\ }\textbf {\bibinfo {volume}
		{97}},\ \bibinfo {pages} {060501} (\bibinfo {year} {2018})}\BibitemShut
{NoStop}%
\bibitem [{\citenamefont {Schrodi}\ \emph {et~al.}(2018)\citenamefont
	{Schrodi}, \citenamefont {Aperis},\ and\ \citenamefont
	{Oppeneer}}]{Schrodi2018}%
\BibitemOpen
\bibfield  {author} {\bibinfo {author} {\bibfnamefont {F.}~\bibnamefont
		{Schrodi}}, \bibinfo {author} {\bibfnamefont {A.}~\bibnamefont {Aperis}}, \
	and\ \bibinfo {author} {\bibfnamefont {P.~M.}\ \bibnamefont {Oppeneer}},\
}\href {\doibase 10.1103/PhysRevB.98.094509} {\bibfield  {journal} {\bibinfo
	{journal} {Phys. Rev. B}\ }\textbf {\bibinfo {volume} {98}},\ \bibinfo
{pages} {094509} (\bibinfo {year} {2018})}\BibitemShut {NoStop}%
\bibitem [{\citenamefont {Gao}\ \emph {et~al.}(2017)\citenamefont {Gao},
	\citenamefont {Yu}, \citenamefont {Zhou}, \citenamefont {Huang},\ and\
	\citenamefont {Wang}}]{Gao2017}%
\BibitemOpen
\bibfield  {author} {\bibinfo {author} {\bibfnamefont {Y.}~\bibnamefont
		{Gao}}, \bibinfo {author} {\bibfnamefont {Y.}~\bibnamefont {Yu}}, \bibinfo
	{author} {\bibfnamefont {T.}~\bibnamefont {Zhou}}, \bibinfo {author}
	{\bibfnamefont {H.}~\bibnamefont {Huang}}, \ and\ \bibinfo {author}
	{\bibfnamefont {Q.-H.}\ \bibnamefont {Wang}},\ }\href {\doibase
	10.1103/PhysRevB.96.014515} {\bibfield  {journal} {\bibinfo  {journal} {Phys.
			Rev. B}\ }\textbf {\bibinfo {volume} {96}},\ \bibinfo {pages} {014515}
	(\bibinfo {year} {2017})}\BibitemShut {NoStop}%
\bibitem [{\citenamefont {Shishidou}\ \emph {et~al.}(2018)\citenamefont
	{Shishidou}, \citenamefont {Agterberg},\ and\ \citenamefont
	{Weinert}}]{Shishidou2018}%
\BibitemOpen
\bibfield  {author} {\bibinfo {author} {\bibfnamefont {T.}~\bibnamefont
		{Shishidou}}, \bibinfo {author} {\bibfnamefont {D.~F.}\ \bibnamefont
		{Agterberg}}, \ and\ \bibinfo {author} {\bibfnamefont {M.}~\bibnamefont
		{Weinert}},\ }\href {\doibase 10.1038/s42005-018-0006-7} {\bibfield
	{journal} {\bibinfo  {journal} {Commun. Phys.}\ }\textbf {\bibinfo {volume}
		{1}},\ \bibinfo {pages} {8} (\bibinfo {year} {2018})}\BibitemShut {NoStop}%
\bibitem [{\citenamefont {Jandke}\ \emph {et~al.}(2017)\citenamefont {Jandke},
	\citenamefont {Yang}, \citenamefont {Hlobil}, \citenamefont {Engelhardt},
	\citenamefont {Rau}, \citenamefont {Zakeri}, \citenamefont {Gao},
	\citenamefont {Schmalian},\ and\ \citenamefont {Wulfhekel}}]{Jandke2017}%
\BibitemOpen
\bibfield  {author} {\bibinfo {author} {\bibfnamefont {J.}~\bibnamefont
		{Jandke}}, \bibinfo {author} {\bibfnamefont {F.}~\bibnamefont {Yang}},
	\bibinfo {author} {\bibfnamefont {P.}~\bibnamefont {Hlobil}}, \bibinfo
	{author} {\bibfnamefont {T.}~\bibnamefont {Engelhardt}}, \bibinfo {author}
	{\bibfnamefont {D.}~\bibnamefont {Rau}}, \bibinfo {author} {\bibfnamefont
		{K.}~\bibnamefont {Zakeri}}, \bibinfo {author} {\bibfnamefont
		{C.}~\bibnamefont {Gao}}, \bibinfo {author} {\bibfnamefont {J.}~\bibnamefont
		{Schmalian}}, \ and\ \bibinfo {author} {\bibfnamefont {W.}~\bibnamefont
		{Wulfhekel}},\ }\href@noop {} {\enquote {\bibinfo {title} {Unconventional
			pairing in single {F}e{S}e layers},}\ } (\bibinfo {year} {2017}),\ \Eprint
{http://arxiv.org/abs/arXiv:1710.08861} {arXiv:1710.08861} \BibitemShut
{NoStop}%
\bibitem [{\citenamefont {Linscheid}\ \emph {et~al.}(2016)\citenamefont
	{Linscheid}, \citenamefont {Maiti}, \citenamefont {Wang}, \citenamefont
	{Johnston},\ and\ \citenamefont {Hirschfeld}}]{Linscheid2016}%
\BibitemOpen
\bibfield  {author} {\bibinfo {author} {\bibfnamefont {A.}~\bibnamefont
		{Linscheid}}, \bibinfo {author} {\bibfnamefont {S.}~\bibnamefont {Maiti}},
	\bibinfo {author} {\bibfnamefont {Y.}~\bibnamefont {Wang}}, \bibinfo {author}
	{\bibfnamefont {S.}~\bibnamefont {Johnston}}, \ and\ \bibinfo {author}
	{\bibfnamefont {P.~J.}\ \bibnamefont {Hirschfeld}},\ }\href {\doibase
	10.1103/PhysRevLett.117.077003} {\bibfield  {journal} {\bibinfo  {journal}
		{Phys. Rev. Lett.}\ }\textbf {\bibinfo {volume} {117}},\ \bibinfo {pages}
	{077003} (\bibinfo {year} {2016})}\BibitemShut {NoStop}%
\bibitem [{\citenamefont {Kreisel}\ \emph {et~al.}(2017)\citenamefont
	{Kreisel}, \citenamefont {Andersen}, \citenamefont {Sprau}, \citenamefont
	{Kostin}, \citenamefont {Davis},\ and\ \citenamefont
	{Hirschfeld}}]{Kreisel2017}%
\BibitemOpen
\bibfield  {author} {\bibinfo {author} {\bibfnamefont {A.}~\bibnamefont
		{Kreisel}}, \bibinfo {author} {\bibfnamefont {B.~M.}\ \bibnamefont
		{Andersen}}, \bibinfo {author} {\bibfnamefont {P.~O.}\ \bibnamefont {Sprau}},
	\bibinfo {author} {\bibfnamefont {A.}~\bibnamefont {Kostin}}, \bibinfo
	{author} {\bibfnamefont {J.~C.~S.}\ \bibnamefont {Davis}}, \ and\ \bibinfo
	{author} {\bibfnamefont {P.~J.}\ \bibnamefont {Hirschfeld}},\ }\href
{\doibase 10.1103/PhysRevB.95.174504} {\bibfield  {journal} {\bibinfo
		{journal} {Phys. Rev. B}\ }\textbf {\bibinfo {volume} {95}},\ \bibinfo
	{pages} {174504} (\bibinfo {year} {2017})}\BibitemShut {NoStop}%
\bibitem [{\citenamefont {Eschrig}\ and\ \citenamefont
	{Koepernik}(2009)}]{Eschrig2009}%
\BibitemOpen
\bibfield  {author} {\bibinfo {author} {\bibfnamefont {H.}~\bibnamefont
		{Eschrig}}\ and\ \bibinfo {author} {\bibfnamefont {K.}~\bibnamefont
		{Koepernik}},\ }\href {\doibase 10.1103/PhysRevB.80.104503} {\bibfield
	{journal} {\bibinfo  {journal} {Phys. Rev. B}\ }\textbf {\bibinfo {volume}
		{80}},\ \bibinfo {pages} {104503} (\bibinfo {year} {2009})}\BibitemShut
{NoStop}%
\bibitem [{\citenamefont {Hao}\ and\ \citenamefont {Hu}(2014)}]{Hao2014}%
\BibitemOpen
\bibfield  {author} {\bibinfo {author} {\bibfnamefont {N.}~\bibnamefont
		{Hao}}\ and\ \bibinfo {author} {\bibfnamefont {J.}~\bibnamefont {Hu}},\
}\href {\doibase 10.1103/PhysRevX.4.031053} {\bibfield  {journal} {\bibinfo
	{journal} {Phys. Rev. X}\ }\textbf {\bibinfo {volume} {4}},\ \bibinfo {pages}
{031053} (\bibinfo {year} {2014})}\BibitemShut {NoStop}%
\bibitem [{\citenamefont {Zhang}\ \emph {et~al.}(2017)\citenamefont {Zhang},
	\citenamefont {Liu}, \citenamefont {Chen}, \citenamefont {Xie}, \citenamefont
	{He}, \citenamefont {Tang}, \citenamefont {He}, \citenamefont {Li},
	\citenamefont {Jia}, \citenamefont {Rebec}, \citenamefont {Ma}, \citenamefont
	{Yan}, \citenamefont {Hashimoto}, \citenamefont {Lu}, \citenamefont {Mo},
	\citenamefont {Hikita}, \citenamefont {Moore}, \citenamefont {Hwang},
	\citenamefont {Lee},\ and\ \citenamefont {Shen}}]{Zhang2017}%
\BibitemOpen
\bibfield  {author} {\bibinfo {author} {\bibfnamefont {C.}~\bibnamefont
		{Zhang}}, \bibinfo {author} {\bibfnamefont {Z.}~\bibnamefont {Liu}}, \bibinfo
	{author} {\bibfnamefont {Z.}~\bibnamefont {Chen}}, \bibinfo {author}
	{\bibfnamefont {Y.}~\bibnamefont {Xie}}, \bibinfo {author} {\bibfnamefont
		{R.}~\bibnamefont {He}}, \bibinfo {author} {\bibfnamefont {S.}~\bibnamefont
		{Tang}}, \bibinfo {author} {\bibfnamefont {J.}~\bibnamefont {He}}, \bibinfo
	{author} {\bibfnamefont {W.}~\bibnamefont {Li}}, \bibinfo {author}
	{\bibfnamefont {T.}~\bibnamefont {Jia}}, \bibinfo {author} {\bibfnamefont
		{S.~N.}\ \bibnamefont {Rebec}}, \bibinfo {author} {\bibfnamefont {E.~Y.}\
		\bibnamefont {Ma}}, \bibinfo {author} {\bibfnamefont {H.}~\bibnamefont
		{Yan}}, \bibinfo {author} {\bibfnamefont {M.}~\bibnamefont {Hashimoto}},
	\bibinfo {author} {\bibfnamefont {D.}~\bibnamefont {Lu}}, \bibinfo {author}
	{\bibfnamefont {S.-K.}\ \bibnamefont {Mo}}, \bibinfo {author} {\bibfnamefont
		{Y.}~\bibnamefont {Hikita}}, \bibinfo {author} {\bibfnamefont {R.~G.}\
		\bibnamefont {Moore}}, \bibinfo {author} {\bibfnamefont {H.~Y.}\ \bibnamefont
		{Hwang}}, \bibinfo {author} {\bibfnamefont {D.}~\bibnamefont {Lee}}, \ and\
	\bibinfo {author} {\bibfnamefont {Z.}~\bibnamefont {Shen}},\ }\href
{http://dx.doi.org/10.1038/ncomms14468} {\bibfield  {journal} {\bibinfo
		{journal} {Nat. Commun.}\ }\textbf {\bibinfo {volume} {8}},\ \bibinfo {pages}
	{14468} (\bibinfo {year} {2017})}\BibitemShut {NoStop}%
\bibitem [{\citenamefont {Ole\'s}(1983)}]{Oles1983}%
\BibitemOpen
\bibfield  {author} {\bibinfo {author} {\bibfnamefont {A.~M.}\ \bibnamefont
		{Ole\'s}},\ }\href {\doibase 10.1103/PhysRevB.28.327} {\bibfield  {journal}
	{\bibinfo  {journal} {Phys. Rev. B}\ }\textbf {\bibinfo {volume} {28}},\
	\bibinfo {pages} {327} (\bibinfo {year} {1983})}\BibitemShut {NoStop}%
\bibitem [{\citenamefont {Takimoto}\ \emph {et~al.}(2004)\citenamefont
	{Takimoto}, \citenamefont {Hotta},\ and\ \citenamefont
	{Ueda}}]{Takimoto2004}%
\BibitemOpen
\bibfield  {author} {\bibinfo {author} {\bibfnamefont {T.}~\bibnamefont
		{Takimoto}}, \bibinfo {author} {\bibfnamefont {T.}~\bibnamefont {Hotta}}, \
	and\ \bibinfo {author} {\bibfnamefont {K.}~\bibnamefont {Ueda}},\ }\href
{\doibase 10.1103/PhysRevB.69.104504} {\bibfield  {journal} {\bibinfo
		{journal} {Phys. Rev. B}\ }\textbf {\bibinfo {volume} {69}},\ \bibinfo
	{pages} {104504} (\bibinfo {year} {2004})}\BibitemShut {NoStop}%
\bibitem [{\citenamefont {Kuroki}\ \emph {et~al.}(2009)\citenamefont {Kuroki},
	\citenamefont {Usui}, \citenamefont {Onari}, \citenamefont {Arita},\ and\
	\citenamefont {Aoki}}]{Kuroki2009}%
\BibitemOpen
\bibfield  {author} {\bibinfo {author} {\bibfnamefont {K.}~\bibnamefont
		{Kuroki}}, \bibinfo {author} {\bibfnamefont {H.}~\bibnamefont {Usui}},
	\bibinfo {author} {\bibfnamefont {S.}~\bibnamefont {Onari}}, \bibinfo
	{author} {\bibfnamefont {R.}~\bibnamefont {Arita}}, \ and\ \bibinfo {author}
	{\bibfnamefont {H.}~\bibnamefont {Aoki}},\ }\href {\doibase
	10.1103/PhysRevB.79.224511} {\bibfield  {journal} {\bibinfo  {journal} {Phys.
			Rev. B}\ }\textbf {\bibinfo {volume} {79}},\ \bibinfo {pages} {224511}
	(\bibinfo {year} {2009})}\BibitemShut {NoStop}%
\bibitem [{\citenamefont {Yates}\ \emph {et~al.}(2007)\citenamefont {Yates},
	\citenamefont {Wang}, \citenamefont {Vanderbilt},\ and\ \citenamefont
	{Souza}}]{Yates2007}%
\BibitemOpen
\bibfield  {author} {\bibinfo {author} {\bibfnamefont {J.~R.}\ \bibnamefont
		{Yates}}, \bibinfo {author} {\bibfnamefont {X.}~\bibnamefont {Wang}},
	\bibinfo {author} {\bibfnamefont {D.}~\bibnamefont {Vanderbilt}}, \ and\
	\bibinfo {author} {\bibfnamefont {I.}~\bibnamefont {Souza}},\ }\href
{\doibase 10.1103/PhysRevB.75.195121} {\bibfield  {journal} {\bibinfo
		{journal} {Phys. Rev. B}\ }\textbf {\bibinfo {volume} {75}},\ \bibinfo
	{pages} {195121} (\bibinfo {year} {2007})}\BibitemShut {NoStop}%
\bibitem [{\citenamefont {Heil}\ \emph {et~al.}(2014)\citenamefont {Heil},
	\citenamefont {Sormann}, \citenamefont {Boeri}, \citenamefont {Aichhorn},\
	and\ \citenamefont {von~der Linden}}]{Heil2014}%
\BibitemOpen
\bibfield  {author} {\bibinfo {author} {\bibfnamefont {C.}~\bibnamefont
		{Heil}}, \bibinfo {author} {\bibfnamefont {H.}~\bibnamefont {Sormann}},
	\bibinfo {author} {\bibfnamefont {L.}~\bibnamefont {Boeri}}, \bibinfo
	{author} {\bibfnamefont {M.}~\bibnamefont {Aichhorn}}, \ and\ \bibinfo
	{author} {\bibfnamefont {W.}~\bibnamefont {von~der Linden}},\ }\href
{\doibase 10.1103/PhysRevB.90.115143} {\bibfield  {journal} {\bibinfo
		{journal} {Phys. Rev. B}\ }\textbf {\bibinfo {volume} {90}},\ \bibinfo
	{pages} {115143} (\bibinfo {year} {2014})}\BibitemShut {NoStop}%
\bibitem [{\citenamefont {Haule}\ \emph {et~al.}(2010)\citenamefont {Haule},
	\citenamefont {Yee},\ and\ \citenamefont {Kim}}]{Haule2010}%
\BibitemOpen
\bibfield  {author} {\bibinfo {author} {\bibfnamefont {K.}~\bibnamefont
		{Haule}}, \bibinfo {author} {\bibfnamefont {C.-H.}\ \bibnamefont {Yee}}, \
	and\ \bibinfo {author} {\bibfnamefont {K.}~\bibnamefont {Kim}},\ }\href
{\doibase 10.1103/PhysRevB.81.195107} {\bibfield  {journal} {\bibinfo
		{journal} {Phys. Rev. B}\ }\textbf {\bibinfo {volume} {81}},\ \bibinfo
	{pages} {195107} (\bibinfo {year} {2010})}\BibitemShut {NoStop}%
\bibitem [{Upp()}]{UppSC}%
\BibitemOpen
\href@noop {} {}\bibinfo {note} {The Uppsala Superconductivity (UppSC) code
	provides a package to selfconsistently solve the anisotropic, multiband, and
	full-bandwidth Eliashberg equations for frequency-even and odd
	superconductivity mediated by phonons or spin-fluctuations on the basis of
	\textit{ab initio} calculated input.}\BibitemShut {Stop}%
\bibitem [{\citenamefont {Aperis}\ \emph {et~al.}(2015)\citenamefont {Aperis},
	\citenamefont {Maldonado},\ and\ \citenamefont {Oppeneer}}]{Aperis2015}%
\BibitemOpen
\bibfield  {author} {\bibinfo {author} {\bibfnamefont {A.}~\bibnamefont
		{Aperis}}, \bibinfo {author} {\bibfnamefont {P.}~\bibnamefont {Maldonado}}, \
	and\ \bibinfo {author} {\bibfnamefont {P.~M.}\ \bibnamefont {Oppeneer}},\
}\href {\doibase 10.1103/PhysRevB.92.054516} {\bibfield  {journal} {\bibinfo
	{journal} {Phys. Rev. B}\ }\textbf {\bibinfo {volume} {92}},\ \bibinfo
{pages} {054516} (\bibinfo {year} {2015})}\BibitemShut {NoStop}%
\bibitem [{\citenamefont {Bekaert}\ \emph {et~al.}(2018)\citenamefont
	{Bekaert}, \citenamefont {Aperis}, \citenamefont {Partoens}, \citenamefont
	{Oppeneer},\ and\ \citenamefont {Milo\ifmmode \check{s}\else
		\v{s}\fi{}evi\ifmmode~\acute{c}\else \'{c}\fi{}}}]{Bekaert2018}%
\BibitemOpen
\bibfield  {author} {\bibinfo {author} {\bibfnamefont {J.}~\bibnamefont
		{Bekaert}}, \bibinfo {author} {\bibfnamefont {A.}~\bibnamefont {Aperis}},
	\bibinfo {author} {\bibfnamefont {B.}~\bibnamefont {Partoens}}, \bibinfo
	{author} {\bibfnamefont {P.~M.}\ \bibnamefont {Oppeneer}}, \ and\ \bibinfo
	{author} {\bibfnamefont {M.~V.}\ \bibnamefont {Milo\ifmmode \check{s}\else
			\v{s}\fi{}evi\ifmmode~\acute{c}\else \'{c}\fi{}}},\ }\href {\doibase
	10.1103/PhysRevB.97.014503} {\bibfield  {journal} {\bibinfo  {journal} {Phys.
			Rev. B}\ }\textbf {\bibinfo {volume} {97}},\ \bibinfo {pages} {014503}
	(\bibinfo {year} {2018})}\BibitemShut {NoStop}%
\bibitem [{\citenamefont {Schrodi}\ \emph {et~al.}(2019)\citenamefont
	{Schrodi}, \citenamefont {Aperis},\ and\ \citenamefont
	{Oppeneer}}]{Schrodi2018_2}%
\BibitemOpen
\bibfield  {author} {\bibinfo {author} {\bibfnamefont {F.}~\bibnamefont
		{Schrodi}}, \bibinfo {author} {\bibfnamefont {A.}~\bibnamefont {Aperis}}, \
	and\ \bibinfo {author} {\bibfnamefont {P.~M.}\ \bibnamefont {Oppeneer}},\
}\href {\doibase 10.1103/PhysRevB.99.184508} {\bibfield  {journal} {\bibinfo
	{journal} {Phys. Rev. B}\ }\textbf {\bibinfo {volume} {99}},\ \bibinfo
{pages} {184508} (\bibinfo {year} {2019})}\BibitemShut {NoStop}%
\bibitem [{\citenamefont {Sprau}\ \emph {et~al.}(2017)\citenamefont {Sprau},
	\citenamefont {Kostin}, \citenamefont {Kreisel}, \citenamefont {B{\"o}hmer},
	\citenamefont {Taufour}, \citenamefont {Canfield}, \citenamefont {Mukherjee},
	\citenamefont {Hirschfeld}, \citenamefont {Andersen},\ and\ \citenamefont
	{Davis}}]{Sprau2017}%
\BibitemOpen
\bibfield  {author} {\bibinfo {author} {\bibfnamefont {P.~O.}\ \bibnamefont
		{Sprau}}, \bibinfo {author} {\bibfnamefont {A.}~\bibnamefont {Kostin}},
	\bibinfo {author} {\bibfnamefont {A.}~\bibnamefont {Kreisel}}, \bibinfo
	{author} {\bibfnamefont {A.~E.}\ \bibnamefont {B{\"o}hmer}}, \bibinfo
	{author} {\bibfnamefont {V.}~\bibnamefont {Taufour}}, \bibinfo {author}
	{\bibfnamefont {P.~C.}\ \bibnamefont {Canfield}}, \bibinfo {author}
	{\bibfnamefont {S.}~\bibnamefont {Mukherjee}}, \bibinfo {author}
	{\bibfnamefont {P.~J.}\ \bibnamefont {Hirschfeld}}, \bibinfo {author}
	{\bibfnamefont {B.~M.}\ \bibnamefont {Andersen}}, \ and\ \bibinfo {author}
	{\bibfnamefont {J.~C.~S.}\ \bibnamefont {Davis}},\ }\href {\doibase
	10.1126/science.aal1575} {\bibfield  {journal} {\bibinfo  {journal}
		{Science}\ }\textbf {\bibinfo {volume} {357}},\ \bibinfo {pages} {75}
	(\bibinfo {year} {2017})}\BibitemShut {NoStop}%
\bibitem [{\citenamefont {Zhang}\ \emph {et~al.}(2016)\citenamefont {Zhang},
	\citenamefont {Lee}, \citenamefont {Moore}, \citenamefont {Li}, \citenamefont
	{Yi}, \citenamefont {Hashimoto}, \citenamefont {Lu}, \citenamefont
	{Devereaux}, \citenamefont {Lee},\ and\ \citenamefont {Shen}}]{Zhang2016}%
\BibitemOpen
\bibfield  {author} {\bibinfo {author} {\bibfnamefont {Y.}~\bibnamefont
		{Zhang}}, \bibinfo {author} {\bibfnamefont {J.~J.}\ \bibnamefont {Lee}},
	\bibinfo {author} {\bibfnamefont {R.~G.}\ \bibnamefont {Moore}}, \bibinfo
	{author} {\bibfnamefont {W.}~\bibnamefont {Li}}, \bibinfo {author}
	{\bibfnamefont {M.}~\bibnamefont {Yi}}, \bibinfo {author} {\bibfnamefont
		{M.}~\bibnamefont {Hashimoto}}, \bibinfo {author} {\bibfnamefont {D.~H.}\
		\bibnamefont {Lu}}, \bibinfo {author} {\bibfnamefont {T.~P.}\ \bibnamefont
		{Devereaux}}, \bibinfo {author} {\bibfnamefont {D.-H.}\ \bibnamefont {Lee}},
	\ and\ \bibinfo {author} {\bibfnamefont {Z.-X.}\ \bibnamefont {Shen}},\
}\href {\doibase 10.1103/PhysRevLett.117.117001} {\bibfield  {journal}
{\bibinfo  {journal} {Phys. Rev. Lett.}\ }\textbf {\bibinfo {volume} {117}},\
\bibinfo {pages} {117001} (\bibinfo {year} {2016})}\BibitemShut {NoStop}%
\bibitem [{\citenamefont {Ge}\ \emph {et~al.}(2019)\citenamefont {Ge},
	\citenamefont {Yan}, \citenamefont {Zhang}, \citenamefont {Agterberg},
	\citenamefont {Weinert},\ and\ \citenamefont {Li}}]{Ge2019}%
\BibitemOpen
\bibfield  {author} {\bibinfo {author} {\bibfnamefont {Z.}~\bibnamefont
		{Ge}}, \bibinfo {author} {\bibfnamefont {C.}~\bibnamefont {Yan}}, \bibinfo
	{author} {\bibfnamefont {H.}~\bibnamefont {Zhang}}, \bibinfo {author}
	{\bibfnamefont {D.}~\bibnamefont {Agterberg}}, \bibinfo {author}
	{\bibfnamefont {M.}~\bibnamefont {Weinert}}, \ and\ \bibinfo {author}
	{\bibfnamefont {L.}~\bibnamefont {Li}},\ }\href {\doibase
	10.1021/acs.nanolett.9b00135} {\bibfield  {journal} {\bibinfo  {journal}
		{Nano Lett.}\ }\textbf {\bibinfo {volume} {19}},\ \bibinfo {pages} {2497 }
	(\bibinfo {year} {2019})}\BibitemShut {NoStop}%
\bibitem [{\citenamefont {Tang}\ \emph {et~al.}(2016)\citenamefont {Tang},
	\citenamefont {Liu}, \citenamefont {Zhou}, \citenamefont {Li}, \citenamefont
	{Ding}, \citenamefont {Li}, \citenamefont {Zhang}, \citenamefont {Li},
	\citenamefont {Song}, \citenamefont {Ji}, \citenamefont {He}, \citenamefont
	{Wang}, \citenamefont {Ma},\ and\ \citenamefont {Xue}}]{Tang2016}%
\BibitemOpen
\bibfield  {author} {\bibinfo {author} {\bibfnamefont {C.}~\bibnamefont
		{Tang}}, \bibinfo {author} {\bibfnamefont {C.}~\bibnamefont {Liu}}, \bibinfo
	{author} {\bibfnamefont {G.}~\bibnamefont {Zhou}}, \bibinfo {author}
	{\bibfnamefont {F.}~\bibnamefont {Li}}, \bibinfo {author} {\bibfnamefont
		{H.}~\bibnamefont {Ding}}, \bibinfo {author} {\bibfnamefont {Z.}~\bibnamefont
		{Li}}, \bibinfo {author} {\bibfnamefont {D.}~\bibnamefont {Zhang}}, \bibinfo
	{author} {\bibfnamefont {Z.}~\bibnamefont {Li}}, \bibinfo {author}
	{\bibfnamefont {C.}~\bibnamefont {Song}}, \bibinfo {author} {\bibfnamefont
		{S.}~\bibnamefont {Ji}}, \bibinfo {author} {\bibfnamefont {K.}~\bibnamefont
		{He}}, \bibinfo {author} {\bibfnamefont {L.}~\bibnamefont {Wang}}, \bibinfo
	{author} {\bibfnamefont {X.}~\bibnamefont {Ma}}, \ and\ \bibinfo {author}
	{\bibfnamefont {Q.-K.}\ \bibnamefont {Xue}},\ }\href {\doibase
	10.1103/PhysRevB.93.020507} {\bibfield  {journal} {\bibinfo  {journal} {Phys.
			Rev. B}\ }\textbf {\bibinfo {volume} {93}},\ \bibinfo {pages} {020507}
	(\bibinfo {year} {2016})}\BibitemShut {NoStop}%
\end{thebibliography}

%

\appendix

\section{Computational aspects}\label{appMethodDetails}

The tight-binding models employed in this work are taken from Refs.\,\cite{Eschrig2009,Hao2014} and give rise to matrix elements obeying $a_{\mathbf{k}n}^p\in\mathbb{R}$  for all $p$, $\mathbf{k}$, and $n$, hence $a_{\mathbf{k}n}^{p\, *} =a_{\mathbf{k}n}^{p}$. As a consequence, some numerical calculations can be made more efficient and an analytical analysis becomes easier.

When calculating the imaginary part of the bare susceptibility we confine ourselves to the irreducible part of the BZ in momentum $\mathbf{q}$, since the result must respect the tetragonal symmetry of the electronic dispersion. This includes, but is not restricted to inversion symmetry:
\begin{align}
{\rm Im}\left(\big[\chi^0_{{\bf q}}(\omega)\big]_{st}^{pq}\right) = {\rm Im}\left(\big[\chi^0_{-{\bf q}}(\omega)\big]_{st}^{pq}\right) . \label{inversionsym}
\end{align}
Due to the matrix elements being real we can immediately write the orbital symmetries
\begin{align}
{\rm Im}\left(\big[\chi^0_{{\bf q}}(\omega)\big]_{st}^{pq}\right) = {\rm Im}\left(\big[\chi^0_{{\bf q}}(\omega)\big]_{pt}^{sq}\right) ~~~~~~~~ \nonumber \\
= {\rm Im}\left(\big[\chi^0_{{\bf q}}(\omega)\big]_{sq}^{pt}\right) = {\rm Im}\left(\big[\chi^0_{{\bf q}}(\omega)\big]_{pq}^{st}\right) . \label{orbitalsym}
\end{align}
Combining Eqs.\,(\ref{inversionsym}) and (\ref{orbitalsym}) leads to further simplifications on the frequency axis, {
\begin{align}
& {\rm Im}\left(\big[\chi^0_{{\bf q}}(\omega)\big]_{st}^{pq}\right) = -\pi \sum_{n,n',\mathbf{k}} a_{{\bf k}n'}^t a_{{\bf k}n'}^{q } a_{{\bf k}+{\bf q}n}^p a_{{\bf k}+{\bf q}n}^{s }  \nonumber \\
& ~~~~~~~~ \times \left[n_{\rm F}(\xi_{{\bf k}n'}) - n_{\rm F}(\xi_{{\bf k}+{\bf q}n})\right] \delta\big( \xi_{{\bf k}+{\bf q}n} - \xi_{{\bf k}n'} + \omega \big) \nonumber \\
&
 = \pi \sum_{n,n',\mathbf{k}'} a_{{\bf k}'n}^p a_{{\bf k}'n}^{s} a_{\mathbf{k}'+\mathbf{q}n'}^t a_{\mathbf{k}'+\mathbf{q}n'}^{q}  \nonumber \\
& ~~~~~~~~ \times \left[n_{\rm F}(\xi_{{\bf k}'n}) - n_{\rm F}(\xi_{\mathbf{k}'+\mathbf{q}n'})\right] \delta\big( \xi_{\mathbf{k}'+\mathbf{q}n'} - \xi_{{\bf k}'n}  - \omega \big) \nonumber \\
&
= -{\rm Im}\left(\big[\chi^0_{{\bf q}}(-\omega)\big]_{st}^{pq}\right) , \label{freqsym}
\end{align}
where we used ${\bf k} = {\bf k}' - {\bf q}$ in the first step and replaced ${\bf k}' \rightarrow {\bf k}$ in the second step. This expression simply means that the system's linear response respects causality. We hence calculate the imaginary susceptibilities only for $\omega<0$.

The delta function in Eq.\,(\ref{bubbleimag}) is approximated by a Gaussian, where we make use of an adaptive smearing method to obtain reasonable broadenings \cite{Yates2007}. Written explicitly, we approximate
\begin{align}
\delta\big( \xi_{{\bf k}+{\bf q}n'} - \xi_{{\bf k}n} + \omega \big) \simeq    \frac{1}{\sqrt{\pi}W_{\mathbf{k}}} e^{-(\xi_{{\bf k}+{\bf q}n'} - \xi_{{\bf k}n} + \omega)^2/W_{\mathbf{k}}^2} ,
\end{align}
with the broadening matrix
\begin{eqnarray}
W_{\mathbf{k}} = \alpha \cdot \left| \frac{\partial }{\partial \mathbf{k}} \big( \xi_{{\bf k}+{\bf q}n'} - \xi_{{\bf k}n}\big) \right| \Delta k ~,
\end{eqnarray}
that adapts to the electronic velocities. The parameter $\alpha$ can be chosen close to unity and $\Delta k$ is the spacing of the momentum grid \cite{Yates2007}.

As stated in the main text, the real bare susceptibility is found by using a Kramers-Kronig relation. The integration bounds of $\pm\infty$, see Eq.\,(\ref{bubblereal}), are truncated as follows:
\begin{align}
{\rm Re}\left(\big[\chi^0_{{\bf q}}(\omega)\big]_{st}^{pq}\right) = \frac{1}{\pi} \mathcal{P} \int_{-\infty}^{\infty} \frac{\omega'\mathrm{d}\omega'}{\omega'^2-\omega^2} {\rm Im}\left(\big[\chi^0_{{\bf q}}(\omega)\big]_{st}^{pq}\right) \nonumber \\
+ \frac{\omega}{\pi} \mathcal{P} \int_{-\infty}^{\infty} \underbrace{\frac{\mathrm{d}\omega'}{\omega'^2-\omega^2} {\rm Im}\left(\big[\chi^0_{{\bf q}}(\omega)\big]_{st}^{pq}\right)}_{\mathrm{odd~in~}\omega'} \nonumber \\
= \frac{1}{\pi} \mathcal{P} \int_{-\infty}^{\infty} \frac{\omega'\mathrm{d}\omega'}{\omega'^2-\omega^2} {\rm Im} \left(\big[\chi^0_{{\bf q}}(\omega)\big]_{st}^{pq}\right) \nonumber \\
= \frac{2}{\pi} \mathcal{P} \int_{-\mathcal{C}}^{0} \frac{\omega'\mathrm{d}\omega'}{\omega'^2-\omega^2} {\rm Im}\left(\big[\chi^0_{{\bf q}}(\omega)\big]_{st}^{pq}\right) .\label{susceptIntegration}
\end{align}
All momentum and orbital symmetries valid for the imaginary part translate directly to the real part. The cutoff $\mathcal{C}>0$ must be chosen such that no high-frequency information is lost. Since the delta function in Eq.\,(\ref{bubbleimag}) peaks at frequencies $\omega=\xi_{{\bf k}n}-\xi_{{\bf k}+{\bf q}n'}$ we keep all contributions by choosing $\mathcal{C}>2 \, \underset{\mathbf{k},n}{\max}\, |\xi_{\mathbf{k}n}|$. Note that $\mathcal{C}$ is not in any way related to the truncation parameter $\omega_{\mathrm{cut}}$, which we extensively use in the main text.

Let us now turn to characteristic properties of the bare susceptibility. The real part in Eq.\,(\ref{bubblereal}) can be rewritten analytically by inserting Eq.\,(\ref{bubbleimag}):
\begin{align}
&{\rm Re}\left(\big[\chi^0_{{\bf q}}(\omega)\big]_{st}^{pq}\right) = - \!\! \sum_{n,n',\mathbf{k}} a_{{\bf k}n}^s a_{{\bf k}n}^{p \,*} a_{{\bf k}+{\bf q}n'}^q a_{{\bf k}+{\bf q}n'}^{t \,*} \left[n_{\rm F}(\xi_{{\bf k}n}) \right. \nonumber \\
& ~~~ \left. - n_{\rm F}(\xi_{{\bf k}+{\bf q}n'})\right] \mathcal{P} \int_{-\infty}^{\infty} \frac{\mathrm{d}\omega'}{\omega'-\omega} \delta\big( \xi_{{\bf k}+{\bf q}n'} - \xi_{{\bf k}n} + \omega' \big) \nonumber \\
& = - \!\! \sum_{n,n',\mathbf{k}} a_{{\bf k}n}^s a_{{\bf k}n}^{p \,*} a_{{\bf k}+{\bf q}n'}^q a_{{\bf k}+{\bf q}n'}^{t \,*}     \frac{n_{\rm F}(\xi_{{\bf k}+{\bf q}n'}) - n_{\rm F}(\xi_{{\bf k}n})}{\xi_{{\bf k}+{\bf q}n'} -\xi_{{\bf k}n} +\omega+i\delta}  \,. \label{alternativeRealBubble}
\end{align}
Such a form is used e.g.\ in Ref.\,\cite{Graser2009}. It is well known that for $\mathbf{q}\rightarrow0$ and vanishing frequencies the susceptibility is equal to the density of states at the Fermi level, provided that also $T\rightarrow0$. It can be shown from Eq.\,(\ref{alternativeRealBubble}) that we recover this limit.  Considering $\chi^{0,\mathrm{stat}}_{\mathbf{q}}$ from Eq.\,(\ref{statbubble}) we find
\begin{align}
\chi^{0,\mathrm{stat}}_{\mathbf{q}} \overset{\mathbf{q}\rightarrow0}{= \! =}& -\frac{1}{2}\sum_{s} \sum_{n,n',\mathbf{k}} a_{{\bf k}n}^s \underbrace{\sum_{p,s} a_{{\bf k}n}^{p \,*} a_{{\bf k}n'}^p}_{\delta_{nn'}} a_{{\bf k}n'}^{s \,*} \nonumber \\
& \times   \frac{n_{\rm F}(\xi_{{\bf k}n'}) - n_{\rm F}(\xi_{{\bf k}n})}{\xi_{{\bf k}n'} -\xi_{{\bf k}n} + i\delta } \nonumber \\
= & -\frac{1}{2} \sum_{n,\mathbf{k}} \underset{n'\rightarrow n}{\mathrm{lim}} \frac{n_{\rm F}(\xi_{{\bf k}n'}) - n_{\rm F}(\xi_{{\bf k}n})}{\xi_{{\bf k}n'} -\xi_{{\bf k}n} + i\delta } \,.
\end{align}
At zero temperature one gets $\chi_0\equiv2\chi^{0,\mathrm{stat}}_{\mathbf{q}}\rightarrow\sum_nN_n(0)$, where $N_n(0)$ is the band-resolved density of states at the Fermi level \cite{Graser2009}. Note that we do {\em not} use $a_{\mathbf{k}n}^p\in\mathbb{R}$ in the above calculation, i.e.\ the result holds for general tight-binding models.

We plot $\chi_0$ as function of temperature in Fig.\,\ref{dosTdep} for both bulk FeSe (red line) and the monolayer case (blue line).
\begin{figure}[t!]
	\centering
	\includegraphics[width=0.9\columnwidth]{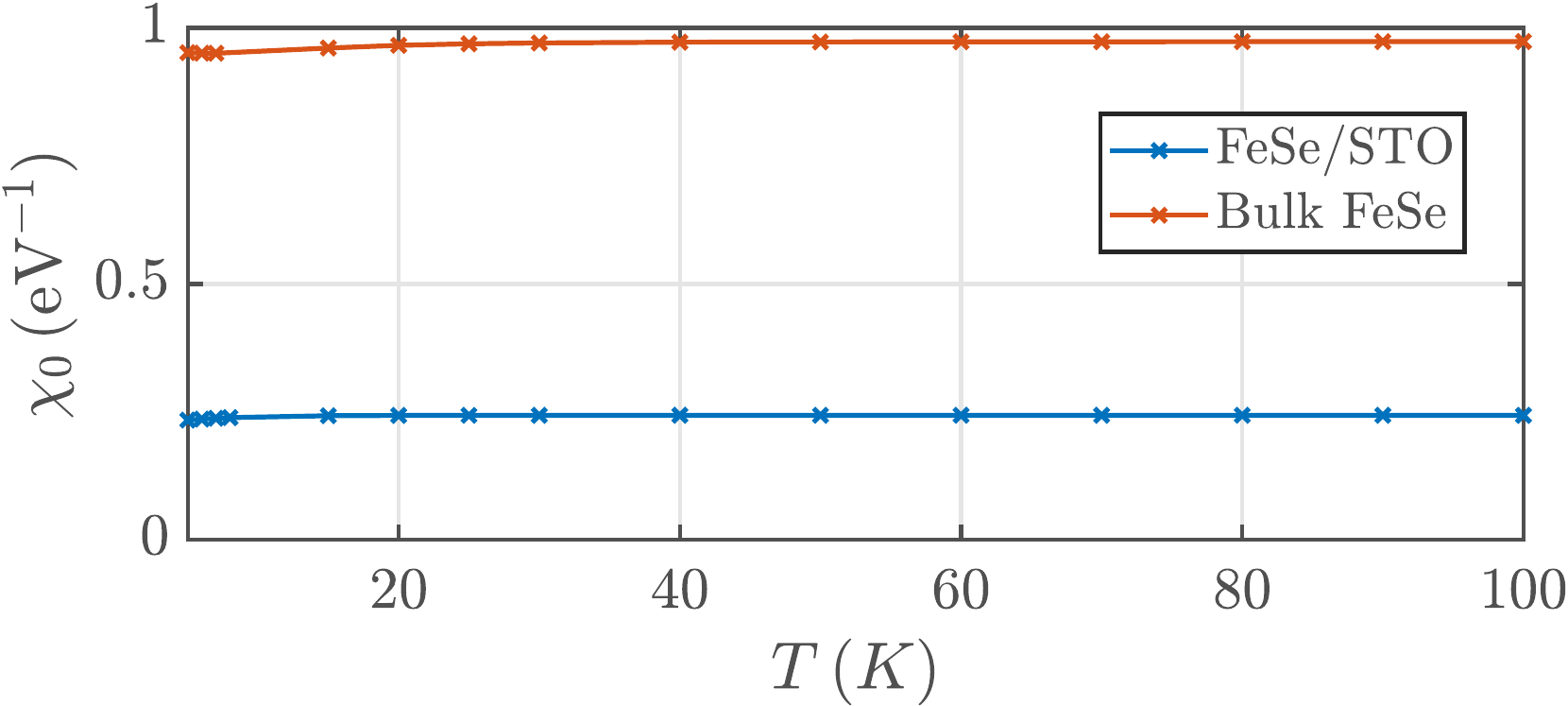}
	\caption{Temperature dependence of $\chi_0$ for FeSe/STO (blue line) and bulk FeSe (red line). The dispersions used are as introduced in the main text, from Refs.\,\cite{Hao2014} and \cite{Eschrig2009}, respectively.}
	\label{dosTdep}
\end{figure}
No significant changes for temperatures $T\in[5,100]\,\mathrm{K}$ are observed for both electronic dispersions. We hence assume that thermal broadening effects do not noticeably alter the bare susceptibility found from Eqs.\,(\ref{bubbleimag}) and (\ref{bubblereal}). Therefore, taking $T=5\,\mathrm{K}$, we calculate the susceptibility only once for given electron energies and use the result also for larger temperatures. A distinction with respect to $T$ enters therefore when transforming the interaction kernels from real to Matsubara frequencies, Eq.\,(\ref{bandkernTrunc}).

For RPA susceptibilities we keep only the imaginary parts from solutions to Eqs.\,(\ref{rpaspin}) and (\ref{rpacharge}), since the real parts are not needed for subsequent calculations. The momentum and frequency symmetries are similar to the imaginary bare susceptibility. Mapping the four-rank tensors to simple matrices, as mentioned in the main text, is a two-step procedure, which we explain here briefly by using $\big[U^S\big]^{pq}_{st}$ as an example. First, the orbital indices within the tensors must be rearranged according to $\big[U^S\big]^{pq}_{st}\rightarrow\big[U^S\big]^{qp}_{ts}$. Note, that this transformation is nontrivial since it does not fall into any orbital symmetry of the bare susceptibility or the Stoner tensors, compare Eqs.\,(\ref{stonerparam}) and (\ref{orbitalsym}). The second step is the actual mapping, which is not uniquely defined. In our implementation we use
\begin{align}
&\forall(i,j)\in[1,L_o^2]: ~ \hat{U}^S_{ij} = \big[U^S\big]^{qp}_{ts} ,
\end{align}
where $L_o$ is the number of orbitals and
\begin{align}
&q=\lceil i/L_o\rceil ~,~ p=1+(i-1)\mod L_o \, ,\nonumber \\
&t=\lceil j/L_o\rceil ~,~ s=1+(j-1)\mod L_o  \, .
\end{align}
Note that this mapping is unidirectional, i.e.\ from knowing $p,q,s,t$ one cannot solve for the associated $i,j$ analytically. One can, however, at any time convert the matrices back to four-rank tensors numerically, so the mapping can be considered invertible.

For implementing Eqs.\,(\ref{orbitalkernP}) and (\ref{orbitalkernM}) we employ again the tensor-matrix mapping as discussed above. The Coulomb terms are present only for $\big[V^{(-)}_{{\bf q}}(\omega)\big]^{pq}_{st}$, since this is the kernel used for off-diagonal Green's function elements. As is common practice we assume the Coulomb interaction to be taken into account in the tight-binding model, which is why no additional terms appear in the coupling used for diagonal elements of the Green's function. In real-frequency space the terms $U^S/2$ and $U^C/2$ in Eq.\,(\ref{orbitalkernM}) are needed for double-counting. As we show below, no such contributions enter the Matsubara space calculation. Inserting the Coulomb terms into Eq.\,(\ref{bandkern}) and using $ a_{{\bf k}n}^{p}\in\mathbb{R}$ gives
\begin{align}
&\frac{1}{2}\sum_{\mathbf{k}}\sum_{stpq} a_{{\bf k}n}^{t \,*} a_{{\bf k}n}^{s \,*} \big[U^S + U^C\big]^{tq}_{ps} a^p_{{\bf k}-\mathbf{q}n'}a^q_{{\bf k}-\mathbf{q}n'} 
= \nonumber \\
& ~~~\frac{1}{2}\sum_{\mathbf{k}}\sum_{stpq} a_{{\bf k}n}^{t} a_{{\bf k}n}^{s} \big[U^S + U^C\big]^{tq}_{ps} a^p_{{\bf k}-\mathbf{q}n'}a^q_{{\bf k}-\mathbf{q}n'} ~.
\end{align}
In Eq.\,(\ref{matsubarakern}) we need to explicitly take the imaginary part of the above, which is identically zero; hence the Coulomb terms do not enter into the kernels used for the Eliashberg equations.

The integral in Eq.\,(\ref{matsubarakern}) is treated similarly as in Eq.\,(\ref{susceptIntegration}). We take only the negative frequency axis into account and set the lower integration bound to $-\mathcal{C}<-2\, \underset{\mathbf{k},n}{\max} \, |\xi_{\mathbf{k}n}|$:
\begin{align}
&{\rm Re}\left(\big[V^{(\pm)}_{{\bf q}}(iq_m)\big]_{nn'}\right) = \frac{1}{\pi}\mathcal{P} \int_{-\infty}^{\infty}\mathrm{d}\omega \left( \frac{\omega}{\omega^2+q_m^2} \right. \nonumber\\
& ~~~~~~~~~~~~~~~~~~~~~~~~~~~~~~\left. + \frac{iq_m}{\omega^2+q_m^2} \right) {\rm Im} \left( \big[V^{(\pm)}_{{\bf q}}(\omega)\big]_{nn'} \right)\nonumber\\
& ~~~~~~~= \frac{2}{\pi}\mathcal{P} \int_{-\mathcal{C}}^{0}\mathrm{d}\omega \frac{\omega}{\omega^2+q_m^2} {\rm Im}\left( \big[V^{(\pm)}_{{\bf q}}(\omega)\big]_{nn'} \right) .
\end{align}
When calculating Eq.\,(\ref{bandkernTrunc}) we replace $\mathcal{C}$ by the truncation cutoff $\omega_{\mathrm{cut}}$ in the above.

The kernels in Matsubara space are even in $q_m$ and, as all quantities considered in this work except the matrix elements, tetragonal in $\mathbf{q}$. Additionally, ${\rm Re}\left(\big[V^{(\pm)}_{{\bf q}}(iq_m)\big]_{nn'}\right)$ is invariant under exchanging the band indices, which is a symmetry directly translated from the real-frequency kernel. Below we show that the latter is invariant under exchanging $n\leftrightarrow n'$:
\begin{align}
&\big[V^{(\pm)}_{{\bf q}}(\omega)\big]_{n'n} = \sum_{\mathbf{k}}\sum_{stpq} a_{{\bf k}n'}^{t \,*} a_{{\bf k}n'}^{s \,*}   \big[V^{(\pm)}_{{\bf q}}(\omega)\big]^{pq}_{st}  a^p_{{\bf k}-\mathbf{q}n}a^q_{{\bf k}-\mathbf{q}n}  \nonumber \\
& ~~~~~ 
= \sum_{\mathbf{k}}\sum_{stpq} a^{p \,*}_{{\bf k}-\mathbf{q}n}a^{q \,*}_{{\bf k}-\mathbf{q}n}  \big[V^{(\pm)}_{{\bf q}}(\omega)\big]^{pq}_{st}  a_{{\bf k}n'}^{t} a_{{\bf k}n'}^{s} \nonumber\\
& ~~~~~ 
=  \sum_{\mathbf{k}'}\sum_{stpq} a^{p \,*}_{{\bf k}'n}a^{q \,*}_{{\bf k}'n}   \big[V^{(\pm)}_{{\bf q}}(\omega)\big]^{pq}_{st}  a_{\mathbf{k}'+\mathbf{q}n'}^{t} a_{\mathbf{k}'+\mathbf{q}n'}^{s} ,
\end{align}
where we used $a_{{\bf k}n}^{p}\in\mathbb{R}$ and $\mathbf{k}=\mathbf{k}'+\mathbf{q}$.
Due to inversion symmetry of $\big[V^{(\pm)}_{{\bf q}}(\omega)\big]_{n'n}$ and $\big[V^{(\pm)}_{{\bf q}}(\omega)\big]^{pq}_{st}$ we can write
\begin{align}
\big[V^{(\pm)}_{{\bf q}}(\omega)\big]_{n'n} = \sum_{\mathbf{k}}\sum_{stpq} a^{q \,*}_{{\bf k}n}a^{p \,*}_{{\bf k}n}  \big[V^{(\pm)}_{{\bf q}}(\omega)\big]^{pq}_{st}   a_{\mathbf{k}-\mathbf{q}n'}^{s} a_{\mathbf{k}-\mathbf{q}n'}^{t}  ,
\end{align}
where we also rename $\mathbf{k}'$ to $\mathbf{k}$. Reshuffling the orbital dummy indices then gives
\begin{align}
\big[V^{(\pm)}_{{\bf q}}(\omega)\big]_{n'n} =& \sum_{\mathbf{k}}\sum_{stpq} a^{t \,*}_{{\bf k}n}a^{s \,*}_{{\bf k}n}   \big[V^{(\pm)}_{{\bf q}}(\omega)\big]^{st}_{pq}   a_{\mathbf{k}-\mathbf{q}n'}^{p} a_{\mathbf{k}-\mathbf{q}n'}^{q} \nonumber \\
= & \big[V^{(\pm)}_{{\bf q}}(\omega)\big]_{nn'} .
\end{align}

The Eliashberg Eqs.\,(\ref{z})-(\ref{phi}) are solved iteratively with a convergence criterion of $10^{-9}$ as threshold for the maximal absolute change in all three functions. Always taking more than 2000 points on the Matsubara axis we are confident to be converged in the number of frequencies. Note that a much larger number is partially employed (and needed), depending on the cutoff used for calculating the interaction kernels, see Secs.\,\ref{scCoupling} and \ref{scBasicProps}. In the calculations presented we adjust the number of frequencies as required depending on $\omega_{\mathrm{cut}}$. For increasing the numerical performance we employ a Fourier convolution scheme in momenta and frequencies. Since the BZ symmetry of the order parameter is \textit{a priori} not known one needs to be careful with the initialization before starting the iterative loop. If the initial guess does not obey the favored symmetry the algorithm might not converge or give a zero-solution. For all results presented in the main text we tested several alternative form factors as initial guess and did not find any nonzero solutions different from the ones presented. This shows that the symmetries discussed in the main text are the only possible scenarios for the gap within the associated setups.

\section{Susceptibilities of FeSe/STO}\label{appSuscept}

In Fig.\,\ref{ML_hao_setup}, panels (c) and (f), of the main text we show  the dynamic bare and spin susceptibilities respectively. As is apparent from this graph, dominant contributions of both quantities appear at wave vector $\mathbf{q}=M$. Due to the magnitudes of $\chi_{\mathbf{q}}^{0,\mathrm{dyn}}$ and $\chi_{\mathbf{q}}^{S,\mathrm{dyn}}$ in close vicinity of this momentum, the low frequency region is not resolved very clearly. For this purpose we re-plot both functions in Fig.\,\ref{zoomedSuscept}, zoomed into the low energy regime and with modified contrast. In this way the distinct branches below the onset of the Stoner continuum are better visible.

\begin{figure}[h!]
	\centering
	\includegraphics[width=1\columnwidth]{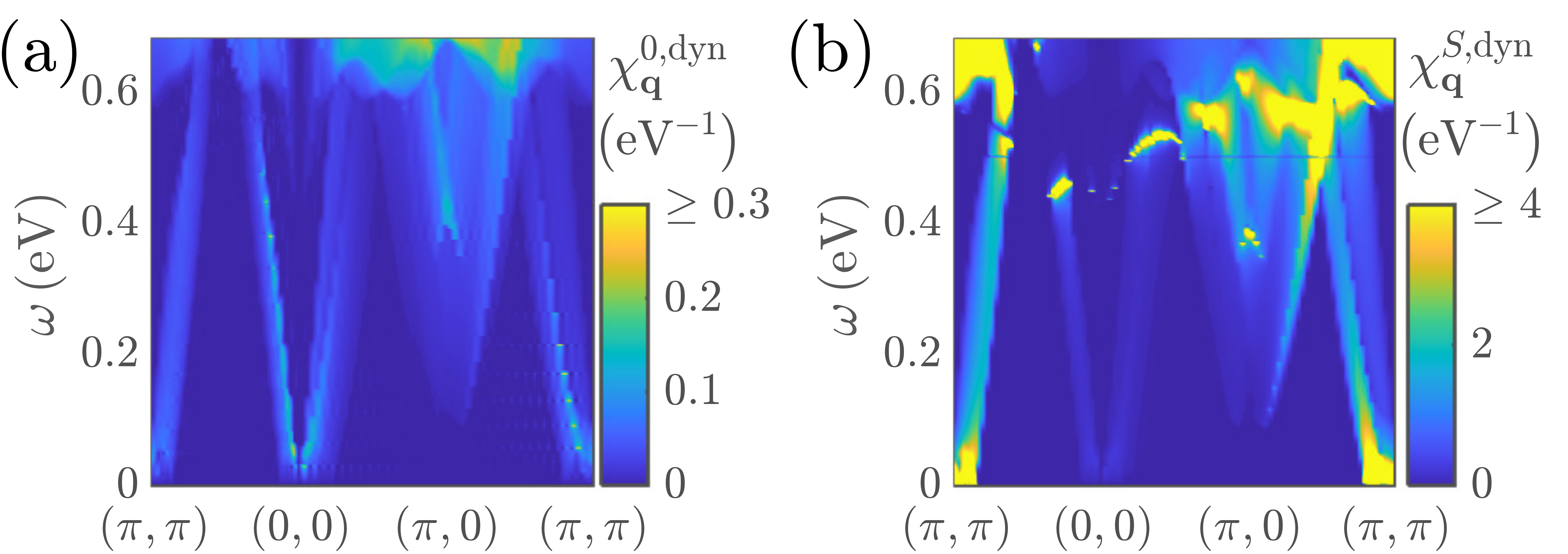}
	\caption{Zoom into the low energy region for susceptibilities of FeSe/STO. (a) Dynamic bare susceptibility as calculated from Eq.\,(\ref{dynbubble}). (b) Spin susceptibility computed from Eq.\,(\ref{dynamicSpin}) as function of frequencies and momenta.}
	\label{zoomedSuscept}
\end{figure}

\section{Modifying the dispersion for FeSe/STO}\label{appModifyTB}

In Section \ref{scInfluenceTB} we show effects on our results for FeSe/STO when changing either the size of the FS pockets or the distance between hole and electron bands. The former is achieved by introducing a rigid chemical potential $\mu$, such that $\xi_{\mathbf{k}n}\rightarrow\xi_{\mathbf{k}n}-\mu$. This operation commutes with all constituents of the Hamiltonian, hence the matrix elements are not affected. On the other hand, when shifting only the hole bands by a nonrigid $\delta\mu$ we need to recalculate $a_{\mathbf{k}n}^p$. Consider the initial kinetic term $\hat{H}_0$, which is diagonalized as $\hat{H}_0 = \hat{a} \hat{\xi} \hat{a}^{\dagger}$, or $\hat{a}^{\dagger} \hat{H}_0 \hat{a} =  \hat{\xi}$. For simplicity we omit the momentum dependence here and write the eigenvalues and eigenvectors in matrix notation. Now we add a selective shift on both sides, which only affects the hole bands:
\begin{align}
\hat{a}^{\dagger}\hat{H}_0\hat{a} - \delta\mu \begin{pmatrix}
\hat{1}_h & \\ & \hat{0}_e
\end{pmatrix} = \hat{\xi}-\delta\mu \begin{pmatrix}
\hat{1}_h & \\ & \hat{0}_e
\end{pmatrix} . \label{partialshift}
\end{align}
In Eq.\,(\ref{partialshift}) we take the band ordering such that first all hole bands, and then all electron bands are listed. This is indicated by subscripts $h$ and $e$, denoting the respective subspaces. On the right hand side we get a modified dispersion $\hat{\xi}'$, which are the eigenvalues of a Hamiltonian $\hat{H}_0'\neq\hat{H}_0$. To find the corresponding eigenvectors we calculate the new Hamiltonian as
\begin{align}
\hat{H}_0' = \hat{H}_0 - \hat{a} \delta\mu \begin{pmatrix}
\hat{1}_h & \\ & \hat{0}_e
\end{pmatrix} \hat{a}^{\dagger} = \hat{a}' \hat{\xi}' \hat{a}'^{\dagger} ~, \label{hamtilde}
\end{align}
which leads to the modified eigenvectors $\hat{a}'$.

\begin{figure}[tb!]
	\centering
	\includegraphics[width=1\columnwidth]{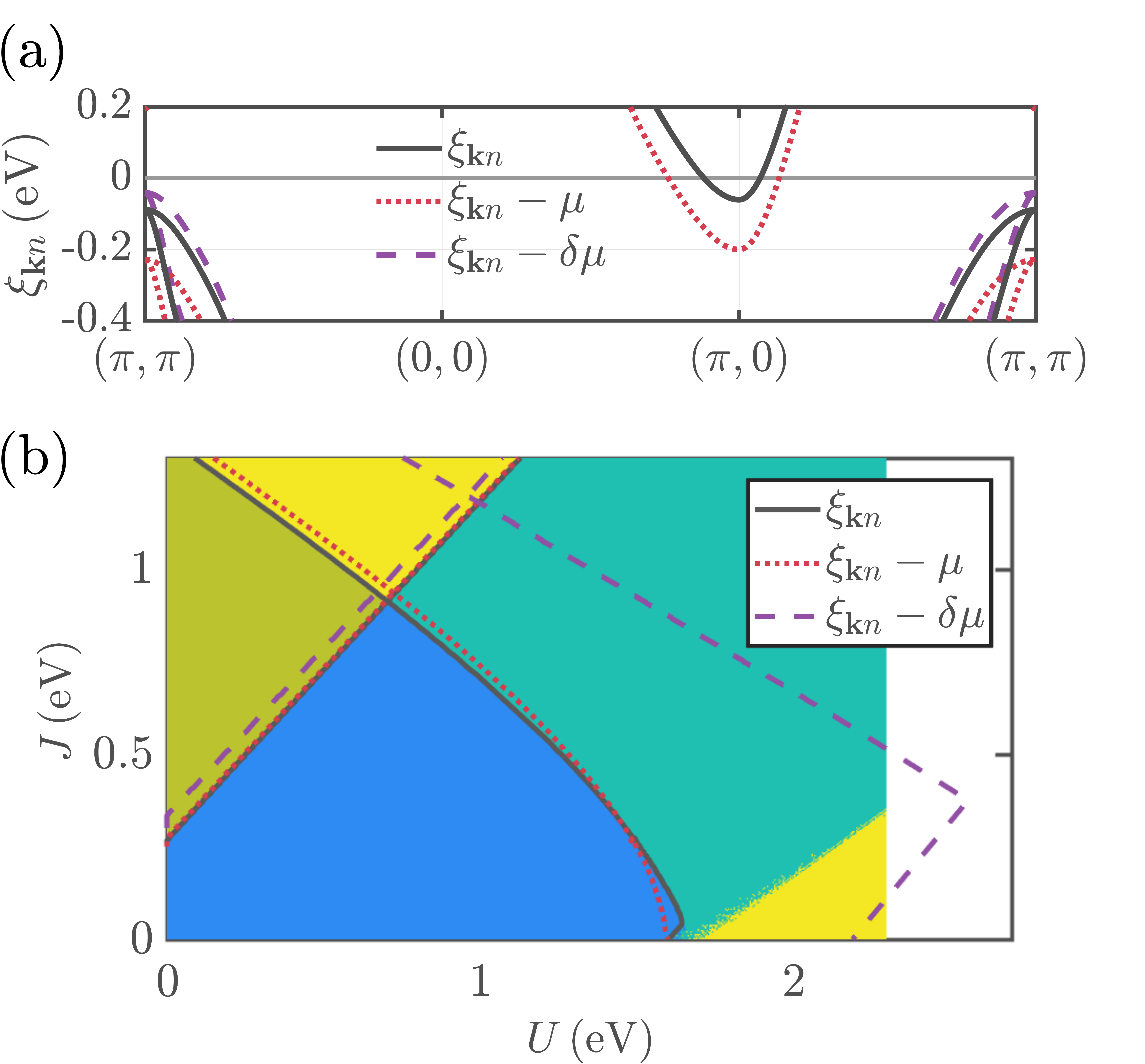}
	\caption{(a) Electronic energies along high symmetry lines of the unfolded BZ. The initial dispersions $\xi_{\mathbf{k}n}$ as used in Sec.\,\ref{scMonolayerSupercond} are shown as solid black curves. Shifting the bands rigidly to lower energies by $\mu=140\,\mathrm{meV}$ results in the red dotted lines. The dashed purple lines show the effect of bringing the hole bands closer to the Fermi level by $\delta\mu=-48\,\mathrm{meV}$. (b) Phase diagram of allowed values for $U$ and $J$, replotted from Fig.\,\ref{ML_hao_setup}(e). The red dotted (purple dashed) boundaries correspond to artificial modifications of the electronic dispersions, as indicated in panel (a).}
	\label{modifiedTB}
\end{figure} 

To summarize, we realize a relative shift of the hole bands by diagonalizing $\hat{H}_0$, which gives the matrix elements $\hat{a}$. These are used to calculate $\hat{H}_0'$ from Eq.\,(\ref{hamtilde}), which again needs to be made diagonal to have access to the desired dispersion $\hat{\xi}'$ and the associated $\hat{a}'$.

The discussions in Sec.\,\ref{scInfluenceTB} are made for energies as plotted in Fig.\,\ref{modifiedTB}(a). Starting from the initial $\xi_{\mathbf{k}n}$ shown in solid black, we modify in two different ways. The influence of the FS pocket size is studied by rigidly shifting the energies as $\xi_{\mathbf{k}n}-\mu$, shown in dotted red, with $\mu=140\,\mathrm{meV}$. To study changes with the distance between electron and hole bands we use $\delta\mu=-48\,\mathrm{meV}$, shown in dashed purple. We choose these two modifications of the tight-binding model, since the effects we want to study are perfectly decoupled in this way.

Both rigid and nonrigid shifts produce changes in the bare susceptibilities due to the evaluation of Fermi-Dirac functions in Eq.\,(\ref{bubbleimag}). This in turn leads to altered boundaries in the $(U,J)$ phase diagram, as we sketch in Fig.\,\ref{modifiedTB}(b). As seen from the red and dotted curves, a rigid shift by $\mu$ introduces slight changes in the boundary for magnetic order, but the effect is rather minor. Contrarily, bringing electron and hole bands closer together allows for significantly larger values of $U$ and $J$ as can be seen from the purple dashed lines. When performing our selfconsistent calculations in $(U,\omega_{\mathrm{cut}})$ space, Fig.\,\ref{ML_rigid_phasediag}, we adjust our parameter choices according to the changed boundaries of Fig.\,\ref{modifiedTB}(b).

\end{document}